\documentclass[12pt]{article}
\usepackage{amsmath}
\usepackage{graphicx,psfrag,epsf}
\usepackage{enumerate}
\usepackage{natbib}
\usepackage{url} 

\newcommand{\blind}{0}

\usepackage{amsmath,amssymb}
\usepackage{amsthm}
\usepackage{multirow}
\usepackage{natbib}
\usepackage{bm}
\usepackage{hyperref}
\usepackage[all]{xy}
\usepackage{graphicx,caption, subcaption}
\usepackage{float}
\usepackage{color}

\newtheorem{thm}{Theorem}[section] 
\newtheorem{lemma}{Lemma}[section] 

\newtheorem{definition}{Definition}[section]
\newcommand{\bed}{\begin{definition}}
\newcommand{\eed}{\end{definition}}

\newcommand{\rom}[1]{\uppercase\expandafter{\romannumeral #1\relax}}

\newcommand{\eps}{\epsilon}
\usepackage{bbm}

\newcommand{\bitem}{\begin{itemize}}
\newcommand{\eitem}{\end{itemize}}

\newcommand{\goto}{\rightarrow}

\newcommand{\beqn}{\begin{equation}}
\newcommand{\eeqn}{\end{equation}}
\newcommand{\balign}{\begin{align}}
\newcommand{\ealign}{\end{align}}

\newcommand{\tr}{\mathrm{tr}}
\usepackage{amssymb}
\newcommand{\beq}{\begin{equation}}
\newcommand{\eeq}{\end{equation}}

\newcommand{\diag}{\mathrm{diag}}

\allowdisplaybreaks

\usepackage{enumitem}
\usepackage{makecell}

\usepackage{listings}
\begin{document}
	
\def\spacingset#1{\renewcommand{\baselinestretch}%
	{#1}\small\normalsize} \spacingset{1}


\if0\blind
{
  \title{\bf Optimal Network Pairwise Comparison}
  \author{Jiashun Jin\thanks{
    The authors gratefully acknowledge \textit{the support of NSF grants DMS-1943902 and DMS-2015469.}}\hspace{.2cm}\\
    Department of Statistics, Carnegie Mellon University\\
    and \\
    Zheng Tracy Ke\\
    Department of Statistics, Harvard University\\
    and\\
    Shengming Luo \\
    Department of Statistics, Carnegie Mellon University\\
    and\\
    Yucong Ma \\
    Department of Statistics, Harvard University
    }
  \maketitle
} \fi

\if1\blind
{
	\bigskip
	\bigskip
	\bigskip
	\begin{center}
		{\LARGE\bf  Optimal Network Pairwise Comparison}
	\end{center}
	\medskip
} \fi

\bigskip
\begin{abstract} 
We are interested in the problem of two-sample network hypothesis testing: given two networks with the same set of nodes, we wish to  test  whether the underlying Bernoulli probability matrices of the two networks are the same or not. We propose Interlacing Balance Measure (IBM)  as a new two-sample testing approach.  We consider the {\it Degree-Corrected Mixed-Membership (DCMM)} model for undirected networks, where we allow severe degree heterogeneity, mixed-memberships, 
flexible sparsity levels, and weak signals.  In such a broad setting, 
how to find a test that has a tractable limiting null and 
optimal testing performances is a challenging problem. 
We show that IBM is such a test:  in a broad  DCMM setting with only mild regularity conditions,  
IBM has  $N(0,1)$ as the limiting null  and achieves the optimal 
phase transition.  

While the above  is for undirected networks, IBM is a unified approach and is directly implementable for directed networks.   
For a broad directed-DCMM (extension of DCMM for directed networks) setting, we show that 
IBM has $N(0, 1/2)$ as the limiting null  and continues to achieve the optimal phase transition. 
We have also applied  IBM to the Enron email network and a 
gene co-expression network, with interesting results. 
\end{abstract}

\noindent%

\noindent%


\tableofcontents

\spacingset{1.5} 

\section{Introduction} \label{sec:intro}  
We are interested in the problem of pairwise network comparison or two-sample network hypothesis testing: given two independent  (directed or undirected) networks for the same set of $n$ nodes, how to test 
whether the underlying network structures are the same.  

The problem is of interest both in theory and in practice, with applications 
in network analysis, neuroscience, cancer research, and case-control studies, among others.   
In dynamic network analysis \citep{Liu927} and 
network change-point detection \citep{Yao2021}, we frequently need to test 
whether the underlying structures of two networks constructed using two disjoint time intervals 
are the same.  In cancer research and case control studies,  it is 
conventional to use the gene co-expression data to construct two binary networks, one for the 
control group and the other for the diseased group, and it is of interest to 
test whether the underlying structures of the two networks are the same \citep{segerstolpe2016single}.  
In neuroscience,  how to test the similarity of two brain graphs is an active 
research topic in the interdisciplinary area of neuroscience, statistics, and 
machine learning; see \cite{tang2017b} for application examples on the topics of neurowiring and 
neuroimaging.

Real social networks frequently have severe degree heterogeneity (i.e., a degree of one node is much 
higher than that of another) and mixed-memberships (i.e., some nodes have nonzero weights in more than $1$ community;  communities are tightly woven clusters of nodes where we have more edges within than between \citep{GirvanNewman}).  Also, the overall sparsity levels may vary significantly from one network to another. 

We are interested in both directed and undirected networks.  
Consider directed networks first.  To capture the above features, 
we adopt the {\it directed Degree-Corrected Mixed-Membership (directed-DCMM}) model.  
For simplicity, we use the terminology of citation networks, but the model is valid for general directed networks. 
Consider a citation network with $n$ nodes.  For two authors $i$ and $j$, in the occurrence of $i$ citing $j$, we say that {\it $i$ is a  citer and $j$ is a citee}.  
Let $A$ be the adjacency matrix of the network, where 
\begin{equation} \label{model1a} 
\mbox{$A_{ij}  = 1$ if node $i$ has ever cited node $j$ and $A_{ij} = 0$ otherwise}. 
\end{equation} 
Conventionally,  we do not count self citations, so all the diagonal entries of $A$ are $0$.   In directed-DCMM, we assume that  there are $K$ perceivable communities ${\cal C}_1, {\cal C}_2, \ldots, {\cal C}_K$, each of which can be thought as a research area  (e.g., ``Bayes", ``Variable Selection").    
For each $1 \leq i \leq n$, we  let $\theta_i$ and $\zeta_i$ be two positive parameters that model the degree heterogeneity of  node $i$ as a citer and as a citee, respectively. Also, we assume that node $i$ is associated with  two $K$-dimensional mixed-membership weight vectors 
$\pi_i = (\pi_i(1),\pi_i(2),\ldots,\pi_i(K))'$ and $\gamma_i = (\gamma_i(1), \gamma_i(2), \ldots, \gamma_i(K))'$, 
where 
$\pi_i(k)$ and $\gamma_i(k)$ are the weights that node $i$ puts in community $k$ as a citer and as a citee, respectively, $1 \leq k \leq K$. Moreover,  for a $K \times K$ nonnegative matrix $P$ which models the community structures,   
we assume that all off-diagonal entries of $A$ are independent Bernoulli variables satisfying 
\begin{equation} \label{model1b}  
\mathbb{P}(A_{ij}=1) = \theta_i \zeta_j \cdot \pi_i' P \gamma_j  = \sum_{k, \ell = 1}^K \theta_i \zeta_j  
\pi_i(k) P(k, \ell) \gamma_j(\ell), \;\;\;   1 \leq i < j \leq n.
\end{equation}
Let $W \in \mathbb{R}^{n,n}$ be the matrix such that 
$W = A - \mathbb{E}[A]$. Write $\theta=(\theta_1,\theta_2, \ldots,\theta_n)'$, $\zeta=(\zeta_1,\zeta_2, \ldots,\zeta_n)'$,  $\Theta = \diag(\theta_1, \theta_2, \ldots, \theta_n)$,  $Z= \diag(\zeta_1, \zeta_2, \ldots, \zeta_n)$, 
$\Pi = [\pi_1, \pi_2, \ldots, \pi_n]'$,  and $\Gamma  = [\gamma_1, \gamma_2, \ldots, \gamma_n]'$. 
With these notations, we can rewrite Model (\ref{model1b}) as 
\begin{equation} \label{model1c}  
A = \Omega - \diag(\Omega) + W, \qquad \mbox{and}   \qquad \Omega = \Theta \Pi P \Gamma'  Z,  
\end{equation} 
where for short, $\diag(\Omega)$ is the diagonal matrix $\diag(\Omega_{11}, \Omega_{22}, \ldots, \Omega_{nn})$.    
Model (\ref{model1b}) is the directed-DCMM model and (\ref{model1c}) is its equivalent matrix form.

Next, we consider undirected networks. Similarly, let $A$ be the adjacency matrix of 
an undirected network, where $A_{ij}  = 1$ if $i \neq j$ and there is an undirected edge between nodes $i$ and $j$, and $A_{ij}  = 0$ otherwise (similarly, all diagonal entries of $A$ are $0$). 
Undirected networks can be viewed as a special 
case of directed networks, where the matrix $A$ is required to be symmetric. 
Therefore, we can use the same modeling strategy, but we must have $P = P'$, $\Theta = Z$, and  $\Pi = \Gamma$.  In this special case, directed-DCMM reduces to undirected-DCMM or DCMM for short, 
where we assume the upper triangular entries of $A$ are independent Bernoulli variables satisfying (note that $\Theta, \Pi$ are as above and $P$ is symmetric) 
\begin{equation} \label{model2} 
A =  \Omega - \diag(\Omega) + W, \qquad   \Omega = \Theta \Pi P \Pi'  \Theta,  \qquad \mbox{and} \qquad   W = A - \mathbb{E}[A].    
\end{equation}  
\bed 
We call Model (\ref{model1a})-(\ref{model1c}) the directed-DCMM for directed networks, and Model (\ref{model2}) the  DCMM for undirected networks. In both models, we call $\Omega$ the Bernoulli probability matrix. 
\eed 

For both directed-DCMM and DCMM, we need a mild identifiability condition;  
 see Section \ref{sec:main}. For analysis, we use $n$ as the driving asymptotic parameter, and allow $(\Theta, \Pi, P, \Gamma, Z)$ to vary with $n$ and so to accommodate severe degree heterogeneity, 
mixed-memberships, flexible sparsity levels, and weak signals. 
See Section \ref{sec:main} for more discussions.  

DCMM was proposed earlier (e.g., see \cite{Mixed-SCORE}).  
DCMM  includes the Degree Corrected Block model (DCBM) \citep{DCBM}, 
Mixed-Membership Stochastic Block Model (MMSBM) \citep{airoldi2009mixed}, and 
Stochastic Block Model (SBM) \citep{holland1983stochastic} as special cases. In DCBM, 
we do not allow mixed-memberships so that all weight vectors $\pi_i$ are degenerate (i.e., 
only one nonzero entry, which is $1$). In MMSBM, we do not model degree heterogeneity, with $\theta_1 = \theta_2 = \ldots = \theta_n$. In SBM, we do not allow either mixed-memberships  or 
degree heterogeneity. Directed-DCMM can also be viewed as an extension of 
directed-DCBM \citep{SCC-JiJin, DCMM-Ji}.

Now, consider two independent networks for the same set of $n$ nodes. 
Let $A$ and $\widetilde{A}$ be the two adjacency matrices, 
and let $\Omega$ and $\widetilde{\Omega}$ be the two 
Bernoulli probability matrices, respectively. Assume either that 
both $A$ and $\widetilde{A}$ satisfy the DCMM model or that both $A$ and $\widetilde{A}$ satisfy 
the directed-DCMM model.  The problem of pairwise comparison is to test 
\begin{equation} \label{CP1a} 
H_0: \Omega  = \widetilde{\Omega}, \qquad \mbox{vs}. \qquad  H_1:  \Omega    \neq \widetilde{\Omega}. 
\end{equation} 
Our primary interest is to find  non-parametric pairwise comparison approaches  that 
(1). work for a broad setting with only mild conditions where we allow 
 severe degree heterogeneity, mixed-memberships,  flexible sparsity levels, and weak signals,  
 (2).  have a tractable null distribution (so a testing $p$-value can be easily computed) and are optimal in testing power, and (3).   are directly implementable for both directed  and undirected networks. 
%
We assess the optimality by phase transition. Phase transition is a well-known theoretical framework for assessing optimality \citep{DJ15}. It is closely related to the classical minimax framework, 
but may offer additional insight in many cases.

\subsection{Literature review and our contributions} \label{subsec:review} 
Spectral approach is an interesting approach to network pairwise comparison. 
\cite{ghosh2019} studied a two-sample testing problem for 
inhomogeneous random graphs and proposed an interesting spectral approach. 
The paper considered a setting 
where we have two undirected random graph models, ${\cal P}$ and ${\cal Q}$, 
and for each model, we observe $m$ independently  
realized networks. The goal is to test whether 
${\cal P} = {\cal Q}$ or not.   Translated to our setting ($m=1$), their approach uses $\|A - \widetilde{A}\|$ as the test statistic for pairwise comparison, 
where $A$ and $\widetilde{A}$ are adjacency matrices of two independent 
networks.  The main challenges of this approach are two fold. First,  the null distribution 
of $\|A - \widetilde{A}\|$ depends on unknown parameters in $\Omega$ and $\widetilde{\Omega}$ 
and is hard to derive, even with the most recent techniques in Random Matrix Theory.  
Second,  the test statistic aims to estimate $\|\Omega - \widetilde{\Omega}\|$ (which is 
$0$ if and only if the null is true) but such an estimate is likely to be inconsistent in the presence 
of severe degree heterogeneity,  where $n \theta_{max} / \|\theta\|_1$ and 
$n \tilde{\theta}_{max} / \|\tilde{\theta}\|_1$ may tend to $\infty$ rapidly (e.g., \cite{bandeira2016sharp,fan2020asymptotic}).    Here,  
$\theta_{max} = \max\{\theta_1, \theta_2, \ldots, \theta_n\}$, $\tilde{\theta}_{max} =\max \{\tilde{\theta}_1, \tilde{\theta}_2, \ldots, \tilde{\theta}_n\}$ and $\|\cdot\|_1$ denotes the $\ell_1$-norm.   
Note also that the main focus of  \cite{ghosh2019} 
is on information lower bound. Therefore, their settings and primary focuses 
are quite different from ours.

Network comparison is related to the problem of two-sample latent space testing. 
Take the DCMM setting for example.  Viewing the rows of the mixed-membership 
matrix as latent variables,  the two-sample latent space 
testing is to test whether the mixed-membership matrices associated with two 
networks are the same.  
  Among the recent works on two-sample latent space testing,  
 \cite{tang2017a} considered the problem of testing whether two independent finite-dimensional random dot product graphs have the same generating latent positions (up to a rotation), and proposed an interesting eigen-space based 
approach.  The approach was further adapted to a weighted SBM setting by 
\cite{TSTLi2018}, where the limiting null distribution of the testing statistic 
was derived by moment matching methods, assuming the numbers of communities is known.  
Despite the interesting results in this paper, the study was focused on the more specific SBM  
models of undirected networks. For the more general directed-DCMM and DCMM settings where we allow severe degree heterogeneity and mixed memberships, both the limiting null distribution and 
the power of the approach remain  unknown. Note also the problem of two-sample latent space testing  
is different from our problem of two-sample network testing.  For these reasons, it is unclear how to adapt their approach  to our settings.  

Two-sample network testing is also related to one-sample global testing. Given a network with $K$ communities, 
the goal of global testing is to test whether $K = 1$ or $K > 1$. Seemingly, the problem is 
quite different from ours. 
Among existing works on one-sample global testing, \cite{Ery3} studied the problem of testing whether 
a graph is Erd\"os-Renyi or has an unusually dense subgraph,  \cite{bubeck2016testing} and \cite{banerjee2016contiguity} 
studied the problem for the more specific SBM setting where they proposed a signed-triangle approach. 
The approach was further extended by \cite{SP2019} to the much broader DCMM settings. 
See also \cite{Yuan2018TestHyper} for hypergraph global testing. 
Despite the interesting progress in these works, 
the problem of one-sample global testing is different from 
the problem of two-sample pairwise comparison considered here, so it remains unclear how to adapt those ideas to our settings.

Network comparison is also related to the change-point detection in dynamic networks \citep{wang2018optimal, Liu927, Yao2021}), but the goal and settings of these papers are different from ours,  and it is unclear how to extend their ideas to our setting.

It is a non-trivial task to find an optimal two-sample testing procedure that 
works for both DCMM and directed-DCMM models under only mild regularity conditions (where we allow severe degree heterogeneity, 
mixed memberships, flexible sparsity levels and weak signals).  
There are many challenges. To name a few:    
(1). For many test statistics, the limiting null distribution may depend on the (large number of) unknown parameters in a complicated way and is not tractable, (2). A test  statistic may  work well in some 
settings (e.g., networks that are very sparse or without severe degree heterogeneity) but 
lack power in others, (3). To find a test that achieves the optimal phase transition, we need to find 
an upper bound and a matching lower bound. 
This is non-trivial even for 
narrower settings.


In this paper, we propose {\it Interlacing Balance Measure (IBM)} as 
a new approach to pairwise comparison.  Let $A$ and $\widetilde{A}$ 
be the adjacency matrices of two independent networks  (either both are directed or both are undirected)
under consideration, and let $A^* = A - \widetilde{A}$. 
We recognize that  the positive and negative entries in  $A^*$ should be {\it balanced}  in some sense when the null is true, 
and {\it unbalanced} otherwise. Therefore, an appropriately designed balance 
measure will have power for differentiating an alternative from a null. 

We explain how IBM overcomes the challenges above. 
First,  we design IBM in a way so that its mean is $0$ when the null is true and is strictly positive 
otherwise. Moreover, we find a convenient estimator for the variance of IBM, which is uniformly consistent 
for all directed-DCMM and DCMM settings considered here, with only mild regularity conditions.   Using this estimator to standardize  IBM and denoting 
the resultant testing statistic by $\psi_n$, we show that 
for all directed-DCMM and DCMM settings considered here, 
$\psi_n \goto N(0, a)$, where $a=1$ for DCMM and $a=1/2$ for directed-DCMM.  
This way, we have derived an explicit limiting null and so have overcome the first challenge. 
 
For the second challenge,  we let  $\Omega$ and $\widetilde{\Omega}$ 
be the Bernoulli probability matrices of the two networks, respectively,  
and let $\lambda_1$, $\tilde{\lambda}_1$, and $\delta_1$ be the largest singular 
value of  $\Omega$, $\widetilde{\Omega}$,  and $(\Omega - \widetilde{\Omega})$, respectively.  
It turns out that the power of $\psi_n$ depends on 
$\delta_1^2 / (\lambda_1 + \tilde{\lambda}_1)$,   
which can be viewed as the Signal-to-Noise Ratio (SNR) of $\psi_n$.  We show that for all directed-DCMM and DCMM settings (where only mild regularity conditions are imposed), 
$\psi_n \goto \infty$ in probability,   
as long as $\delta_1^2/(\lambda_1 + \tilde{\lambda}_1) \goto \infty$. Therefore,  
the test statistic has asymptotically full power in separating two hypotheses, 
for all parameters in the range of interest. This overcomes the second challenge. 

For the third challenge, we show that the condition 
$\delta_1^2/(\lambda_1+\tilde{\lambda}_1)\goto\infty$
can not be substantially relaxed. 
In fact, for any network with Bernoulli probability matrix $\Omega$, 
we can pair it with another network with Bernoulli probability matrix $\widetilde{\Omega}$ 
so that the $\chi^2$-divergence between two models converges to $0$, once 
$\delta_1^2/(\lambda_1+\tilde{\lambda}_1)\goto 0$. 
This says that the proposed test achieves the optimal phase transition.   This overcomes the third challenge aforementioned.  

In summary, our contributions are as follows.  

{\it \underline{(1). Broadness}}.   We propose IBM as a new test for network comparison that works for a broad setting where we allow severe degree heterogeneity, mixed memberships, flexible sparsity levels, and weak signals, with only mild regularity conditions required. 

{\it \underline{(2). Sharpness}}.  We show the limiting null  of the test statistic is $N(0,1/2)$ and $N(0,1)$ for the directed and undirected cases, respectively, and that the test statistic achieves the optimal phase transition, with an upper bound that matches the lower bound.

{\it \underline{(3). Unified}}.  For both directed and undirected networks, the same test works and attains the optimal phase transition. 

%
As far as we know,  our approach is new and 
 has advantages in all three aspects above.

\subsection{The IBM statistic for signed graphs}  \label{subsec:SGCC} 
Consider two independent networks (directed or undirected) and let $A$ and $\widetilde{A}$  be the adjacency matrices, respectively. Introduce 
\begin{equation} \label{DefineDelta} 
A^* = A - \widetilde{A}.   
\end{equation} 
The entries of $A^*$ take values from $\{-1, 0, 1\}$ and $A^*$ is the adjacency matrix of a signed graph  
\citep{SignedGraph}, where each edge has a weight of either $-1$ or $1$.  A cycle in a graph is a trail where the only repeated vertex is the first and the last vertices (e.g.,  a length-$3$ cycle is a triangle and a length-$4$ cycle is a quadrilateral).   A cycle in the signed graph is called {\it balanced} if 
the product of the weights on all edges of the cycle is positive and {\it unbalanced} otherwise \citep{SignedGraph}. 
Fixing $m \geq 3$,  let $O_m$, $O_m^+$, and $O_m^{-}$ be the number of length-$m$ (unweighted) cycles,  
balanced cycles, and unbalanced cycles, respectively. 
It follows that $\sum_{i_1 i_2 \ldots i_m (dist)} |A_{i_1 i_2}^* A_{i_2 i_3}^* \ldots A_{i_m i_1}^*| = O_m$ and 
$\sum_{i_1 i_2 \ldots i_m (dist)}  A_{i_1 i_2}^* A_{i_2 i_3}^* \ldots A_{i_m i_1}^*  = O_m^+ - O_m^{-}$,  
where `dist' stands for `distinct'. 
Balance checking is of primary interest for signed graphs. Intuitively, 
$O_m^+ - O_m^{-}$ is small (in absolute value) when the null is true and is large otherwise, and so  
can be viewed as an (un-normalized) balance measure \citep{SignedGraph}. 

This motivates {\it Interlacing Balance Measure (IBM}) as a model-free statistic 
as follows.  Fixing $m \geq 2$,  define the order-$m$ IBM statistic by  
($(A^*)'$ denotes the transpose of $A^*$): 
\begin{equation} \label{DefineU2} 
U_n^{(m)} = \sum_{i_1 i_3  \ldots i_{2m-1} (dist), i_2 i_4 \ldots i_{2m} (dist)} 
A_{i_1 i_2}^* (A^*)'_{i_2 i_3}  A_{i_3 i_4}^* (A^*)'_{i_4 i_5}  \ldots A_{i_{2m-1} i_{2m}}^* (A^*)'_{i_{2m} i_1}. 
\end{equation}  
When $A^*$ is symmetric, $U_n^{(m)} = O_{2m}^+ - O_{2m}^{-}$ and reduces to the balance measure aforementioned. For non-symmetric $A^*$,  note that  each term on the right hand side is a product of $2m$ entries {\it alternating}  between the two matrices $A^*$ and $(A^*)'$. 
For these reasons, we call the statistic the Interlacing Balance Measure (IBM).    

To see why IBM is a reasonable idea,  consider two independent directed-DCMM models where 
$A = \Omega - \diag(\Omega) + W$ and 
$\widetilde{A} = \widetilde{\Omega} - \diag(\widetilde{\Omega}) + \widetilde{W}$ for some 
matrices $(\Omega, \widetilde{\Omega}, W, \widetilde{W})$ as in (\ref{model2}), 
and the rank of $\Omega$ and $\widetilde{\Omega}$ are $K$ and $\widetilde{K}$, respectively. 
It follows that $A^*  =  \Delta - \diag(\Delta) + (W - \widetilde{W})$, where $\Delta = \Omega - \widetilde{\Omega}$.  
Let $\delta_k$ be the $k$-th largest singular value of $\Delta$. Note that the rank 
of $\Delta$ is no more than $K + \widetilde{K}$.  
Since the entries of $W - \widetilde{W}$ are independent zero-mean 
random variables, direct calculations show that under mild conditions, 
\begin{equation} \label{EqU1} 
\mbox{$\mathbb{E}[U_n^{(m)}]  = 0$ in the null and  
$\mathbb{E}[U_n^{(m)}]  
\sim \tr((\Delta \Delta')^m) = \sum_{k = 1}^{K+\widetilde{K}}\delta_k^{2m}$ in the alternative}. 
\end{equation} 
The first claim holds, because in (\ref{DefineU2}) we require that $i_1,  i_3,   \ldots,  i_{2m-1}$ are distinct and that  $i_2, i_4,  \ldots,  i_{2m}$ are  distinct; the claim may not hold if such constraints are removed. 
By (\ref{EqU1}), $U_n^{(m)}$ has potential powers  to differentiate the alternative from the null.

The statistic $U_n^{(m)}$ is defined for all $m \geq  2$. We may try to extend the statistic to the case of $m = 1$, but it won't work well in this case; see Remark 2 below.   In this paper, we focus on the IBM statistic with the lowest order (i.e., 
$m = 2$). The study for higher-order IBM is similar but more tedious. When $m = 2$,  the IBM statistic reduces to 
\begin{equation} \label{DefineQ2}
Q_n = \sum_{i_1, i_2 (dist), j_1, j_2 (dist)} A_{i_1 j_1}^* A_{i_2 j_1}^* A_{i_2 j_2}^*  A_{i_1 j_2}^* = 
\sum_{i_1, i_2,  i_3,  i_4 (dist)} A_{i_1 i_2}^* (A^*)'_{i_2 i_3} A_{i_3 i_4}^* (A^*)'_{i_4 i_1}.    
\end{equation} 
Here, we have used the fact that when $m = 2$,  
$A_{i_1 j_1}^* A_{i_2 j_1}^* A_{i_2 j_2}^* A_{i_3 j_2}^* \ldots A_{i_m j_m}^* A_{i_1 j_m}^*$ 
is nonzero only when  $i_1, i_2, \ldots, i_m,  j_1, j_2, \ldots, j_m$ are distinct (this is not necessarily true when $m \geq 3$).

To study the variance of $Q_n$, we define  
\begin{equation} \label{DefineC2} 
C_n = \sum_{i_1, i_2, i_3, i_4 (dist)}  A_{i_1 i_2} (A')_{i_2 i_3} A_{i_3 i_4} (A')_{i_4 i_1}, \;\;  
\widetilde{C}_n = \sum_{i_1, i_2, i_3, i_4 (dist)}  \widetilde{A}_{i_1 i_2} (\widetilde{A}')_{i_2 i_3} \widetilde{A}_{i_3 i_4} (\widetilde{A}')_{i_4i_1}. 
\end{equation} 
We call them the {\it Interlacing Cycle Count (ICC)} statistics, because
when $A$ and $\widetilde{A}$ are symmetric, $C_n$ and $\widetilde{C}_n$ 
are the respective numbers of length-$4$ cycles (quadrilaterals) in the two networks,
and when $A$ and $\widetilde{A}$ are asymmetric, $C_n$ and $\widetilde{C}_n$ 
are the respective numbers of specifically oriented (interlacing) quadrilaterals in the two networks. 
In Section \ref{sec:main}, we show that under mild conditions,  
$\mathrm{Var}(Q_n) \sim  32(\mathbb{E}[C_n + \widetilde{C}_n])$ if both networks are directed, and 
$\mathrm{Var}(Q_n) \sim 64(\mathbb{E}[C_n + \widetilde{C}_n])$ if both networks are undirected. 
This motivates the statistic 
$\psi_n =  (1/8) Q_n / [C_n + \widetilde{C}_n]^{1/2}$.   
Consider the null case first. In Section \ref{sec:main}, we  show that under mild conditions,  
\begin{equation} \label{nullpreview2a} 
\psi_n \goto \left\{
\begin{array}{ll}
N(0, 1/2), &\qquad \mbox{if both networks are directed},  \\
N(0, 1), &\qquad \mbox{if both networks are undirected}, \\ 
\end{array}  
\right.  \qquad \mbox{under $H_0^{(n)}$}.   \\ 
\end{equation} 
The variances of two limiting nulls are different.  This is because for any $i\neq j$, $A^*_{ij}$ and $A^*_{ji}$ are two independent variables for directed networks, and $A^*_{ij}=A^*_{ji}$ for undirected networks. 

Next, consider the alternative case. In Section \ref{sec:main}, we show that under mild conditions,  
\begin{equation} \label{nullpreview2b} 
\mathbb{E}[\psi_n] \approx (1/8)   \tr([\Delta \Delta']^2) / (\mathbb{E}[C_n  + \widetilde{C}_n])^{1/2}.    
\end{equation}  
Let $\delta_k, \lambda_k, \widetilde{\lambda}_k$ be the $k$-th singular value of $\Delta$, $\Omega$, and 
$\widetilde{\Omega}$, respectively.  Assuming $\max\{K, \widetilde{K}\}$ is finite, we can further show that  
$\tr([\Delta \Delta']^2) / (\mathbb{E}[C_n  + \widetilde{C}_n])^{1/2}   \approx \Bigl(\sum_{k = 1}^{K + \widetilde{K}} \delta_k^4\Bigr) / \Bigl(\sum_{k = 1}^K \lambda_k^4 + \sum_{k = 1}^{\widetilde{K}} 
\widetilde{\lambda}_k^4\Bigr)^{1/2} \asymp \delta_1^4 /(\lambda_1 + \widetilde{\lambda}_1)^2$.  
Therefore, we expect to have $\psi_n \goto \infty$ if $\delta_1^2 / (\lambda_1  + \widetilde{\lambda}_1)  \goto \infty$.  
This says that  $\psi_n$  is able to differentiate the alternative from the null once $\delta_1^2 / (\lambda_1 + \widetilde{\lambda}_1) \goto \infty$.

Moreover, in Section \ref{subsec:LB},  we study the minimax lower bounds. We show that the condition $\delta_1^2 / (\lambda_1  + \widetilde{\lambda}_1) \goto \infty$ can not be significantly relaxed, 
and the test statistic achieves the optimal phase transition. 
Therefore, IBM not only has a tractable limiting null, but is also optimal in testing powers. See Section \ref{subsec:LB} 
for details.

We now discuss the computation cost of  $\psi_n$.   For any $n \times n$ matrices $X$ and $Y$, let $X \circ Y$ be the Hadamard product of $X$ and $Y$ (i.e., $(X \circ Y)_{ij} = X_{ij} Y_{ij}, 1 \leq i, j \leq n$), and  let 
$|X|$ be the $n \times n$ matrix where $(|X|)_{ij} = |X_{ij}|, 1 \leq i, j \leq n$.     Introduce 
$q(X) = \text{tr}([XX']^2) -  \text{tr}(XX' \circ XX')-\text{tr}(X'X \circ X'X) + 1_n' |X| 1_n$.  
For a symmetric or asymmetric network, let $\bar{d}$ be the average degree and 
let $d_{max}$ be the maximum degree.  Recall that $A^* = A - \widetilde{A}$.
The next lemma  is proved in the supplement. 
\begin{lemma} 
\label{lemma:comp} 
We have   $\psi_n = q(A^*)/[8 (q(A) + q(\widetilde{A}))^{1/2}]$. 
Moreover,  if we choose to store $X$ and $XX'$ in our code for $X = A^*, A, \widetilde{A}$,  then the computation cost is $O(n^2 \bar{d})$.  If instead of storing the whole matrices 
$X$ and $XX'$,  we  only store the adjacency lists of $X$ and $XX'$ in our code, then the computation cost can be further reduced to $O(n \bar{d}d_{\max})$.
\end{lemma}
 The matrices $X$ and $XX'$ may be very sparse (i.e., most entries are $0$), so 
the overhead of computing $X$ and $XX'$ is large. Therefore,  instead of storing the whole matrix $X$ and $XX'$  in our code, we 
can choose to only store the adjacency lists (i.e., nonzero entries) of $X$ and $XX'$ (when $X$ is stored in the form of adjacency list, the cost of finding the neighbors of any node is $O(1)$ per neighbor), and the resultant computational cost is further reduced. 

{\bf Remark 1}. The statistic $U_n^{(m)}$ may look similar to the statistic of $\tr([A^*(A^*)']^m)$, 
but it is quite different. Consider the case where $A^*$ is symmetric for example.  
In this case, $\tr([A^*(A^*)']^m)  =\tr((A^*)^{2m})= \sum_{i_1, i_2, \ldots, i_{2m}}  A_{i_1i_2}^* A_{i_2i_3}^* \ldots 
A_{i_{2m}i_1}^*$, where unlike $U_n^{(m)}$ as in (\ref{DefineU2}),   the indices $i_1, i_2, \ldots, i_{2m}$   are not required to be distinct.  Therefore, unlike $U_n^{(m)}$,  $\mathbb{E}[\tr([A^*(A^*)']^m)]$ depends on many unknown parameters and is nonzero,  so it is unclear how to 
normalize it to have a tractable limiting null. Also,  the variance of $\tr([A^* (A^*)']^m)$ is  much larger than that of $U_n^{(m)}$, and the  statistic is also less efficient in power.

{\bf Remark 2}.  The statistic $U_n^{(m)}$ is defined for all $m \geq 2$. If we try to extend it to the case of $m = 1$, then 
 $U_n^{(m)}$ reduces to the statistic of $\sum_{i,j} (A_{ij}^*)^2$. In this case,  the statistic has a nonzero mean under the null, so it is unclear how to standardize it to have a tractable limiting null.   Also, the Signal-to-Noise Ratio (SNR) of the statistic turns out to be much smaller than those in the cases of $m \geq 2$, so the statistic is less efficient in power.

{\bf Remark 3}. Our idea is extendable to $m > 2$.  Take $m =3$ for example (see Section~\ref{supp:high-order} of the supplementary material). In this case, we may change the test statistic $\psi_n$ to $\phi_n$, where $\phi_n = U_n^{(3)} / [384 (C_n + \widetilde{C}_n)]^{1/2}$ and 
$C_n = \sum_{i_1, i_3, i_5 (dist), i_2, i_4, i_6 (dist)} A_{i_1 i_2} A_{i_2 i_3} \ldots A_{i_6 i_1}$  
($\widetilde{C}_n$ is similar).   Equivalently,  if we let $q(A) = \mathrm{trace}((A)^6)- 6\cdot \mathrm{trace}(A^2\circ A^4) + 6\cdot \mathrm{trace}((A \circ A \circ A) A^3)  +3\cdot 1_n' (A\circ A)^3 1_n +4\cdot \mathrm{trace}(A^2 \circ A^2 A^2) - 12 \cdot \mathrm{trace}((A \circ A)^2 \circ A^2) + 4\cdot 1_n' (A \circ A  \circ A  \circ A  \circ A  \circ A) 1_n$, then  $\psi_n = q(A^*) / [384 (q(A) + q(\widetilde{A})]^{1/2}$.  With such a formula, the computational costs for $m=3$ is the same as that of $m = 2$.  The limiting null and power analysis of $\phi_n$ are similar to that of $\psi_n$, but 
technically much more involved. For a finite $n$,  $\phi_n$ is slower in convergence to the limiting null and in 
computation (e.g., $4$ times slower when $n = 5000$),  but may have a better power in some cases (it is unclear whether we can have uniform power improvement).

{\bf Remark 4}.  In the one-sample undirected network setting, researchers (e.g.,  \cite{bubeck2016testing, banerjee2016contiguity, SP2019}  used cycle count approaches to test whether $K = 1$  or $K > 1$ ($K$: number of communities). To extend their ideas to our setting, we face challenges. For example, in the undirected network case,  we may compute the cycle count statistics $C_m$ and $\widetilde{C}_m$ for two networks (similarly as in previous works) and use $T=C_m-\widetilde{C}_m$ as the test statistic.  Unfortunately, this test loses power in many cases.  In face,  since $\mathbb{E}[C_m] \sim  \sum_{k = 1}^K \lambda_k^m$ 
and $\mathbb{E}[\widetilde{C}_m] = \sum_{k = 1}^{\widetilde{K}} \widetilde{\lambda}_k^m$, the test loses power when $\sum_{k = 1}^K \lambda_k^m = \sum_{k = 1}^{\widetilde{K}} \widetilde{\lambda}_k^m$, but this can be an easy-to-test case as we may have $(\lambda_1, \lambda_2, \ldots, \lambda_K) \neq 
(\widetilde{\lambda}_1, \widetilde{\lambda}_2, \ldots, \widetilde{\lambda}_{\widetilde{K}})$.  In the directed networks case,   there are multiple ways to define a cycle as the edges have directions, and it is unclear which cycle count approach may give rise to optimal tests.

{\bf Remark 5}. An alternative test is $T_n = s_1^2(A^*) /[s_1(A) + s_1(\widetilde{A})]$, where $s_1(\cdot)$ denotes the largest singular value. For simplicity, consider the DCMM case where $A, \widetilde{A}$ are symmetrical.   When the null is true,  we can show (e.g., \cite{SCORE}) $T_n \leq C \log(n) \theta_{max} \|\theta\|_1 / \|\theta\|^2$ with high probability.  For cases where $\theta_{max} \leq c_0 \theta_{min}$, $\theta_{max} \|\theta\|_1 / \|\theta\|^2  \leq C$,  so we may use $T_n$ if you are satisfied with something crude. However, in the presence of 
severe degree heterogeneity, $\theta_{max} / \theta_{min}$ may grow to $\infty$ rapidly as $n$ diverge, and the test is far from optimal (note also that the limiting null of $T_n$ is not explicit and may depend on unknown parameters).  In this paper, $\theta_{max} = \max\{\theta_1, \ldots, \theta_n\}$ and $\theta_{min} = \min\{\theta_1, \ldots, \theta_n\}$.

\subsection{Content} 
Section~\ref{sec:main} presents the optimality of IBM for both directed and undirected cases.   Section~\ref{sec:real} studies real-data applications, and Section~\ref{sec:numer} studies simulations. Section~\ref{sec:discuss} concludes the paper with discussions.  Proofs of the theorems and lemmas are in the supplement.

\section{Main results} \label{sec:main} 
We start by discussing the identifiability for the two models, 
directed-DCMM and DCMM.  We then present the optimality of IBM under the DCMM model for undirected networks, where 
we analyze the limiting null and the power of IBM in Section \ref{subsec:DCMM}, and
present our results on minimax lower bound and phase transition in Section \ref{subsec:LB}. 
The optimality of IBM under the directed-DCMM models for directed networks is in Section \ref{subsec:dDCMM}. 

\subsection{The identifiability for directed-DCMM and DCMM models} \label{subsec:ident} 
Consider a directed-DCMM model (\ref{model1a})-(\ref{model1c}) with $K$ communities, where $\pi_i$ and $\gamma_i$ are the membership vectors of node $i$ as a citer and a citee, respectively. 
Fix $1 \leq i \leq n$ and $1 \leq k \leq K$.  We call node $i$ a {\it pure node} of community $k$ as a citer (or as a citee) if $\pi_i$ (or $\gamma_i$) is a 
degenerate weight vector (i.e., one entry is  $1$, others are $0$). 
We call a non-negative $K \times K$ matrix $P$ double stochastic if the sum of  each row and each column is $1$. 
Lemma \ref{lemma:iden1} discusses identifiability of  directed-DCMM and is proved in the supplement

\begin{lemma}\label{lemma:iden1}
For any $\Omega = \Theta \Pi P \Gamma' Z$ as in Model \eqref{model1a}-\eqref{model1c} where $P$ is fully indecomposable,
\spacingset{1}\footnote{We call $P$ fully indecomposable if we do not have permutation matrices $S_1, S_2$ such that 
$S_1 P S_2$ is a $2 \times 2$ block matrix where the upper right block is $0$.}\spacingset{1.75}
we can always re-parametrize the model so that $P$ is doubly stochastic and $\|\theta\| = \|\zeta\|$. Conversely, if $\Omega = \Theta \Pi P \Gamma' Z$  where $P$ is non-singular, fully indecomposable  and doubly stochastic,  $\|\theta\| = \|\zeta\|$, and each community has at least one node which is pure both as a citer and as a citee, then $(\Theta, \Pi, P, \Gamma, Z)$ are uniquely determined by $\Omega$  
\end{lemma}
For DCMM, Lemma \ref{lemma:iden1} still applies, as DCMM is a special directed-DCMM, but existing works have suggested other choices of identifiability conditions.    
See Lemma \ref{lemma:iden2} for example, which is proved in \cite{Mixed-SCORE}.  
\begin{lemma}\label{lemma:iden2}
In model (\ref{model2}),  if  $P$ is non-singular,  irreducible, and has unit diagonal entries, and if each community has at least one pure node,  
then the model is identifiable. 
\end{lemma}
To be consistent with literature works, we choose to use the conditions of Lemma \ref{lemma:iden2} for 
DCMM models.  Note that two sets of conditions 
are the same except that one requires $P$ to be doubly stochastic and the other 
requires $P$ to have unit diagonal entries.

\subsection{The limiting null and power of IBM for DCMM models} \label{subsec:DCMM}   
Consider two independent undirected networks on the same set of $n$ nodes and both satisfy the DCMM model (\ref{model2}). Let $A$ and $\widetilde A$ be the adjacency matrices of two networks,   $\Omega$ and $\widetilde{\Omega}$ be the Bernoulli probability matrices, and $K$ and $\widetilde{K}$ be the numbers of communities, respectively. Recall that  
$\Omega = \Theta \Pi P \Pi' \Theta$  and 
$\widetilde{\Omega} = \widetilde{\Theta} \widetilde{\Pi} \widetilde{P} \widetilde{\Pi}' \widetilde{\Theta}$, where 
$\Theta  = \diag(\theta_1, ..., \theta_n)$ and $\Pi = [\pi_1, \pi_2, \ldots, \pi_n]'$;  similar for $(\widetilde{\Theta}, \widetilde{\Pi})$. 
We assume three mild conditions as follows. 
\begin{equation} \label{cond-theta}
\max_{1\leq i\leq n}\{\theta_i + \tilde\theta_i \} \goto 0,   \qquad 
 \|\theta\|  \goto\infty, \qquad \|\tilde{\theta}\| \goto\infty. 
\end{equation} 
The last two conditions are necessary. The first condition is mainly for simplicity  and can be relaxed. In fact, the IBM test statistic decomposes into the sum of several terms, 
so the total variance is the sum of many variance terms. With such a condition, one of these terms dominates in variance, so the total variance equals approximately to the variance of this particular term and has a succinct form. Without such a condition, a succinct form is hard to obtain.  Same discussion for (\ref{D-cond-theta}) below. 
We also assume 
\begin{equation} \label{cond-P}
\bigl\|\bigl(\|\theta\|^{-2}  \Pi'\Theta^2\Pi\bigr)^{-1} \bigr\| \leq C,  \qquad \bigl\|\bigl(\|\tilde{\theta}\|^{-2} \widetilde \Pi'\widetilde\Theta^2\widetilde\Pi\bigr)^{-1} \bigr\| \leq C,   \qquad \max\{\|P\|, \|\widetilde{P}\|\}   \leq C.   
\end{equation} 
Here, as the $K \times K$ matrices $\|\theta\|^{-2}  \Pi'\Theta^2\Pi$ and $\|\tilde{\theta}\|^{-2} \widetilde \Pi'\widetilde\Theta^2\widetilde\Pi$  are properly scaled, the first two  are only  
 mild conditions on the community balance.   The last one is also a mild condition.

Consider the null hypothesis first.  Recall that for undirected networks,  the test statistic is  $\psi_n = Q_n / (8 [C_n + \widetilde{C}_n]^{1/2})$.  Under the null, $(\Omega, \Theta, \Pi, P) = (\widetilde{\Omega}, \widetilde{\Theta}, \widetilde{\Pi}, \widetilde{P})$, and $C_n$ and $\widetilde{C}_n$ have the same distribution. 
The following lemma is proved in the supplement.

\begin{thm}
({\it Null behavior (DCMM)}).  \label{thm:null}  
Consider the pairwise comparison problem as in (\ref{CP1a}),  
where both networks satisfy the DCMM model (\ref{model2}).  Suppose conditions \eqref{cond-theta}-\eqref{cond-P} hold.  If the null hypothesis holds, then as $n \goto \infty$,  we have (1). $\mathbb{E}[Q_n]  = 0$, and $\mathrm{Var}(Q_n)=   128 [1+o(1)]  \cdot \mathbb{E}[C_n]$, (2). $\mathbb{E}[C_n] =  \tr(\Omega^4) + O(\|\theta\|^4\|\theta\|_4^4)$,  $\mathrm{Var}({C}_n) \leq C\tr(\Omega^4) (1+\|\theta\|_3^6)$, and $C_n / \mathbb{E}[C_n] \stackrel{p}{\goto} 1$, and (3).  $\psi_n \goto N(0, 1)$ in law.  
\end{thm} 
Theorem  \ref{thm:null} shows that the limiting null of $\psi_n$ is $N(0,1)$. In practice, 
once we obtain the testing score $\psi_n$ for a pair of networks, 
we can use $\mathbb{P}(N(0,1) \geq \psi_n)$ to approximate the $p$-value; see 
our real-data analysis in Section \ref{sec:real}. Now, fix $0 < \alpha < 1$ and let $z_{\alpha}$ be the  $(1-\alpha)$-quantile of $N(0, 1)$. 
Consider the  IBM test where we 
\begin{equation} \label{SGCCtest1} 
\mbox{reject the null}   \qquad \Longleftrightarrow   \qquad \psi_n \geq   z_{\alpha}.  
\end{equation} 
As $n \goto \infty$, by Theorem   \ref{thm:null}, the Type I error of the  test converges to $\alpha$ as expected. 

We now analyze the power.  As before, let $\Delta = \Omega - \widetilde{\Omega}$, and let   
 $\lambda_k$, $\tilde{\lambda}_k$, and $\delta_k$ be the $k$-th largest (in magnitude) 
eigenvalue of $\Omega$, $\widetilde{\Omega}$, and $\Delta$, respectively.  Since $\Omega$ and $\widetilde{\Omega}$ are non-negative matrices,  by Perron's theorem \citep{HornJohnson},  
$\lambda_1 > 0$ and $\tilde{\lambda}_1  > 0$.  The following theorem is proved in the supplement. 

\begin{thm}\label{thm:alt}
({\it Power analysis (DCMM)}).  
Consider the pairwise comparison problem (\ref{CP1a}) where both networks satisfy the DCMM model \eqref{model2} where conditions \eqref{cond-theta}-\eqref{cond-P} hold. 
Assume that $\delta_1^2 / (\lambda_1 + \widetilde{\lambda}_1)  \goto \infty$ as $n\goto\infty$ and that $K$ and $\widetilde K$ are  fixed. Then, (1). $\mathbb{E}[{Q}_n] = \tr(\Delta^4) + o\big(\|\theta + \tilde{\theta}\|^2\cdot\tr(\Delta^2)\big) = (1 + o(1))  \tr(\Delta^4)$,  and 
$\mathrm{Var}(Q_n) \leq C \big(\tr(\Omega^4) + \tr(\widetilde{\Omega}^4)  + [\tr(\Delta^2)]^3\big) \leq C\bigl(\|\theta\|^8 + \|\tilde{\theta}\|^8 + [\tr(\Delta^2)]^3\bigr)$, and (2).  $\psi_n\to\infty$, in probability. Therefore, for any fixed $\alpha$,  the power of the IBM test defined in (\ref{SGCCtest1})  goes to  $1$ as $n \goto \infty$.  
%
\end{thm}
{\bf Remark 6}. By Theorems \ref{thm:null}-\ref{thm:alt},  if we let the level $\alpha$ of the IBM test in (\ref{SGCCtest1}) depend on $n$   (i.e., $\alpha =  \alpha_n$)  and let $\alpha_n$ tend to $0$ sufficiently slow, then  the Type I error  of the test $\goto 0$, and the power  of the test $\goto 1$, so the sum of Type I and Type II errors $\goto 0$.     

{\bf Remark 7}. The main condition of Theorem \ref{thm:alt} is $\delta_1^2/(\lambda_1 + \tilde{\lambda}_1) \goto \infty$. 
If we relax it  to $\delta_1^2/(\lambda_1 + \tilde{\lambda}_1) \goto c_0$ for some constant $c_0 > 0$, then the power of the IBM test converges to a number in $(0, 1)$. 
If we further relax it to $\delta_1^2/(\lambda_1 + \tilde{\lambda}_1) \goto 0$, 
then we are in the {\it impossibility region}, where we can find many pairs of DCMM 
models that are asymptotically inseparable (so the sum of Type I and Type II errors of any test is $\geq (1 + o(1))$).  See Section \ref{subsec:LB} for details, where 
we discuss the minimax lower bound and phase transition.

The main condition of Theorem \ref{thm:alt} is $\delta_1^2 / (\lambda_1 + \tilde{\lambda}_1) \goto \infty$.  The 
condition has a simple form but is not completely obvious, so it is worthy to explain such a condition, especially the connection between the term  $\delta_1^2 / (\lambda_1 + \tilde{\lambda}_1)$  and  {\it Signal-to-Noise Ratio} of the IBM test statistic $Q_n$. The Signal-to-Noise ratio of $Q_n$ is $\mathbb{E}[Q_n]/\mathrm{SD}(Q_n)$, 
but we do not have an explicit formula for it.  We introduce a proxy by 
\begin{equation} \label{DefineSNR} 
\mathrm{SNR} = (1/8)\cdot \tr(\Delta^4) / \sqrt{\tr(\Omega^4) + \tr(\widetilde{\Omega}^4)}. 
\end{equation} 
We have the more challenging {\it weak signal}  case and less challenging {\it strong signal}  case: 
\begin{equation} \label{twocases} 
\left\{ 
\begin{array}{ll}  
\mbox{Case 1 (weak signal)}:   & \qquad [\tr(\Delta^2)]^3 \leq \tr(\Omega^4) + \tr(\widetilde{\Omega}^4),  \\
\mbox{Case 2 (strong signal)}:  & \qquad [\tr(\Delta^2)]^3 >  \tr(\Omega^4) + \tr(\widetilde{\Omega}^4).  \\ 
\end{array} 
\right. 
\end{equation}
Consider the weak signal case (Case 1) first. In this case,   
By Theorem \ref{thm:alt},  $\mathbb{E}[Q_n] \sim \tr(\Delta^4)$ and $\mathrm{Var}(Q_n) \leq C [\tr(\Omega^4) + \tr(\widetilde{\Omega}^4)]$. Therefore,  $\mathrm{SNR} \asymp \mathbb{E}[Q_n]/\mathrm{SD}(Q_n)$, so the definition of SNR in (\ref{DefineSNR}) is reasonable.  
Recall that $\Delta = \Omega - \widetilde{\Omega}$, where $\mathrm{rank}(\Omega) = K$ and 
$\mathrm{rank}(\widetilde{\Omega}) = \widetilde{K}$.  Let $r = \mathrm{rank}(\Delta)$; note that $r \leq K + \widetilde{K}$.   
It is seen that 
$\mathrm{SNR} = (1/8) \Bigl(\sum_{k = 1}^r \delta_k^4\Bigr)  / \Bigl[(\sum_{k = 1}^K \lambda_k^4)  + (\sum_{k = 1}^{\widetilde{K}} \tilde{\lambda}_k^4)\Bigr]^{1/2}$.  
By Condition (\ref{cond-theta}),  $\|\theta\| \goto \infty$, and $\|\tilde{\theta}\| \goto \infty$, 
 and we can show that $\lambda_1 \asymp \|\theta\|^2$, 
$\tilde{\lambda}_1 \asymp \|\tilde{\theta}\|^2$.  
This says that $\delta_1 \goto \infty$ once we assume $\delta_1^2/(\lambda_1 + \tilde{\lambda}_1) \goto \infty$ as in Theorem \ref{thm:alt}. 
Combining these with elementary calculations, it follows that if $K + \widetilde{K}$ is bounded, then 
\begin{equation} \label{DefineSNR2} 
\mathrm{SNR} \asymp  \delta_1^4 / (\lambda_1 + \tilde{\lambda}_1)^2. 
\end{equation} 
Therefore,  in the weak signal case, $[\delta_1^2 / (\lambda_1 + \tilde{\lambda}_1)]^2 \asymp \mathrm{SNR} \asymp \mathbb{E}[Q_n] /
\mathrm{SD}(Q_n)$,  and the main condition of  $\delta_1^2 / (\lambda_1 + \tilde{\lambda}_1) \goto \infty$ in Theorem 
\ref{thm:alt} is the same as the condition of $\mathrm{SNR} \goto \infty$. This explains why the test has asymptotically full power. 

We now discuss the strong signal case (Case 2).  In this case, the variance of the IBM test statistic $Q_n$ may be larger than that of the weak signal case (which is $\asymp  [\tr(\Omega^4) + \tr(\widetilde{\Omega}^4)]$),  but the mean of $Q_n$ is also much larger that that of the weak signal case. As a result,  this is a less challenging case for pairwise comparison. In fact, 
by similar arguments, 
$(\lambda_1 + \tilde{\lambda}_1) \asymp (\|\theta\|^2 + \|\tilde{\theta}\|^2) \goto \infty$. Therefore, by the condition of  $\delta_1^2 / (\lambda_1 + \tilde{\lambda}_1) \goto \infty$, we have  $\delta_1 \goto \infty$.  
Note that $\mathbb{E}[Q_n] / \mathrm{SD}(Q_n) \asymp \tr(\Delta^4) / [\tr(\Delta^2)]^{3/2} \asymp 
\delta_1$.   Therefore, as long as  $\delta_1^2 / (\lambda_1 + \tilde{\lambda}_1) \goto \infty$, the IBM test statistic also has   asymptotically full power in this case.   
   
{\bf Remark 8}.  Our idea is readily extendable to the case where we have multiple independent samples for each of the models. Consider the DCMM case for simplicity, where $A$ and $\widetilde{A}$ are the average of $n$ and $m$ independent adjacency matrix from two DCMM models where the  
Bernoulli probability matrices are $\Omega$, and $\widetilde{\Omega}$, respectively. In this setting, we can similarly write $A = \Omega - \diag(\Omega) + W$ and $\widetilde{A} = \widetilde{\Omega} - \diag(\widetilde{\Omega}) + \widetilde{W}$,  
where the only difference is,  $W$ and $\widetilde{W}$ are matrices of  
(scaled) centered Binomial instead of centered Bernoulli.  In this setting, we expect to have similar results as in Section 2.1,  where the SNR is at the order of $\delta_1^2 /[(\lambda_1 / \sqrt{n})  + (\tilde{\lambda}_1/\sqrt{m})]$, while that for our setting is $\delta_1^2 / (\lambda_1 +  \tilde{\lambda}_1)$.  
The optimality of the test statistic also implies the optimal sample complexity.  

\subsection{The minimax lower bound for DCMM}    \label{subsec:LB} 
The key in the lower bound analysis is to find  the least favorable configuration. 
A standard approach is to use randomization: 
fixing a Bernoulli probability matrix  $\Omega = \Theta \Pi P \Pi' \Theta$  for a DCMM model with $K$ communities, 
we construct another Bernoulli probability matrix $\widetilde{\Omega}$ using $\Omega$ and 
randomization.  In detail,  let $\sigma=(\sigma_1, \ldots, \sigma_n)$ be iid Radermacher variables.  
We construct a randomized Bernoulli probability matrix $\widetilde{\Omega} = \widetilde{\Omega}(\sigma)$ 
as follows. Recall that $\Pi  \in \mathbb{R}^{n, K}$.  We construct a new community $(K+1)$ and randomly move part of the weights of community $K$ to this new community.  In detail, 
introduce a matrix $\check\Pi(\sigma) \in\mathbb{R}^{n, K+1}$ where  
for $1 \leq i \leq n$, $\check{\Pi}_{i \ell}(\sigma) = \Pi_{i\ell}$ if $1 \leq \ell \leq K-1$, 
$\check{\Pi}_{i \ell}(\sigma) = \Pi_{iK}\cdot (1 + \sigma_i)/2$ if $\ell=K$, and 
$\check{\Pi}_{i \ell}(\sigma) = \Pi_{iK}\cdot (1 -\sigma_i)/2$ if $\ell=K+1$.  
For a small positive number $\epsilon_n$,  
\[
\mbox{write}\quad 
P = \left(\begin{matrix} P_0& \alpha\\\alpha' & 1
\end{matrix}\right) \in\mathbb{R}^{K,K}, \qquad \mbox{and let}\quad 
\check{P} = \left(\begin{matrix} P_0 & \alpha & \alpha \\
\alpha' & 1+\eps_n & 1-\eps_n \\
\alpha' & 1-\eps_n & 1 + \eps_n 
\end{matrix}\right) \in \mathbb{R}^{K+1, K+1}.  
\]
Introduce two diagonal matrices $D = \diag(1, \dots, 1, \sqrt{1+\eps_n}, \sqrt{1+\eps_n})$ and $G\in\mathbb{R}^{n\times n}$ where $G_{ii} = \sum_{k=1}^{K-1} \Pi_{ik} + \sqrt{1+\eps_n} \cdot\Pi_{iK}$, for $1\leq i\leq n$. Let $\widetilde{P}=D^{-1} \check{P}D^{-1}$, $\widetilde{\Pi} = G^{-1}\check\Pi D$,  and $\widetilde{\Theta}=\Theta G$. Our construction for $\widetilde{\Omega}  = \widetilde{\Omega}(\sigma)$ is 
\beq \label{RMM-DCMM}
\widetilde{\Omega}(\sigma) =  \widetilde{\Theta}\widetilde{\Pi} \widetilde{P}\widetilde{\Pi}'\widetilde{\Theta}. 
\eeq 

Let ${\cal M}_{n,0}(K)$ be all Bernoulli probability matrices $\Omega$ for DCMM models with $K$ communities where $\Omega = \Theta \Pi P  \Pi' \Theta$ as in (\ref{model2}) and the conditions of Lemma \ref{lemma:iden2} hold.   
Given a positive sequence $\{\beta_n\}_{n=1}^{\infty}$ and $K\geq 1$, $0<c_0 <1$,  define a class of  DCMM models by 
\begin{align*}
{\cal M}_n(\beta_n, K,  c_0) =  \bigl\{ \Omega \in {\cal M}_{n,0}(K): \theta_{\max} \leq K\beta_n,\; \|\theta\|\geq \beta_n^{-1}, \\     c_0K^{-1}\|\theta\|^2\leq \|\Omega\|\leq c_0^{-1}K\|\theta\|^2, \; \|P\|_{\max}\leq  c_0^{-1} \bigr\},  
\end{align*}
where $\|P\|_{\max}$ is the maximum element of $P$. 
The following  is proved in the supplement. 
\begin{thm}({\it Least favorable configuration for  DCMM}).  \label{thm:LF} 
Fix $K\geq 1$,  $c_0 \in (0,1)$  and a positive sequence $\{\beta_n\}_{n=1}^{\infty}$ such that $\beta_n = o(1)$. Given any sequence $\{\Omega_n\}_{n=1}^\infty$ with $\Omega_n\in \mathcal{M}_n(\beta_n, K, c_0)$, let $\theta(\Omega_n)$ denote the vector of degree parameters. 
We construct a sequence $\{\widetilde\Omega_n\}_{n=1}^{\infty}$ as in \eqref{RMM-DCMM},
 where $\epsilon_n$ satisfies that $\epsilon_n\cdot\|\theta(\Omega_n)\|\to 0$. 
\begin{itemize}
\item With probability $1-o(1)$, $\widetilde{\Omega} = \widetilde\Omega_n(\sigma)\in\mathcal{M}_n(\beta_n, K+1,  c_0)$,   and $\delta_1^2/(\lambda_1 + \tilde{\lambda}_1) \goto 0$, where $\delta_1, \lambda_1, \tilde{\lambda}_1$ are the first eigenvalue of $(\Omega_n - \widetilde{\Omega}_n), \Omega_n$, and $\widetilde{\Omega}_n$, respectively.  
\item Consider a null case and an alternative case as follows. 
For the null case, we generate two $n \times n$ network adjacency matrices $A$ and $\widetilde{A}$ independently with the same probability matrix $\Omega_n$. For the alternative case, we generate $A$ in the same way but generate $\widetilde{A}$ from the random-membership DCMM associated with $\widetilde{\Omega}_n$  as in \eqref{RMM-DCMM}, independently of $A$.  As $n\to\infty$, the $\chi^2$-distance between these two models tends to $0$. 
\end{itemize}
\end{thm}

Once we have the least favorable configuration, we can obtain a minimax lower bound. 
Similar as before, let $\Delta = \Omega - \widetilde{\Omega}$,  $r, K, \tilde{K}$ be the ranks of 
$\Delta$, $\Omega$ and $\widetilde{\Omega}$, respectively, and $\delta_k$, $\lambda_k$, and $\tilde{\lambda}_k$ be the $k$-th largest eigenvalue (in magnitude) of $\Delta$, $\Omega$, and $\widetilde{\Omega}$, respectively. Recall that $\mathrm{SNR} = (1/8) (\sum_{k = 1}^r \delta_k^4) / [(\sum_{k = 1}^K \lambda_k^4) + (\sum_{k = 1}^{\widetilde{K}} \tilde{\lambda}_k^4)]^{1/2}$. 
When $(K+\widetilde{K})$ is finite, it holds that $\mathrm{SNR} \asymp [\delta_1^2 /(\lambda_1 + \tilde{\lambda}_1)]^2$.  
Define the class of DCMM model pairs for the null by 
\begin{equation}
\mathcal{S}_n^*(\beta_n, K, c_0) = \{(\Omega,\widetilde{\Omega})\in \mathcal{M}_n(\beta_n, K, c_0)\times\mathcal{M}_n(\beta_n, K, c_0): \Omega = \widetilde{\Omega}\}, 
\end{equation}
and define a class of DCMM model pairs for the alternative case by 
\begin{equation}
\mathcal{S}_n(\beta_n,\rho_n, K, \widetilde{K}, c_0) = \{(\Omega,\widetilde{\Omega})\in \mathcal{M}_n(\beta_n, K, c_0)\times\mathcal{M}_n(\beta_n, \widetilde{K}, c_0):  \delta_1^2 \geq \rho_n(\lambda_1 + \tilde{\lambda}_1)\}. 
\end{equation}
Within this class, we can find model pairs where  $\delta_1^2 / (\lambda_1 + \tilde{\lambda}_1) \goto 0$, which are asymptotically inseparable as in Theorem \ref{thm:LF}. 
The following theorem is proved in the supplement. 
\begin{thm}(Minimax lower bound for DCMM).  \label{thm:LB}
	For any given $K\geq 1$,  $c_0 \in (0,1)$, and positive sequences $\{\beta_n\}_{n=1}^{\infty}$ and $\{\rho_{n}\}_{n=1}^{\infty}$ such that $\beta_n = o(1)$ and $\rho_n = o(1)$, we have
	\begin{equation}
	\label{def:risk2}
	\inf_{\psi}\bigg\{ \sup_{(\Omega,\widetilde{\Omega})\in\mathcal{S}_n^*(\beta_n, K, c_0)}\mathbb{P}(\psi = 1)
	+ \sup_{(\Omega,\widetilde{\Omega})\in\mathcal{S}_n(\beta_n, \rho_n, K, K+1, c_0)}\mathbb{P}(\psi = 0)\bigg\}\goto1
	\end{equation}
	as $n\goto\infty$, where the infimum is taken over all possible tests $\psi$.  
\end{thm}

Combining Theorems \ref{thm:null}-\ref{thm:LB}, we have the following phase transition. 
Consider a sequence of DCMM model pairs indexed by $n$,  where for each pair, $\Omega$ and $\widetilde{\Omega}$ are the Bernoulli probability matrices, 
respectively.  Consider the pairwise comparison problem where we test  $H_0^{(n)}: \Omega = \widetilde{\Omega}$ versus $H_1^{(n)}: \Omega \neq 
\widetilde{\Omega}$.   Recall that $\Delta = \Omega - \widetilde{\Omega}$ and $\delta_1$, $\lambda_1$, $\tilde{\lambda}_1$ are the 
largest eigenvalues (in magnitude) of $\Delta, \Omega, \widetilde{\Omega}$, respectively.

{\it Possibility}. When $\delta_1^2/(\lambda_1 +   \tilde{\lambda}_1)  \goto \infty$, the two models are asymptotically separable,   
and the sum of Type I and Type II errors of the IBM test  $\goto 0$. 

{\it Impossibility}.  When $\delta_1^2/(\lambda_1 +  \tilde{\lambda}_1)  \goto 0$, two models are not always asymptotically separable. 
In fact, for each $\Omega$, we can pair it with an $\widetilde{\Omega}$ such that $\delta_1^2/(\lambda_1 +   \tilde{\lambda}_1)  \goto 0$ 
and the $\chi^2$-divergence between the two models $\goto 0$. Therefore, for any test, 
the sum of Type I and Type II errors is $\geq 1 + o(1)$ (see also Remark 7).    
%
%
%

\subsection{Optimality of the IBM test for directed-DCMM}  \label{subsec:dDCMM} 
 We study the IBM test for directed networks. IBM uses the same test statistic for undirected and directed networks, but the analysis of directed networks is quite different from that of undirected networks. In Theorems~\ref{thm:null-d}-\ref{thm:LB-d} below, we study the limiting null, power, and optimality of IBM in the directed-DCMM setting. 

Consider two independent {\it directed} networks on the same set of $n$ nodes that satisfy the directed-DCMM model (\ref{model1a})-(\ref{model1c}). Let $A$ and $\widetilde A$ be the two adjacency matrices,  let $\Omega$ and $\widetilde{\Omega}$ be the two Bernoulli probability matrices, and let $K$ and $\widetilde{K}$ be the two numbers of communities, respectively. Recall that  
$\Omega = \Theta \Pi P \Gamma' Z$  and 
$\widetilde{\Omega} = \widetilde{\Theta} \widetilde{\Pi} \widetilde{P} \widetilde{\Gamma}' \widetilde{Z}'$, where 
$\Theta  = \diag(\theta_1, ..., \theta_n)$, $\Pi = [\pi_1, \pi_2, \ldots, \pi_n]'$, 
$\Gamma = [\gamma_1, \gamma_2, \ldots, \gamma_n]'$, and $Z =  \diag(\zeta_1, ..., \zeta_n)$; similar for $(\widetilde{\Theta}, \widetilde{\Pi}, \widetilde{\Gamma}, \widetilde{Z})$.  We assume the identifiability conditions of Lemma~\ref{lemma:iden1} hold, so  $\|\theta\|=\|\zeta\|$ and $\|\tilde{\theta}\|=\|\tilde{\zeta}\|$.  
We impose the following regularity conditions :
\begin{equation} \label{D-cond-theta}
\max_{1\leq i\leq n}\{\theta_i + \tilde\theta_i \} \goto 0, \qquad 
\max_{1\leq i\leq n}\{\zeta_i + \tilde\zeta_i\}\goto 0, \qquad 
 \|\theta\|  \goto\infty, \qquad \|\tilde{\theta}\| \goto\infty,  
\end{equation}
\begin{equation} \label{D-cond-P}
\|(\|\theta\|^{-2}  \Pi'\Theta^2\Pi)^{-1} \| \leq C, \;\;  \|(\|\zeta\|^{-2} \Gamma'Z^2\Gamma)^{-1}\| \leq C, \;\;  \| PP'\| \leq C,    \;\; \min_{1\leq k\leq K}\{P_{kk}\}\geq C,  
\end{equation} 
\begin{equation} \label{D-cond-P-alt}
\|(\|\tilde{\theta}\|^{-2} \widetilde \Pi'\widetilde\Theta^2\widetilde\Pi)^{-1} \| \leq C, \;\;  \|(\|\tilde{\zeta}\|^{-2}\widetilde\Gamma'\widetilde Z^2\widetilde\Gamma)^{-1}\| \leq C, \;\;  \| \widetilde P\widetilde P'\| \leq C, \;\;     \min_{1\leq k\leq \widetilde{K}}\{\widetilde{P}_{kk}\}\geq C.
\end{equation} 
These conditions are mild: they are similar to (\ref{cond-theta})-(\ref{cond-P}) in Section \ref{subsec:DCMM}, but are slightly more complicated as the directed-DCMM has  more parameters than DCMM. 
The condition $\min_{1\leq k\leq K}\{P_{kk}\}\geq C$ is not needed in the undirected-DCMM, because the identifiability condition in Lemma~\ref{lemma:iden2} already yields $P_{kk}=1$ for $1\leq k\leq K$ (similar for $\widetilde{P}$).

Consider the limiting null distribution first. Recall that the IBM test statistic is 
$\psi_n = (1/8) Q_n/[C_n + \widetilde C_n]^{1/2}$.  
Under the null, $\Omega = \widetilde{\Omega}$, and so $(\Theta, \Pi, P, \Gamma, Z) = (\widetilde\Theta, \widetilde\Pi, \widetilde P,\widetilde \Gamma, \widetilde Z)$ by our identifiability conditions.  Especially,  $C_n$ and $\widetilde C_n$ have the same distribution.  
\begin{thm}\label{thm:null-d} ({\it Null behavior (directed-DCMM)}). 
Consider the pairwise comparison problem where both networks satisfy the directed-DCMM model (\ref{model1a})-(\ref{model1c}). Suppose Conditions \eqref{D-cond-theta}-\eqref{D-cond-P} and the identifiability conditions of Lemma \ref{lemma:iden1} hold.  As $n\goto\infty$, under $H_0$,  we have (1).  $\mathbb{E}[Q_n] = 0$, and $\mathrm{Var}(Q_n)  = 64 [1+o(1)] \cdot\mathbb{E}[C_n]$, (2).  $\mathbb{E}[C_n] = \tr([\Omega\Omega']^2) + O(\|\theta\|^4\|\zeta\|_4^4 + \|\zeta\|^4\|\theta\|_4^4) \asymp \|\theta\|^8$,  $\mathrm{Var}(C_n) \leq  C\|\theta\|^8(1+\|\zeta\|_3^6+\|\theta\|_3^6)$, and  $C_n/\mathbb{E}[C_n] \goto 1$ in probability, and (3).  $\psi_n\goto N(0, 1/2)$ in law.
%
\end{thm}
Theorem \ref{thm:null-d} is proved in the supplement. Compared with Theorem \ref{thm:null},   $\psi_n \goto N(0, 1/2)$   (and so 
$\sqrt{2} \psi_n \goto N(0,1)$) for the directed case here, 
 and $\psi_n \goto N(0, 1)$ for the undirected case there. This is because the variances of $Q_n$ in two cases are different by a factor of $2(1 + o(1))$.  As before, fix $0 < \alpha < 1$ and let $z_{\alpha}$ be the $(1-\alpha)$-quantile of $N(0,1)$. Consider the IBM test where we 
reject the null if and only if $\sqrt{2} \psi_n\geq z_\alpha$. 
  As $n\goto\infty$, by Theorem~\ref{thm:null}, the Type I error of the IBM test $\goto\alpha$ as expected. 

We now analyze the power.  Similarly, let $\Delta = \Omega - \widetilde\Omega$ and $r  = \mathrm{rank}(\Delta)$.  It is seen that $r \leq K + \widetilde{K}$. 
Since $\Omega$,  $\widetilde{\Omega}$, and $\Delta$ are asymmetric, the eigenvalues are not necessarily real, so it is more convenient to consider the singular values.  To abuse the notation a little bit, let $\delta_k$, $\lambda_k$, and $\tilde{\lambda}_k$,  be the $k$-th singular value of $\Delta$, $\Omega$ and $\widetilde\Omega$, respectively.   The following theorem is proved in the supplement. 
\begin{thm}\label{thm:alt-d}
({\it Power analysis for directed-DCMM}).  
Consider pairwise comparison problem where both networks satisfy the directed-DCMM model (\ref{model1a})-(\ref{model1c}), where conditions \eqref{D-cond-theta}-\eqref{D-cond-P} and the identifiability conditions in Lemma \ref{lemma:iden1} hold. Assume $\delta_1^2 / (\lambda_1 + \widetilde{\lambda}_1)  \goto \infty$ as $n\goto\infty$ and for fixed $K$ and $\widetilde K$, we have (1).  $\mathbb{E}[{Q}_n] = \tr([\Delta'\Delta]^2)  + o\big((\|\theta+\tilde{\theta}\|^2)\cdot\tr(\Delta'\Delta)\big)  = (1 + o(1)) \cdot  \tr([\Delta \Delta']^2)$ 
 and $\mathrm{Var}(Q_n) \leq  C\big(\tr([\Omega \Omega']^2) + \tr([\widetilde{\Omega} \widetilde{\Omega}']^2)  + [\tr(\Delta'\Delta)]^3\big) \leq  C\big(\|\theta\|^8 + \|\tilde{\theta}\|^8 + [\tr(\Delta'\Delta)]^3\big)$, and (2).   $\psi_n\to \infty$, in probability. Therefore, for any fixed $0 < \alpha < 1$ and the IBM test where we reject 
the null if and only if $\sqrt{2} \psi_n \geq z_{\alpha}$, the power of the test goes to $1$.   
%
%
%
\end{thm} 
Similarly,   the main condition of Theorem \ref{thm:alt-d} is $\delta_1^2 / (\lambda_1 + \widetilde{\lambda}_1)  \goto \infty$. To interpret, define
\begin{equation} \label{DefineSNR-d} 
\mathrm{SNR} = [1/(4\sqrt{2})]\cdot \tr([\Delta \Delta']^2) / \bigl[\tr([\Omega \Omega']^2) + \tr([\widetilde{\Omega} \widetilde{\Omega}']^2)\bigr]^{1/2}.   
\end{equation} 
We have two cases,  the weak signal case where $[\tr(\Delta'\Delta)]^3 \leq  \tr([\Omega \Omega']^2) + \tr([\widetilde{\Omega} \widetilde{\Omega}']^2)$ and the strong signal case where $[\tr(\Delta'\Delta)]^3 >   \tr([\Omega \Omega']^2) + \tr([\widetilde{\Omega} \widetilde{\Omega}']^2)$.    
By similar arguments, we have (a) in the weak signal case, $[\delta_1^2/(\lambda_1 + \tilde{\lambda}_1)]^2 \asymp \mathrm{SNR} 
\asymp \mathbb{E}[Q_n] / \mathrm{SD}(Q_n)$, and  the main condition of $\delta_1^2/ (\lambda_1 + \tilde{\lambda}_1) \goto \infty$ in Theorem \ref{thm:alt-d} is equivalent to that of $\mathrm{SNR} \goto \infty$, and (b) in the strong signal case, $\mathbb{E}[Q_n] / \mathrm{SD}(Q_n) \asymp \delta_1 \goto \infty$ as long as $\delta_1^2/(\lambda_1 + 
\tilde{\lambda}_1) \goto \infty$.

We now study the lower bound. Similarly, the goal is to show that within the class of all model pairs satisfying $\delta_1^2/(\lambda_1 + \tilde{\lambda}_1) \goto 0$, there exist pairs where the two models within the pair are asymptotically 
inseparable (i.e., the $\chi^2$-divergence $\goto 0$).  In detail, let ${\cal M}^{dir}_{n,0}(K)$ be all Bernoulli matrices $\Omega$ for  directed-DCMM models with $K$ communities where $\Omega = \Theta \Pi P \Gamma' Z$ and (\ref{model1a})-(\ref{model1c}) hold.   
Given a positive sequence $\{\beta_n\}_{n=1}^{\infty}$, an integer $K\geq 1$, and constants $0<c_0, c_1, c_2 <1$, we define a class of directed-DCMM models by 
\[
{\cal M}^{dir}_n(\beta_n, K, {\bf c}) =   \left\{\begin{array}{l}
\Omega \in {\cal M}^{dir}_{n,0}(K),\;   \max\{\theta_{\max},\zeta_{\max}\}  \leq K\beta_n,\;  \min\{\|\theta\|,\|\zeta\|\} \geq K^{-1}\beta_n^{-1},\\
P_{KK}=1,\; c_0K^{-1}\|\theta\|\|\zeta\| \leq \|\Omega\| \leq c_0^{-1}K\|\theta\|\|\zeta\|,\;  \|P\|_{\max}\leq c^{-1}_1 K,  \\
\|\theta\circ\pi^{(K)}\|\geq c_2 K^{-1}2^{-K/2}\|\theta\|,\;\; \|\zeta\circ\gamma^{(K)}\|\geq c_2 K^{-1}2^{-K/2}\|\zeta\|  
\end{array}
\right\},  
\]
where for short, ${\bf c} = (c_0 , c_1, c_2)$.  
Similar as before, let $\Delta = \Omega - \widetilde{\Omega}$, let $r, K, \tilde{K}$ be the rank of 
$\Delta$, $\Omega$ and $\widetilde{\Omega}$, respectively, and let  $\delta_k$, $\lambda_k$, and $\tilde{\lambda}_k$ be the $k$-th largest singular value of $\Delta$, $\Omega$, and $\widetilde{\Omega}$, respectively.  Define the class of directed-DCMM model pairs for the null case by 
\[
\mathcal{S}_n^{dir*}(\beta_n, K, {\bf c}) = \bigl\{(\Omega,\widetilde{\Omega})\in \mathcal{M}_n(\beta_n, K,  {\bf c})\times\mathcal{M}_n(\beta_n, K,  {\bf c}):\; \Omega = \widetilde{\Omega}\bigr\}, 
\]
and define a class of diercted-DCMM model pairs for the alternative case by 
\[
\mathcal{S}^{dir}_n(\beta_n,\rho_n, K, \widetilde{K}, {\bf c}) = \bigl\{(\Omega,\widetilde{\Omega})\in \mathcal{M}^{dir}_n(\beta_n, K,  {\bf c})\times\mathcal{M}^{dir}_n(\beta_n, \widetilde{K},  {\bf c}):\;  \delta_1^2 \geq \rho_n(\lambda_1 + \tilde{\lambda}_1)\bigr\}. 
\]
The following theorem is proved in the supplement. 
\begin{thm}(Minimax lower bound (directed-DCMM)).  \label{thm:LB-d}
	Fix $K\geq 1$, $c_0,c_1,c_2\in (0,1)$, and positive sequences $\{\beta_n\}_{n=1}^{\infty}$ and $\{\rho_{n}\}_{n=1}^{\infty}$ such that $\beta_n = o(1)$ and $\rho_n = o(1)$, we have
	\begin{equation}
	\label{def:risk1}
	\inf_{\psi}\bigg\{ \sup_{(\Omega,\widetilde{\Omega})\in\mathcal{S}_n^{dir*}(\beta_n, K, {\bf c})}\mathbb{P}(\psi = 1)
	+ \sup_{(\Omega,\widetilde{\Omega})\in\mathcal{S}^{dir}_n(\beta_n, \rho_n, K, K+1, {\bf c})}\mathbb{P}(\psi = 0)\bigg\}\goto1
	\end{equation}
	as $n\goto\infty$, where the infimum is taken over all possible tests $\psi$.  
\end{thm}
Combining Theorems  \ref{thm:null-d}-\ref{thm:LB-d}, we have the following phase transition. 
Consider a sequence of directed-DCMM model pairs indexed by $n$,  where for each pair, $\Omega$ and $\widetilde{\Omega}$ are the Bernoulli probability matrices, 
respectively. Recall that $\Delta = \Omega - \widetilde{\Omega}$ and 
$\mathrm{SNR} = (1/4\sqrt{2})  \tr([\Delta \Delta']^2) / [(\tr([\Omega \Omega']^2) + \tr([\widetilde{\Omega} \widetilde{\Omega}']^2)]^{1/2}$. 
In the pairwise comparison problem, we test  $H_0^{(n)}: \Omega = \widetilde{\Omega}$ versus $H_1^{(n)}: \Omega \neq 
\widetilde{\Omega}$.   We have the following phase transition. 

{\it Possibility}. When $\delta_1^2/(\lambda_1 + \tilde{\lambda}_1)  \goto \infty$, the two models are asymptotically separable,    and the sum of Type I and Type II errors of the IBM test  $\goto 0$.  

{\it Impossibility}.  When $\delta_1^2/(\lambda_1 + \tilde{\lambda}_1)  \goto  0$, the two models are not always asymptotically separable. 
    In fact, for each $\Omega$, we can pair it with an $\widetilde{\Omega}$ such that 
    $\delta_1^2/(\lambda_1 + \tilde{\lambda}_1)  \goto  0$, and 
     the $\chi^2$-divergence between the two models $\goto 0$. Therefore, for any tests, 
    the sum of Type I and Type II errors is $\geq 1 + o(1)$ (see also Remark 7).  
    
%
%

Our ideas are not tied to the DCMM or directed-DCMM models, and are extendable to  general Bernoulli probability model 
$A = \Omega - \diag(\Omega) + W$, where $\mathrm{rank}(\Omega) = K$. In fact, by SVD, we may write $\Omega = G U D V' H$ and $\widetilde{\Omega} = \widetilde{G} \widetilde{U} \widetilde{D} \widetilde{V}' \widetilde{H}$, where   
$G$ and $H$ are $n \times n$ positive diagonal matrices, 
$U, V$ are $n \times K$ matrices where each row has unit-$\ell_1$ norm, and 
$D = \diag(s_1,  \ldots, s_K)$, consisting nonzero singular values of $\Omega$; similar for $\widetilde{\Omega} = \widetilde{G} \widetilde{U} \widetilde{D} \widetilde{V}' \widetilde{H}$.  Our theorems are readily extendable if we translate the regularity conditions on $(\Theta, \Pi, P, \Gamma, Z)$ and $(\widetilde{\Theta}, \widetilde{\Pi}, \widetilde{P}, \widetilde{\Gamma}, \widetilde{Z})$ above to similar conditions on $(G, U, D, V, H)$ and  $(\widetilde{G}, \widetilde{U}, \widetilde{D}, \widetilde{V}, \widetilde{H})$.

\section{Real-data applications}  \label{sec:real} 
We use IBM to analyze the Enron email network and a gene co-expression network. 

\begin{figure}[tb] 
\centering
\includegraphics[height = 2.5 in, width = 2.84 in]{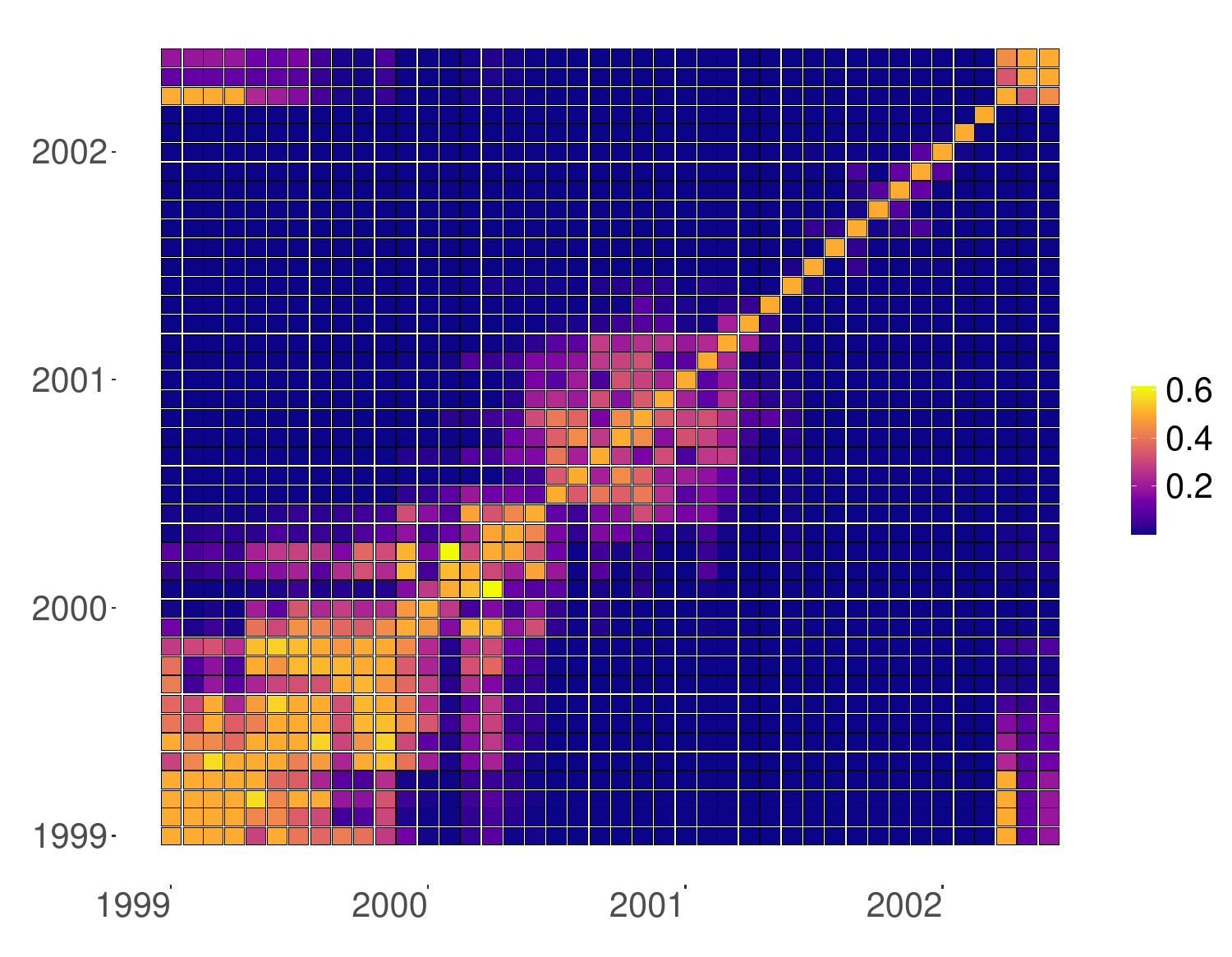}  
\includegraphics[height = 2.5 in, width = 2.84 in, trim=0 0 0 20, clip=true]{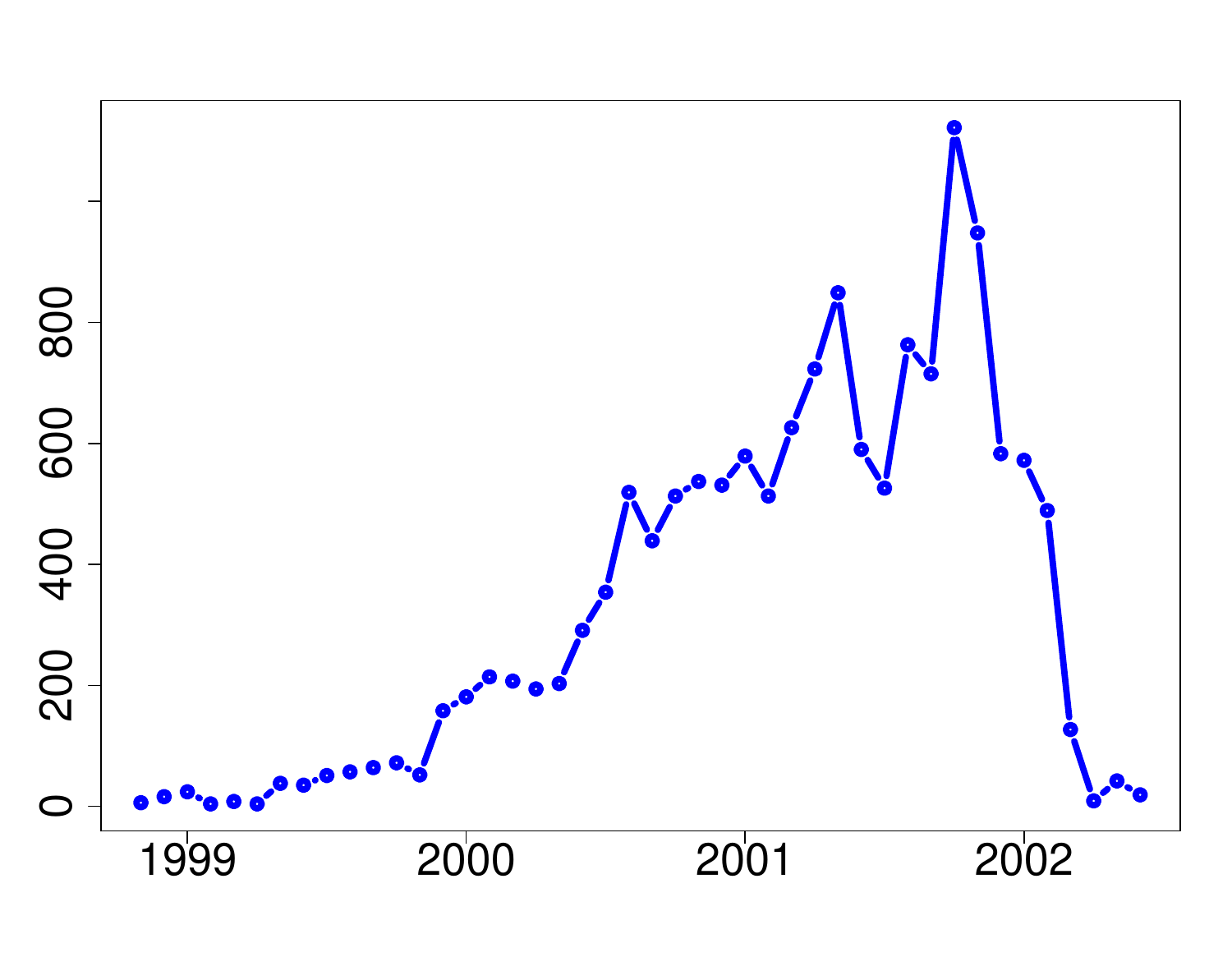}  
\caption{Enron networks (left: p-value heat map;  right: number of edges of each network).}  \label{fig:Enron}
\end{figure}

{\bf Analysis of the Enron email data}.   
The dataset contains the email communication data of $184$ users in a total of $44$ months (November  1999 to  June 2002), and provides valuable information for studying the Enron scandal. 
 For each month during the time period, we construct an undirected and unweighted network, where nodes $i$ and $j$ are connected if and only if they had email communication during that month. This gives us a total of $44$ networks for the same of $184$ nodes, with the number of edges varying from $200$ to $1400$ (see Figure \ref{fig:Enron} (right panel)).  For any $1 \leq i, j \leq 44$ and $i \leq j$, 
we conduct a network comparison  and derive 
an IBM test statistic $\psi_{n; ij}$.  By Theorem \ref{thm:null}, 
the $p$-value of the statistic is approximately $p_{ij} = \mathbb{P}(N(0,1) \geq \psi_{n; ij})$.  
The $p$-values are presented in Figure~\ref{fig:Enron} (left panel)  as a heatmap.  
Note that a darker cell means a smaller $p$-value, suggesting that 
the difference between the two networks being compared is more significant.

The heatmap suggests that  there are two major ``change points'', corresponding to August of 2000 and August of 2001 (note that each time point is a month), respectively. At the first change point, what happened is that the Enron stock hit all-time high of $\$ 90.56$ per share with a market valuation of $70$ billion dollars, indicating an increasing popularity of the company. Note that  the right panel of Figure~\ref{fig:Enron} also suggests a significant increase of email numbers in that month, confirming the sudden change of the email network structure.  
At the second change point of August of 2001,  what happened is that former CEO Jeffrey
Skilling  resigned on August 14th and Kenneth Lay took over.  After that, the scandal was gradually discovered by the public. This explains why the network structures after August 2001 are so different from each other. 
Moreover, on the right panel of Figure~\ref{fig:Enron}, we also find drastic changes around that time point, on the number of email changes. 
 
From the heatmap, we also observe that the monthly email network are very different since late 2001, and the underlying reason is that Enron was undergoing many significant changes. For example, in November of 2001, Enron restated the 3rd quarter earnings and the dynegy deal collapsed. In January of 2002, criminal investigation started and Kenneth Lay resigned (and was implicated in fraud later in February). For finer details of the heatmap, see  Section~\ref{supp:numerical} of the supplementary material.   

\spacingset{1.2}
\spacingset{1.5}

{\bf Analysis of the gene co-expression network}. 
We aim to identify subtle disparity between the gene co-expression networks from healthy and type 2 diabetic donors (T2D) based on transcriptome of thousands of islet cells. Hormone-secreting cells within pancreatic islets of Langerhans play important roles in metabolic homeostasis and disease. We leverage the transcriptome of 2,209 cells from six healthy and four T2D donors profiled using Smart-seq2 protocol in study \citep{segerstolpe2016single}. The cells were broadly categorized based on the transcription profiles into six major types, including exocrine ductal and acinar cells, and endocrine $\beta,\alpha,\gamma$ and $ \delta$ cells (cell types with less than 100 cells are not detailed here), as summarized in Table \ref{tab: sc_seg_cell_type}.

\spacingset{1.2}
\begin{table}[tb]
\begin{center}
\begin{tabular}{r| ccccccc|c}
\hline
cell type & $\beta$ & $\alpha$ &$\gamma$  & $\delta$ & acinar & ductal & other & total\\
\hline
\# Normal (control)& 171 & 443 & 75 & 59 &112 & 135 & 102 & 1097 \\
\# Type 2 diabetic (case) & 99 & 443 & 122 & 55 & 73 & 251 & 69 & 1112\\
\# total & 279 & 886 & 197 & 114 & 185 & 386 & 171 & 2209 \\
\hline
\end{tabular}
\end{center}
\caption{Number of cells in each identified cell types after quality control, categorized by the disease status of donor (normal or type 2 diabetic).  Cell types with less than 100 cells are labeled as ``other".}
\label{tab: sc_seg_cell_type}
\end{table}
\spacingset{1.5}

The transcriptome heterogeneity for healthy (controls) and T2D (cases) individuals is examined by comparing gene co-expression networks in a cell type resolved manner. To construct the gene co-expression network, the raw counts from single-cell RNAseq data are first normalized and log transformed.
A total of 25525 gene expressions were detected. We restrict our analysis to 500 most highly expressed genes selected using ``vst'' method provided by ``Seurat'' toolbox \citep{RN262}, as a convention in Single Cell literature.
Then Spearman correlations between each pair of highly variable genes are calculated separately for healthy and T2D cells and for each cell type. The absolute value of correlation is hard-thresholded at $99\%$ quantile to generate a binary adjacency matrix (the thresholding is used to ensure the difference in average degree does not contribute to the testing significance).  Also,  thresholding at the $99\%$ quantile is equivalent to thresholding the Pearson correlations (in magnitude) at a threshold between $0.4$ to $0.8$ across six different cell types, which is common in practice. 
We view the binary matrix as the adjacency matrix of the gene co-expression network.

\begin{figure}[htb!] 
  \centering
    \includegraphics[width=0.32\textwidth, ,trim=0 0 0 0 ,clip = True]{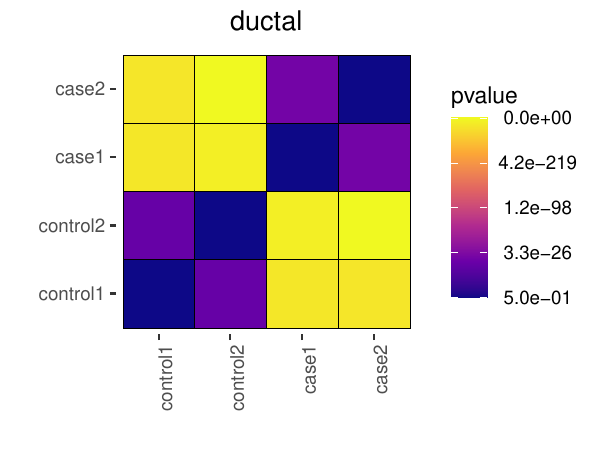}
    \includegraphics[width=0.32\textwidth, ,trim=0 0 0 0 ,clip = True]{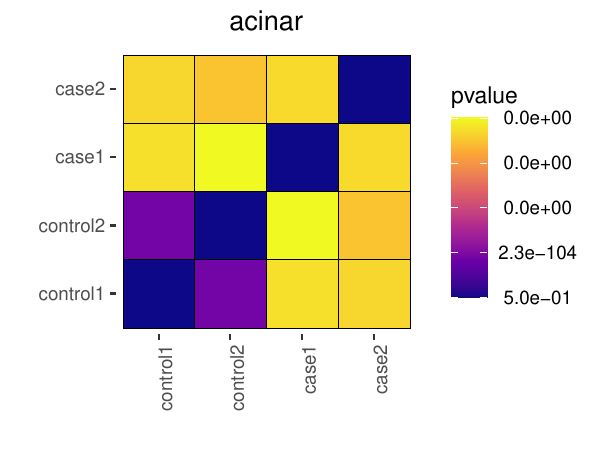}
    \includegraphics[width=0.32\textwidth, ,trim=0 0 0 0 ,clip = True]{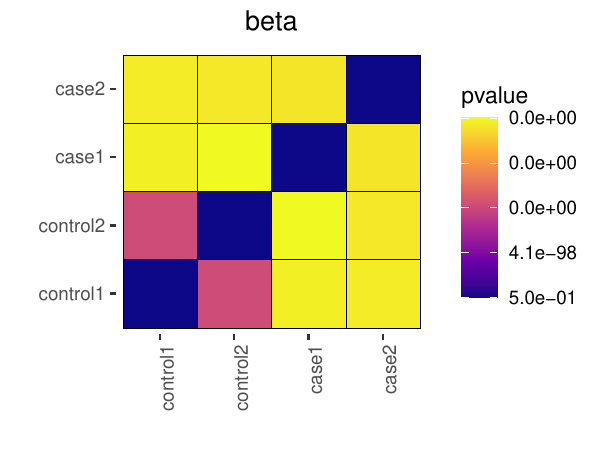}
    \includegraphics[width=0.3\textwidth, ,trim=0 0 100 0 ,clip = True]{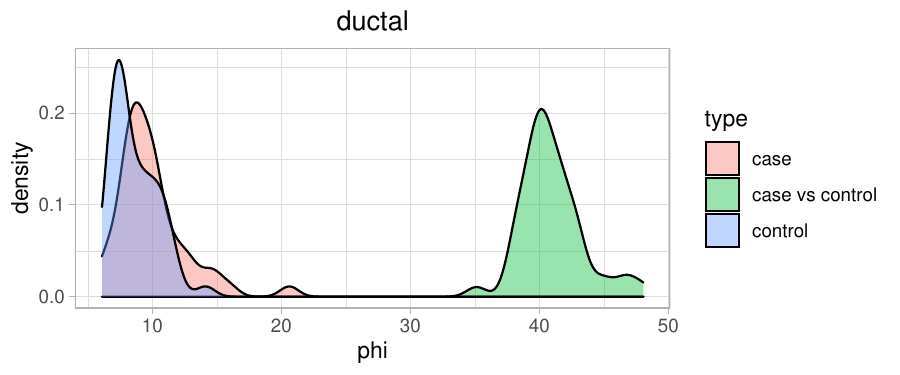}
    \includegraphics[width=0.3\textwidth, ,trim=0 0 100 0 ,clip = True]{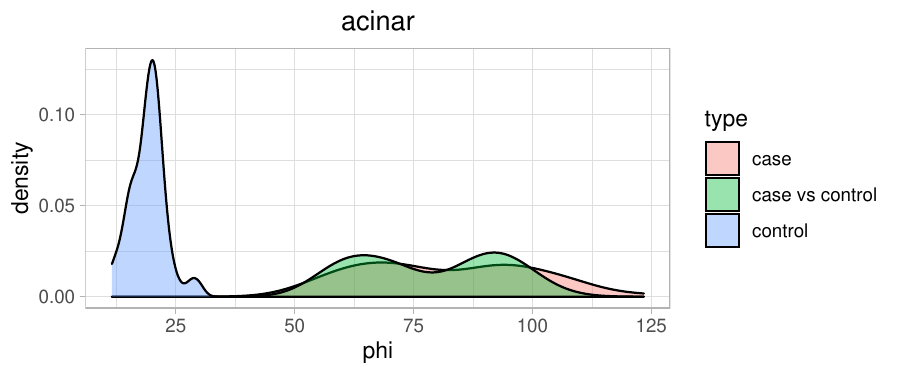}
    \includegraphics[width=0.38\textwidth, ,trim=0 0 0 0 ,clip = True]{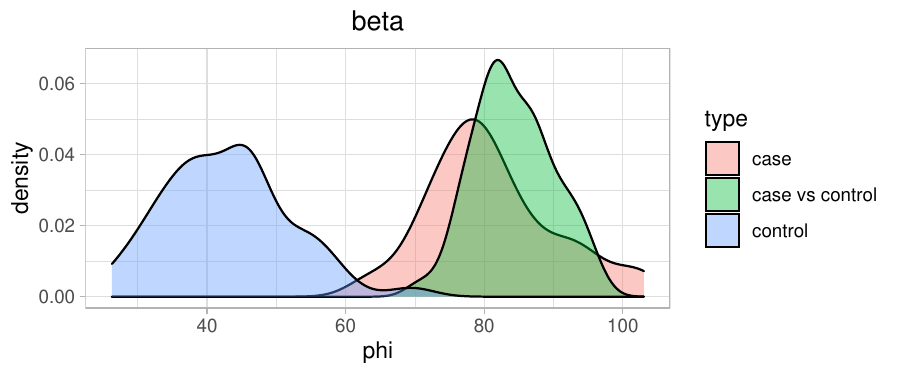}
    \includegraphics[width=0.32\textwidth, ,trim=0 0 0 0 ,clip = True]{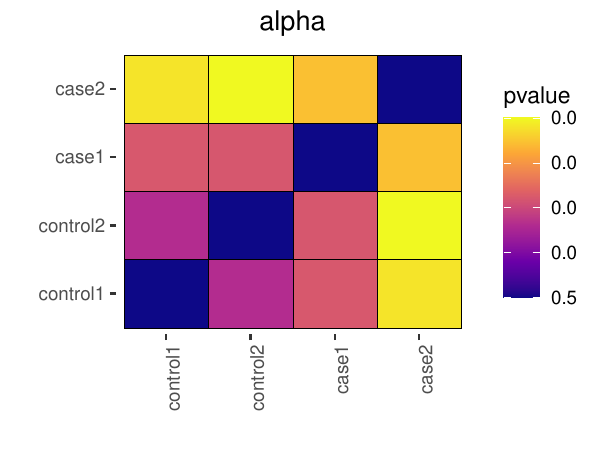}
    \includegraphics[width=0.32\textwidth, ,trim=0 0 0 0 ,clip = True]{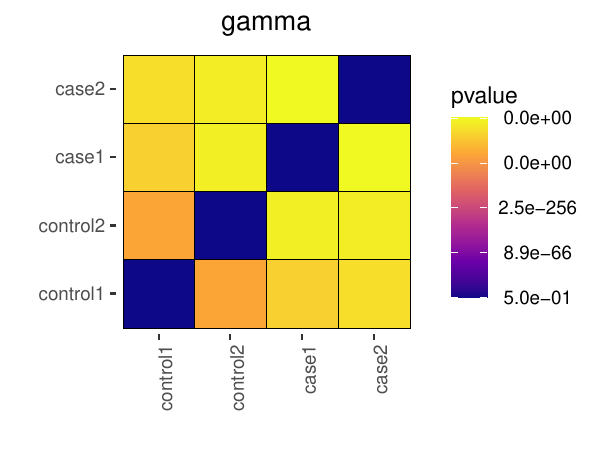}
    \includegraphics[width=0.32\textwidth, ,trim=0 0 0 0 ,clip = True]{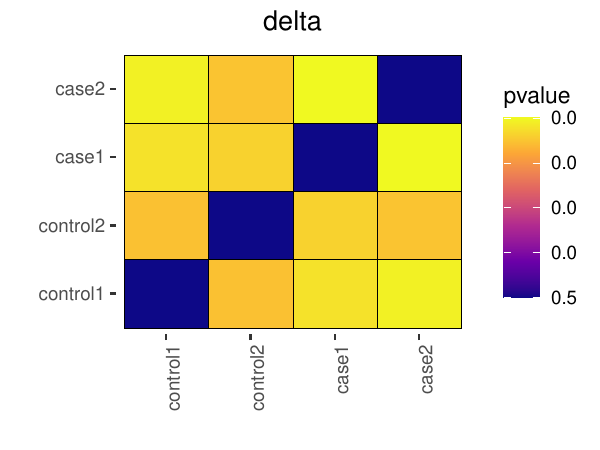}
    \includegraphics[width=0.3\textwidth, ,trim=0 0 100 0 ,clip = True]{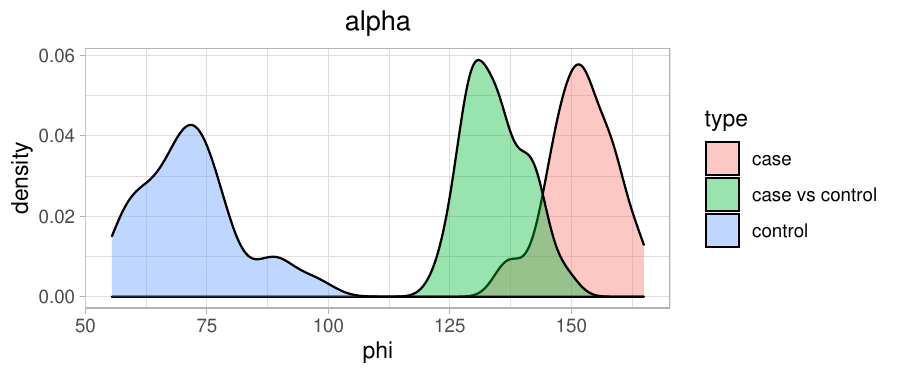}
    \includegraphics[width=0.3\textwidth, ,trim=0 0 100 0 ,clip = True]{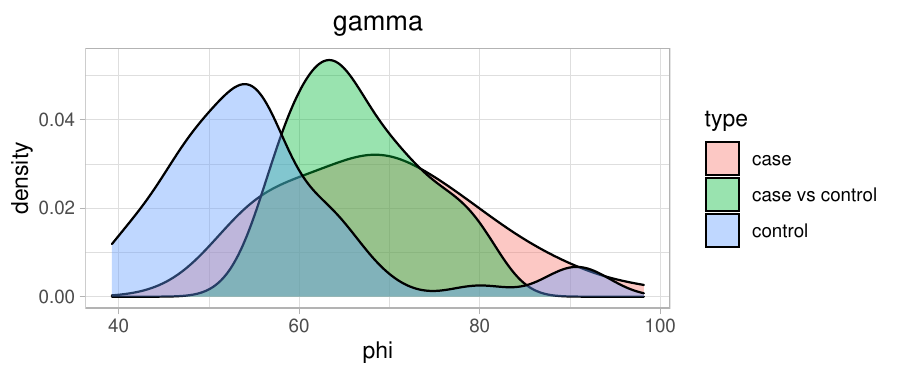}
    \includegraphics[width=0.38\textwidth, ,trim=0 0 0 0 ,clip = True]{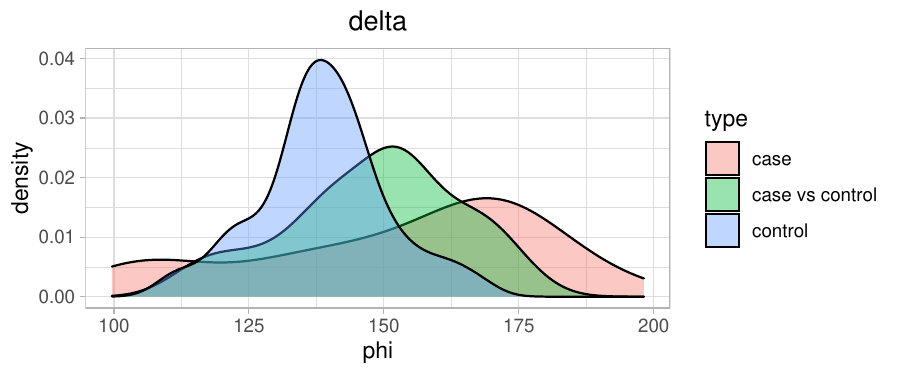}
    \caption{Plots of p-values of test statistics by comparing gene co-expression network between cases and controls for each cell type (for comparison,  cases and controls are randomly split into half and a network is built for each half of cells). Density plots of test statistics values are added for each cell type for three testing objectives, i.e. control   versus control   (control), case  versus case  (case), and case versus control (case v.s. control).
    }
    \label{fig: sc_seg_pvalues}
\end{figure}

To compare the healthy versus T2D network disparity with the within-healthy or within T2D network disparity, the cells {\it are randomly split into halves}  for each of cases and controls.  For each random split,  we construct $4$ networks as above, denoted by Case1, Case2, Control1, and Control2. For each pair of networks, we apply the IBM statistic and obtain a $p$-values similarly as above. The process is repeated for $50$ times and the medians of $p$-values are visualized in Figure \ref{fig: sc_seg_pvalues}. The density plot for IBM test scores  are also provided.

\spacingset{1.2}
\spacingset{1.5}

Among the results from six cell types, ductal cells demonstrate the most remarkable distinction of gene co-expression networks between cases and controls, which results in $p$-values in off-diagonal block orders of magnitude smaller than $p$-values in two diagonal blocks (Figure \ref{fig: sc_seg_pvalues}). This is indicative of significant alterations in gene expression in cell from T2D subjects compared to healthy subjects. Acinar, $\alpha$ and $\beta$ cells uncover similar patterns of prominent distinction between cases and controls, in addition, the two networks built from a random split of cases also express comparable disparity. For $\gamma$ and $\delta$ the disease-dependent effect is less evident regarding network structure.  To be more specific, the number of differentially expressed (DE) genes between the cells from T2D objects (case) and health objects (control)
is 250 for ductal, 50-100 for acinar, $\alpha$ and $\beta$, and less than 10 for $\gamma$ and $\delta$.
The density plots align with the biological observations that ductal cells have the highest number of differentially expressed genes, where the test statistic for case versus control test is much larger than that for the control versus control or case versus case test. Followed by acinar, $\beta$ and $\alpha$ cells where the case vs control test statistic values are greater than control vs control, but not significantly higher than case vs case. For $\gamma$ and $\delta$ cells, there is no evident difference between the case versus control or control versus control.

The results are consistent with the evidence revealed in previous study \citep{segerstolpe2016single} that ductal cells entail the most different differentially expressed (DE) genes between cells from health and T2D donors (250 genes), followed by acinar, $\alpha$ and $\beta$ cells (50-100 DE genes each), while for $\gamma$ and $\delta$ cell less than 10 DE genes are identified. Results also confirm that comparing healthy and T2D gene co-expression networks in a cell-type manner 
uncovers the cell-level heterogeneity associated with the disease and shed light on future functional studies. 

In conclusion, the IBM test statistic performs well in both data sets.  It is useful for identifying change points of dynamic networks, for identifying network pairs with significant differences,  and  for visualizing how a large number of networks are 
different from each other.

\section{Simulations} \label{sec:numer} 
We investigate undirected networks and directed networks in Experiment 1 and Experiment 2, respectively, 
We also compare IBM with spectral methods in Experiment 3.

{\bf Experiment 1: Undirected networks.} 
Given $(n, K, \beta_n, b_n)$, we first generate $\theta_i=\beta_n\times \theta_i^u/\|\theta^u\|$, for $1\leq i\leq n$, where $\theta_i^{u}\stackrel{iid}{\sim} \text{Unif}(2,3)$ and $\beta_n$ controls the $\ell^2$-norm of $\theta$. We then generate $\pi_1,\cdots, \pi_n\stackrel{iid}{\sim} \text{dir}(1,\cdots,1)$, for $1\leq i\leq n$, where $\text{dir}$ is the Dirichlet distribution. Let $P=(1-b_n)\mathbf{I}_K+b_n\mathbf{1}_K\mathbf{1}_K'$ and $\Theta=\diag(\theta_1,\theta_2,\ldots,\theta_n)$ and $\Pi=[\pi_1,\pi_2,\ldots,\pi_n]'$. We construct $\Omega=\Theta\Pi P\Pi'\Theta$. We then generate $\widetilde{\Omega}$. There can be multiple sources of differences between $\Omega$ and $\widetilde{\Omega}$, e.g., different degree parameters, different number of communities, or different mixed memberships. We investigate the three cases separately. 

\textit{Case 1: Different degree parameters}. We let $\widetilde{\Omega}=\widetilde{\Theta}\Pi P\Pi'\widetilde{\Theta}$, where $(\Pi, P)$ are the same as those in $\Omega$ and $\tilde{\theta}$'s are generated as follows: $\tilde{\theta}_i=\beta_n\times \tilde{\theta}_i^u/\|\tilde{\theta}^u\|$, for $1\leq i\leq n$, where $\tilde{\theta}_i^{u}\stackrel{iid}{\sim} 0.95\delta_1+0.05\delta_3$ with $\delta_a$ representing a point mass at $a$. We fix $(n, K)=(1000, 5)$ and let $\beta_n$ range from 6 to 10.5 with a step size 1.5. As $\beta_n$ increases, the network becomes less sparse. For each value of $\beta_n$, we select $b_n$ (the off-diagonal elements of $P$) such that the SNR defined in \eqref{DefineSNR} is fixed at $3.75$.

\textit{Case 2: Different numbers of communities}. We construct $\Omega$ and $\widetilde{\Omega}$ such that the two networks have $K$ and $2K$ communities, respectively. Let  
$P=(1-b_n)\mathbf{I}_K+b_n\mathbf{1}_K\mathbf{1}_K'$ and 
$\widetilde{P}=(1-b_n)\mathbf{I}_{2K}+b_n\mathbf{1}_{2K}\mathbf{1}_{2K}'$. We generate $iid$ samples of $\tilde{\pi}_i\in\mathbb{R}^{2K}$ from $\text{dir}(1,\cdots,1)$, for $1\leq i\leq n$, and construct $\pi_i\in\mathbb{R}^K$ by $\pi_i(k)=\tilde{\pi}_i(2k-1)+\tilde{\pi}(2k)$, for $1\leq k\leq K$, $1\leq i\leq n$. Let $\Theta$ be generated in the same way as before (see the paragraph above Case 1). Let $\Omega=\Theta\Pi P\Pi'\Theta$ and $\widetilde{\Omega}=\Theta\widetilde{\Pi}\widetilde{P}\widetilde{\Pi}'\Theta$. In this construction, each community in $\Omega$ is split into two communities in $\widetilde{\Omega}$.  
Fix $(n, K)=(1000, 2)$. We let $\beta_n$ range from 6 to 15 with a step size 3. For each $\beta_n$, $b_n$ is chosen such that the SNR in \eqref{DefineSNR} is equal to 2.25. 

\textit{Case 3: Different mixed membership vectors}. Fix $(n, K)=(1000, 2)$. We generate $(\Theta, P)$ in the same way as in Case 1. We then generate ${\pi}_i\stackrel{iid}{\sim} \text{dir}(1.6,0.4)$ and $\tilde{\pi}_i\stackrel{iid}{\sim} \text{dir}(1,1)$, $1\leq i\leq n$. Let $\Omega=\Theta\Pi P\Pi'\Theta$ and $\widetilde{\Omega}=\Theta\widetilde{\Pi}P\widetilde{\Pi}'\Theta$. Let $\beta_n$ range from 6 to 15 with a step size 3, where for each value of $\beta_n$ we select $b_n$ such that the SNR is equal to 2.

\spacingset{1.2}
\begin{figure}[tb!]
    \centering
    \includegraphics[height=7.2cm]{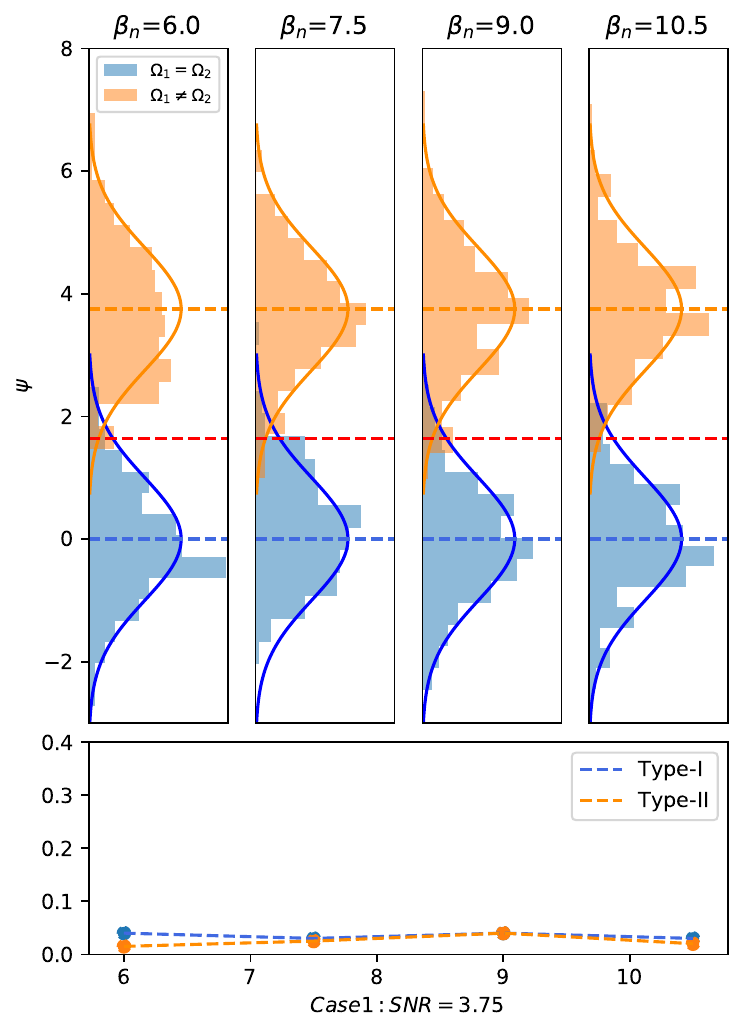}
    \includegraphics[height=7.2cm]{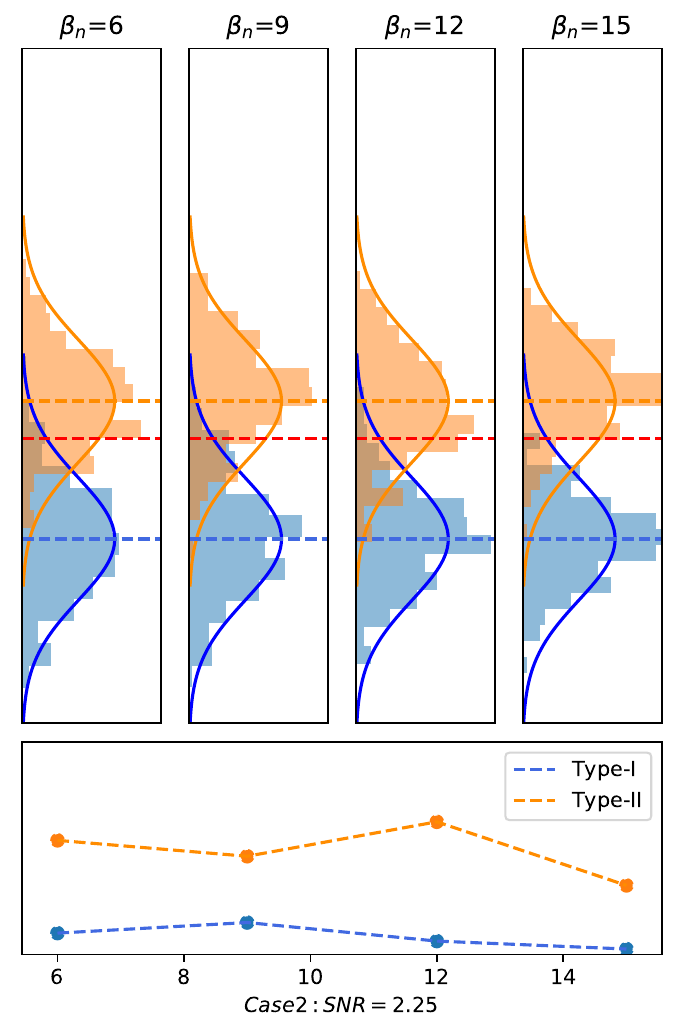}
    \includegraphics[height=7.2cm]{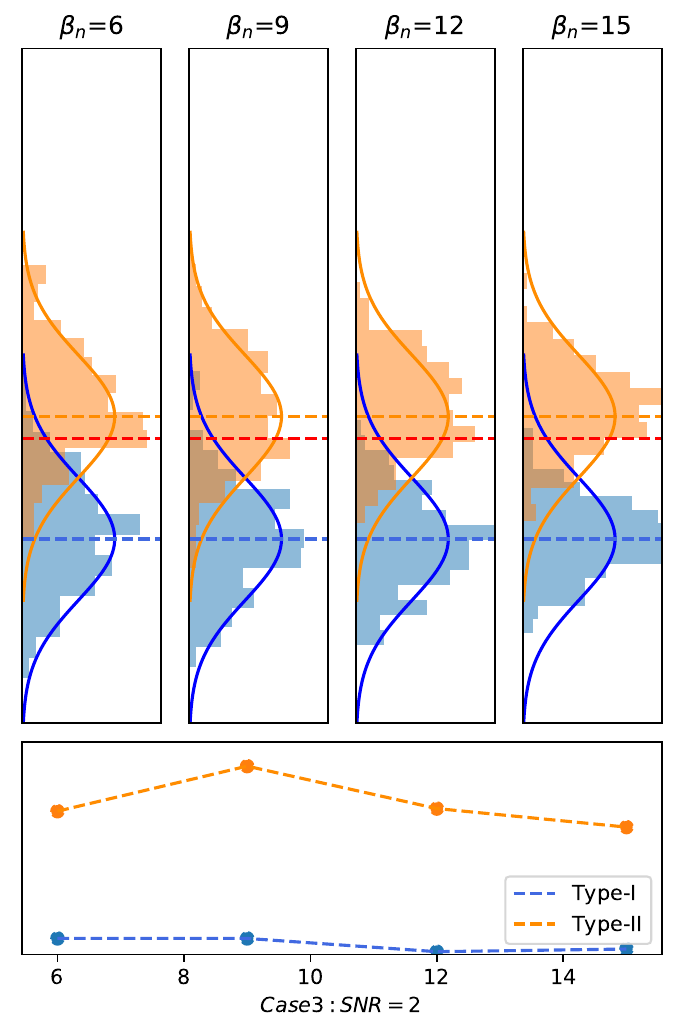}
    \caption{\small The IBM test for undirected networks, where top panels show the histograms of $\psi_n$ and bottom panels show the testing errors. In the alternative, the difference between $\Omega$ and $\widetilde{\Omega}$ lies on degree parameters (Case 1), number of communities (Case 2), and mixed memberships (Case 3), respectively. For each case, $\beta_n$ controls network sparsity. As $\beta_n$ varies, we keep the SNR in \eqref{DefineSNR} unchanged. Orange dashed line:  SNR.  Red dashed line: cut-off of level-95\% IBM test.}  \label{fig:Exp-Undirected}
\end{figure}

\spacingset{1.5}

For each parameter setting, we first generate $(\Omega, \widetilde{\Omega})$ and then generate 400 independent networks, $\{A_t\}_{t=1,\cdots,400}$, from $\Omega$ and 200 independent networks, $\{\widetilde{A}_t\}_{t=1,\cdots,200}$, from $\widetilde{\Omega}$. We use them to construct 200 instances of the null hypothesis, $\{(A_t, A_{t+200})\}_{1\leq t\leq 200}$, and apply the level-95\% IBM test to each instance.   We also construct 200 instances of the alternative hypothesis, $\{(A_t, \widetilde{A}_t)\}_{1\leq t\leq 200}$.  The results are presented in Figure~\ref{fig:Exp-Undirected}.

For a wide range of $\beta_n$ (e.g., $\beta_n\in [6,11]$), the histogram of the test statistic (in blue) fits well with  $N(0,1)$  under the null, and the type-I error is $\approx 5\%$ (for $\beta_n = 15$ in Case 3, the fitting is less well; in this setting, the null standard deviation of the test statistic is smaller than 1, which means that our test is conservative, and so the type-I error is still under control).  
Under the alternative, the histogram of the test statistic (in orange) converges to a normal distribution centered at the SNR (see \eqref{DefineSNR}), and the type-II error is small as long as the SNR is properly large.  From Case 1 to Case 3, we have decreased the SNR  purposely so that it is increasingly more difficult to separate two hypotheses. The type-II error is also increasing. 
These observations validate our theory in Section~\ref{sec:main}.

{\bf Experiment 2: Directed networks.} 
Similarly,  we consider $3$ different cases of $(\Omega, \widetilde{\Omega})$:

\textit{Case 4: Different degree parameters}. Let $\Omega=\Theta\Pi P\Gamma'Z$ and $\widetilde{\Omega}=\widetilde{\Theta}\Pi P\Gamma'\widetilde{Z}$, where $P=(1-b_n)\mathbf{I}_K+b_n\mathbf{1}_K\mathbf{1}_K'$, $\pi_1,\cdots, \pi_n,\gamma_1,\cdots,\gamma_n\stackrel{iid}{\sim}\text{dir}(1,...,1)$, and $\theta_1,\ldots,\theta_n, \zeta_1,\ldots,\zeta_n$ are as follows: We draw $\theta_1^{u},\cdots, \theta_n^{u},\zeta_1^{u},\cdots, \zeta_n^{u}\stackrel{iid}{\sim} \text{Unif}(2,3)$ and $\tilde{\theta}_1^{u},\cdots, \tilde{\theta}_n^{u},\tilde{\zeta}_1^{u},\cdots, \tilde{\zeta}_n^{u}\stackrel{iid}{\sim} 0.95\delta_1+0.05\delta_3$, and let $\theta_i=\beta_n\times \theta_i^u/\|\theta^u\|$, $\zeta_i=\beta_n\times \zeta_i^u/\|\zeta^u\|$, $\tilde{\theta}_i=\beta_n\times \tilde{\theta}_i^u/\|\tilde{\theta}^u\|$, and $\tilde{\zeta}_i=\beta_n\times \tilde{\zeta}_i^u/\|\tilde{\zeta}^u\|$.
Fix $(n, K)=(1000, 5)$ and let $\beta_n$ range from 6 to 10.5 with a step size 1.5. For each $\beta_n$, choose $b_n$ so that the SNR in \eqref{DefineSNR-d} is fixed at $3.9$.

\textit{Case 5: Different numbers of communities}. This setting is similar to that of 
Case 2, where we construct $\Omega$ and $\widetilde{\Omega}$ such that the two networks have $K$ and $2K$ communities, respectively. 
Let $\Omega=\Theta\Pi P\Gamma'Z$ and $\widetilde{\Omega}=\Theta\widetilde{\Pi} \widetilde{P}\widetilde{\Gamma}'Z$. Here, $\Theta$ and $Z$ are generated in the same way as in Case 4,      
$P=(1-b_n)\mathbf{I}_K+b_n\mathbf{1}_K\mathbf{1}_K'$, and 
$\widetilde{P}=(1-b_n)\mathbf{I}_{2K}+b_n\mathbf{1}_{2K}\mathbf{1}_{2K}'$. 
Generate $\tilde{\pi}_1,\cdots, \tilde{\pi}_n,\tilde{\gamma}_1,\cdots, \tilde{\gamma}_n\stackrel{iid}{\sim} \text{dir}(1,\cdots,1)$ and let $\pi_i(k)=\tilde{\pi}_i(2k-1)+\tilde{\pi}_i(2k)$ and $\gamma_i(k)=\tilde{\gamma}_i(2k-1)+\tilde{\gamma}_i(2k)$, for $1\leq k\leq K$ and $1\leq i\leq n$.  Here,  each (incoming or outgoing) community in $\Omega$ is split into two in $\widetilde{\Omega}$.   
We fix $(n, K)=(1000, 2)$, let $\beta_n$ range from 6 to 15 with a step size 3, and choose $b_n$ so that the SNR is fixed at $ 3.2$. 

\textit{Case 6: Different  membership vectors}. Fix $(n, K)=(1000, 2)$. Let $(\Theta, Z, P)$ be the same as in Case 4. Generate ${\pi}_1,\cdots, {\pi}_n,\gamma_1,\cdots,\gamma_n \stackrel{iid}{\sim} \text{dir}(1.6,0.4)$ and then generate $\tilde{\pi}_1,\cdots, \tilde{\pi}_n,\tilde{\gamma}_1,\cdots, \tilde{\gamma}_n\stackrel{iid}{\sim} \text{dir}(1,1)$. Let $\Omega=\Theta\Pi P\Gamma'Z$ and $\widetilde{\Omega}=\Theta\widetilde{\Pi}P\widetilde{\Gamma}'Z$. Let $\beta_n$ range from 6 to 15 with a step size 3. Choose $b_n$ accordingly so the SNR in \eqref{DefineSNR-d} is $3$.

\spacingset{1.2}
\begin{figure}[tb!]
    \centering
    \includegraphics[height=7.2cm]{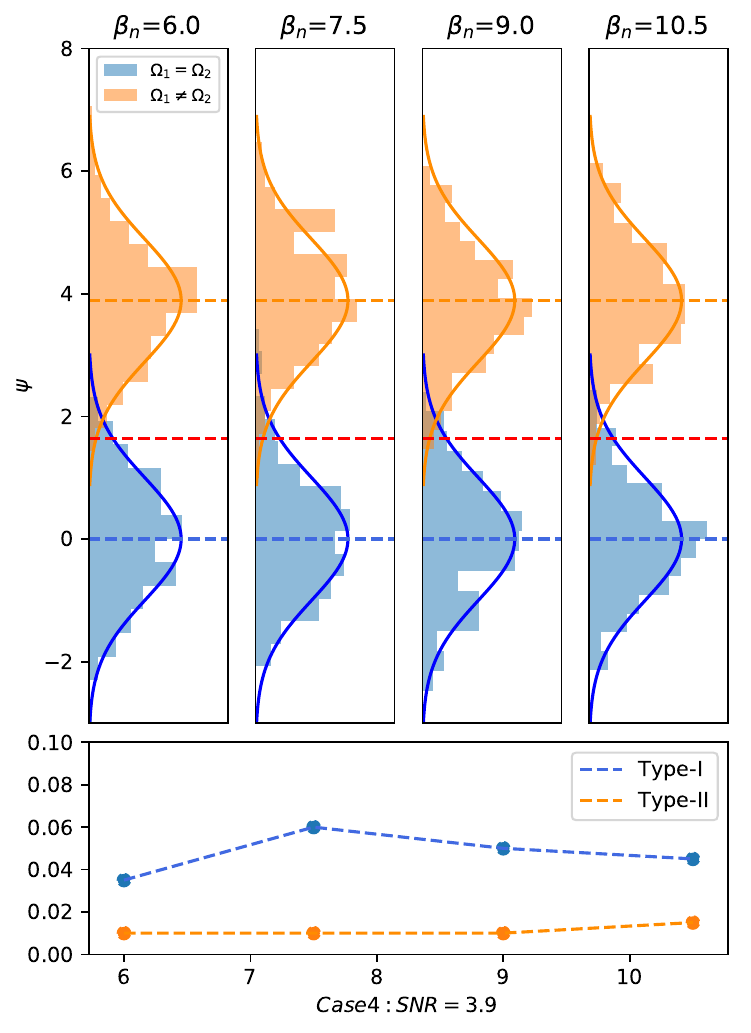}
    \includegraphics[height=7.2cm]{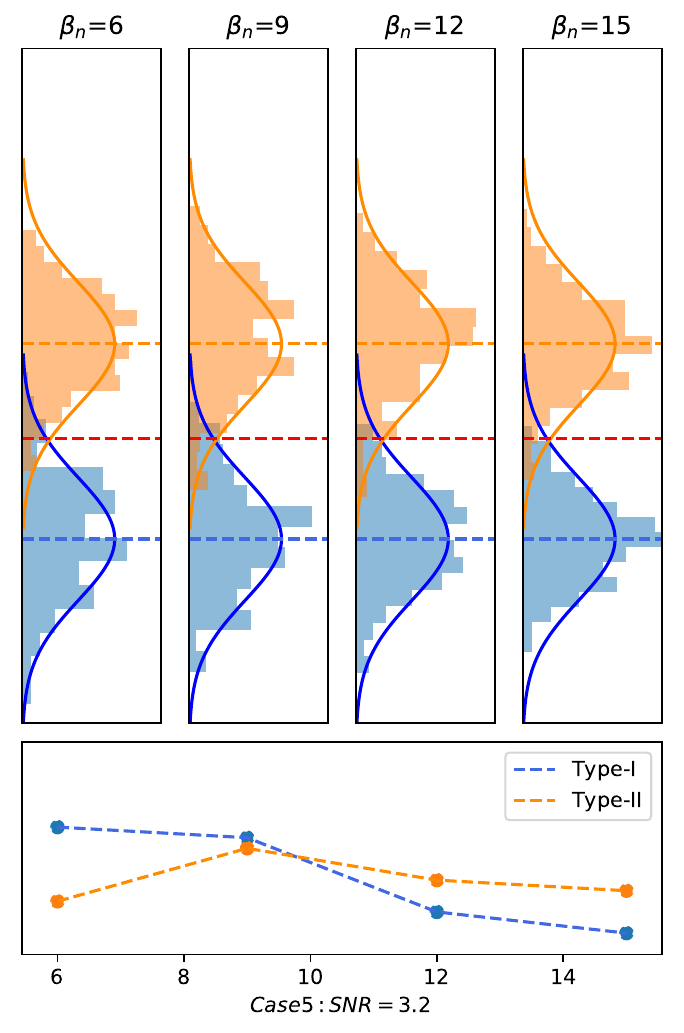}
    \includegraphics[height=7.2cm]{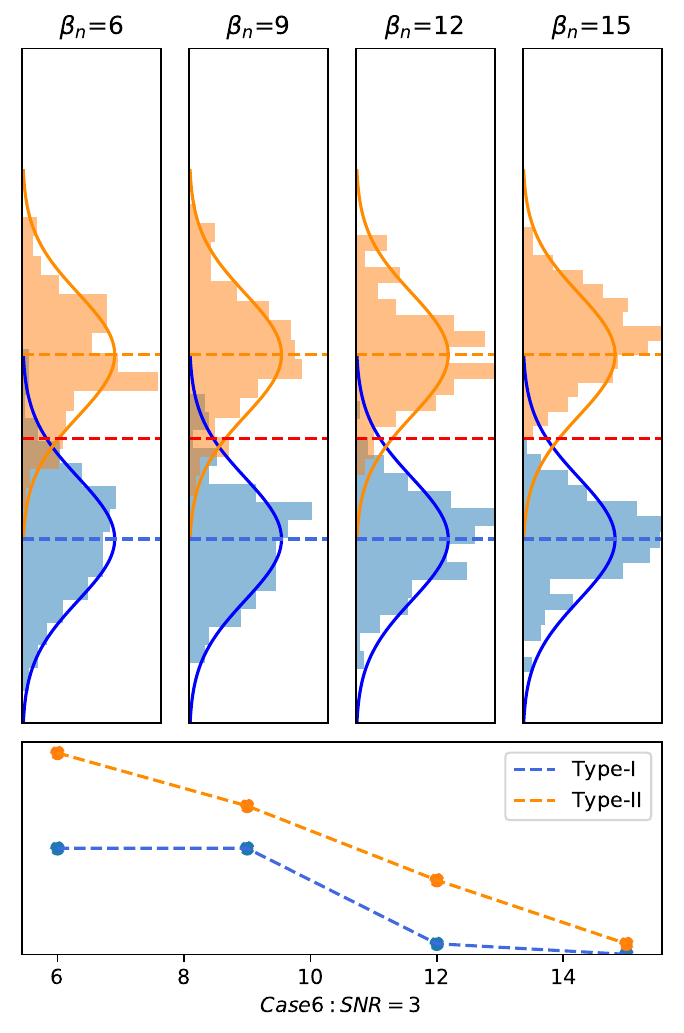}
   \caption{\small The IBM test for directed networks, where top panels show the histograms of $\sqrt{2}\psi_n$ and bottom panels show the testing errors. In the alternative, the difference between $\Omega$ and $\widetilde{\Omega}$ lies on degree parameters (Case 4), number of communities (Case 5), and mixed memberships (Case 6), respectively. For each case, $\beta_n$ controls network sparsity. As $\beta_n$ varies, we keep the SNR in \eqref{DefineSNR-d} unchanged. Orange dashed line:  SNR.  Red dashed line: cut-off of level-95\% IBM test.}  \label{fig:Exp-Directed}
\end{figure}
\spacingset{1.5}
 
For each parameter setting, once $\Omega$ and $\widetilde{\Omega}$ are generated, we then construct 200 pairs of networks under the null hypothesis and 200 pairs under the alternative hypothesis, similarly as in Experiment 1.  The results are in Figure~\ref{fig:Exp-Directed}. Similar to the case of undirected networks, the behaviors of the test statistic both under the null and under the alternative are consistent with our theory. The type-I error is controlled under $ 5\%$ in all settings, and the type-II error is reasonably small. 
When $\beta_n$ is large (e.g., $\beta_n=15$ in Case 5 and Case 6), the variance of the IBM test statistic gets smaller than 1, under both the null  and the alternative; therefore, although our test statistic tends to be conservative, the type-I and type-II errors are even smaller.

{\bf Experiment 3: Comparison with the spectral approach.}  
We compare IBM with the test in \cite{TSTLi2018}. Their test statistic explores the difference between the principal eigen-space of two networks and is defined to be $T_{n,K}= (nK)^{-1} \|(\widetilde{\Xi}\hat{O}-\hat{\Xi})\hat{\Lambda}\|_F^2$, where $\hat{\Xi}$ and $\widetilde{\Xi}$ contain the first $K$ eigenvectors of $A$ and $\widetilde{A}$, respectively, $\hat{\Lambda}$ is a diagonal matrix whose diagonal elements are the $K$ largest eigenvalues (in magnitude) of $A$, and $\hat{O}$ is an orthogonal matrix that minimizes $\|\widetilde{\Xi}O-\hat{\Xi}\|_F^2$.  
They estimated the null distribution of $T_{n,K}$ by assuming that both networks follow an undirected SBM, but this estimate is not valid in our setting, as we consider the more general DCMM. Instead, we use the true mean and standard deviation of $T_{n,K}$ under the null hypothesis (obtained from simulating data sets from the true null model) to standardize it. The resulting test statistic is not practically feasible, but it is still interesting to see its comparison with the IBM test.

\begin{figure}[tb!]
    \centering
    \includegraphics[height=6 cm]{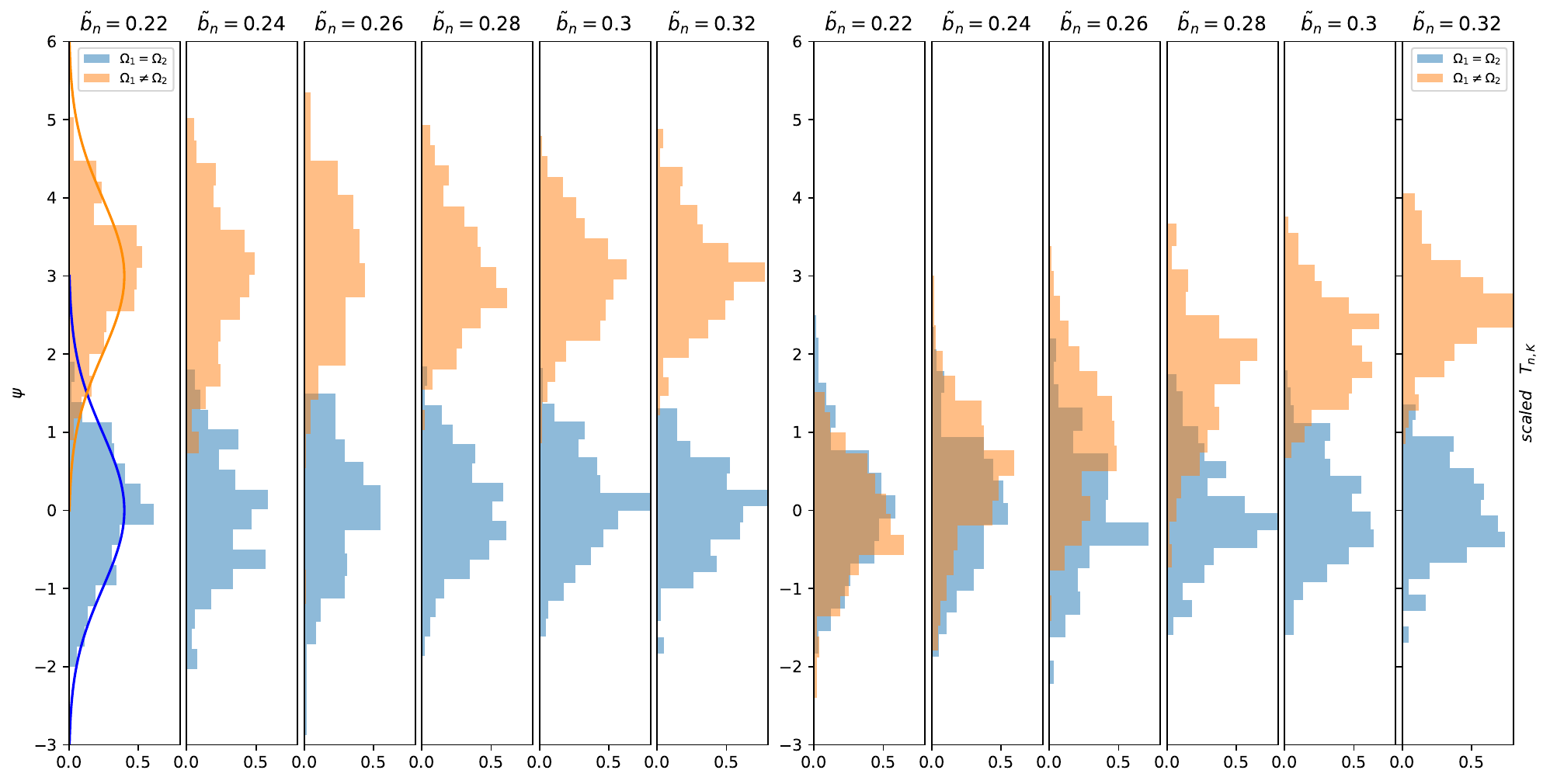}
    \caption{\small Comparison between the IBM test (left) and the eigenspace test (right).}  \label{fig:Exp-Compare}
\end{figure}

Fix $(n,K)=(1000,2)$. We generate $\theta_1^{u},\cdots, \theta_n^{u}\stackrel{iid}{\sim} \text{Unif}(0.9,1.1)$ and let $\theta_i=\beta_n\cdot \theta_i^{u}/\|\theta^u\|$, $1\leq i\leq n$. Let $P=(1-b_n)\mathbf{I}_2+b_n\mathbf{1}_2\mathbf{1}_2'$ and $\widetilde{P}=(1-\tilde b_n)\mathbf{I}_2+\tilde b_n\mathbf{1}_2\mathbf{1}_2'$. Let $\pi_i=(1,0)$ for $i\leq n/2$ and $(0,1)$ for $i> n/2$; let $\tilde\pi_i=(1,0)$ for $i\leq n/2+10$ and $(0,1)$ for $i> n/2+10$. We then construct $\Omega=\Theta\Pi P\Pi'\Theta$ and $\widetilde{\Omega}=\Theta\widetilde{\Pi}\widetilde{P}\widetilde{\Pi}\Theta$. 
The signal to noise ratio of the IBM test is governed by $(\beta_n, b_n, \tilde{b}_n)$, where $\beta_n$ controls network sparsity, and $(b_n, \tilde{b}_n)$ control the difference between two community structure matrices. Fixing $b_n=0.5$, we let $\tilde{b}_n$ range from 0.22 to 0.32, and choose $\beta_n$ coordinately to make the SNR in \eqref{DefineSNR} fixed at 3. For each parameter setting, after $(\Omega, \widetilde{\Omega})$ is generated, we simulate 200 network pairs under the null hypothesis and 200 pairs under the alternative hypothesis. We compare the histogram of the IBM statistic with that of the (ideally standardized) $T_{n,K}$. The results are in Figure~\ref{fig:Exp-Compare}.

We note that $T_{n,K}$ was designed to test $\Pi=\widetilde{\Pi}$. In this experiment, both the difference between $\Pi$ and $\widetilde{\Pi}$ and the difference between $P$ and $\widetilde{P}$ contribute signals. The IBM statistic captures both sources of signals and thus have higher power. 
In contrast, $T_{n,K}$ only captures the signals in $\Pi-\widetilde{\Pi}$. 
When $\tilde{b}_n$ is small, the networks are very sparse (recall that we fix the SNR; hence, a smaller $\tilde{b}_n$ yields a smaller $\beta_n$). It turns out that the signals in $\Pi-\widetilde{\Pi}$ alone are too weak to separate two hypotheses. As $\tilde{b}_n$ increases, the networks get less sparse, and the power of $T_{n,K}$ also increases and gets close to that of IBM.  Note also the IBM has an explicit limiting null,  but the limiting null of $T_{n,K}$ is  unclear under the general DCMM.


\section{Discussion} \label{sec:discuss}
Motivated by applications in social science, genetics and neurosicence, we consider the problem of testing whether the Bernoulli probability matrices of two networks are the same or not. 
We propose the Interlacing Balance Measure (IBM) as a new family of test statistics.  
IBM has several noteworthy advantages: It works for a broad class of DCMM and directed-DCMM settings allowing for different sparsity levels, severe degree heterogeneity, and mixed memberships.  
It has a tractable null distribution, though the models we consider have a large number of  unknown parameters. The explicit limiting null allows us to approximate the $p$-values of the test statistics. IBM is also powerful in separating the null from the alternative, and attains the optimal phase transitions. It is also a unified procedure: the same test statistic can be used for both directed and undirected networks, without modifications. 
We provide sharp theoretical analysis on the limiting null, power, and optimality of the test. We also apply the IBM test to analyze the Enron email network and a gene co-expression network, with interesting discoveries.

\spacingset{1.2}
\spacingset{1.5}

We focus on the DCMM model in this paper, but the method and theory are applicable to other network models where $\Omega$ and $\widetilde{\Omega}$ have low ranks.  In fact, as in Section \ref{subsec:DCMM}, the SNR of the IBM statistic $U_n^{(m)}$ is $(1/8) (\sum_{k = 1}^r \delta_k^4)  / [(\sum_{k = 1}^K \lambda_k^4)  + (\sum_{k = 1}^{\widetilde{K}} \tilde{\lambda}_k^4)]^{1/2}$, which does not depend on the particular form of DCMM. The estimate of the null variance of $U_n^{(m)}$ does not depend on the particular form of DCMM either. Therefore, results about the limiting null and the power of the IBM test are readily extendable to general low-rank network models. Also, for convenience, we assume $K$ and $\widetilde{K}$ are finite in this paper. For  the case where $K$ and $\widetilde{K}$ diverge to $\infty$, the above formula for SNR  is still valid, and our results continue to be true 
with some mild regularity conditions (e.g., $\max_{1 \leq k, m \leq K} \{P_{k,m}\} \leq C$).

The IBM test can be extended in multiple directions. Suppose we are given $N_1$ and $N_2$ independent networks drawn from $\Omega$ and $\widetilde{\Omega}$, respectively. To detect the difference between $\Omega$ and $\widetilde{\Omega}$, a similar IBM statistic can be defined based on $\bar{A}^*=\bar{A}_1-\bar{A}_2$, where $\bar{A}_1$ and $\bar{A}_2$ are the average of adjacency matrices of the $N_1$ and $N_2$ graphs, respectively. We expect that this test will inherit the nice properties of the IBM test and that the phase transitions will also depend on $(N_1, N_2)$.  This idea can be further extended to solve the change point detection problem in dynamic network analysis \citep{wang2018optimal}. In a standard binary segmentation procedure for identifying the change point, it requires to have a statistic that detects the difference between the Bernoulli probability matrices for two nested time blocks $B_1$ and $B_2$. We may construct an IBM-type test statistic from $\bar{A}^*(B_1,B_2)=\bar{A}(B_1)-\bar{A}(B_2)$, where $\bar{A}(B_k)$ is a weighted average of the adjacency matrices in time block $B_k$. The fact that the IBM-type statistics have tractable null distributions will help us design a tuning-free procedure for change-point detection. It is also interesting to study the optimality of this procedure when all the networks in the series are generated from the DCMM model. 

The idea of IBM may also be adapted to the problem of comparing two large covariance matrices of Gaussian ensembles \citep{cai2013two,zhu2017testing}. We note that $U_n^{(m)}$ is an estimate of $\tr((\Omega-\widetilde{\Omega})^{2m})$. This estimate is better than $\tr((A-\widetilde{A})^{2m})$ by removing from the sum those terms with nonzero means. The same idea is also potentially useful for detecting the difference between two covariance matrices. We leave it to future exploration. 

While we focus on testing in this paper,  a related question is how to identify the subset of nodes that are most different between two networks. This is similar to the problem of variable selection,  and we can approach it by creating $p$ node-wise features. For example, fixing a node, we can count the node degree 
and the numbers of $m$-cycles containing the node, and each of these is a node-wise feature. 
We can then select the subset of nodes that are most different between two networks using ideas from the variable selection literature.

\spacingset{1} 
{\small 
\bibliographystyle{chicago}
\bibliography{network} }

\begin{thebibliography}{}

\bibitem[\protect\citeauthoryear{Airoldi, Blei, Fienberg, and Xing}{Airoldi
  et~al.}{2008}]{airoldi2009mixed}
Airoldi, E., D.~Blei, S.~Fienberg, and E.~Xing (2008).
\newblock Mixed membership stochastic blockmodels.
\newblock {\em J. Mach. Learn. Res.\/}~{\em 9}, 1981--2014.

\bibitem[\protect\citeauthoryear{Arias-Castro and Verzelen}{Arias-Castro and
  Verzelen}{2014}]{Ery3}
Arias-Castro, E. and N.~Verzelen (2014).
\newblock Community detection in dense networks.
\newblock {\em Ann. Statist.\/}~{\em 42\/}(3), 940--969.

\bibitem[\protect\citeauthoryear{Bandeira and Van~Handel}{Bandeira and
  Van~Handel}{2016}]{bandeira2016sharp}
Bandeira, A.~S. and R.~Van~Handel (2016).
\newblock Sharp nonasymptotic bounds on the norm of random matrices with
  independent entries.
\newblock {\em Ann. Probab.\/}~{\em 44\/}(4), 2479--2506.

\bibitem[\protect\citeauthoryear{Banerjee}{Banerjee}{2018}]{banerjee2016contiguity}
Banerjee, D. (2018).
\newblock Contiguity and non-reconstruction results for planted partition
  models: the dense case.
\newblock {\em Electron. J. Probab\/}~{\em 23\/}(18), 1--28.

\bibitem[\protect\citeauthoryear{Bubeck, Ding, Eldan, and R{\'a}cz}{Bubeck
  et~al.}{2016}]{bubeck2016testing}
Bubeck, S., J.~Ding, R.~Eldan, and M.~Z. R{\'a}cz (2016).
\newblock Testing for high-dimensional geometry in random graphs.
\newblock {\em Random Struct. Algor.\/}~{\em 49\/}(3), 503--532.

\bibitem[\protect\citeauthoryear{Cai, Liu, and Xia}{Cai
  et~al.}{2013}]{cai2013two}
Cai, T., W.~Liu, and Y.~Xia (2013).
\newblock Two-sample covariance matrix testing and support recovery in
  high-dimensional and sparse settings.
\newblock {\em J. Amer. Statist. Assoc.\/}~{\em 108\/}(501), 265--277.

\bibitem[\protect\citeauthoryear{Donoho and Jin}{Donoho and Jin}{2015}]{DJ15}
Donoho, D. and J.~Jin (2015).
\newblock Higher criticism for large-scale inference: especially for rare and
  weak effects.
\newblock {\em Statist. Sci.\/}~{\em 30\/}(1), 1--25.

\bibitem[\protect\citeauthoryear{Fan, Fan, Han, and Lv}{Fan
  et~al.}{2022}]{fan2020asymptotic}
Fan, J., Y.~Fan, X.~Han, and J.~Lv (2022).
\newblock Asymptotic theory of eigenvectors for random matrices with diverging
  spikes.
\newblock {\em J. Amer. Statist. Assoc.\/}~{\em 117\/}(538), 996--1009.

\bibitem[\protect\citeauthoryear{Foucart and Rauhut}{Foucart and
  Rauhut}{2013}]{Foucart2013}
Foucart, S. and H.~Rauhut (2013).
\newblock {\em A Mathematical Introduction to Compressive Sensing}.
\newblock Birkh\"{a}user Basel.

\bibitem[\protect\citeauthoryear{Ghoshdastidar, Gutzeit, Carpentier, and
  Luxburg}{Ghoshdastidar et~al.}{2020}]{ghosh2019}
Ghoshdastidar, D., M.~Gutzeit, A.~Carpentier, and U.~v. Luxburg (2020).
\newblock Two-sample hypothesis testing for inhomogeneous random graphs.
\newblock {\em Ann. Statist.\/}~{\em 48\/}(4), 2208--2229.

\bibitem[\protect\citeauthoryear{Girvan and Newman}{Girvan and
  Newman}{2002}]{GirvanNewman}
Girvan, M. and M.~Newman (2002).
\newblock Community structure in social and biological networks.
\newblock {\em Proc. Natl. Acad. Sci.\/}~{\em 99\/}(12), 7821--7826.

\bibitem[\protect\citeauthoryear{Hall and Heyde}{Hall and
  Heyde}{2014}]{hall2014martingale}
Hall, P. and C.~C. Heyde (2014).
\newblock {\em Martingale limit theory and its application}.
\newblock Academic press.

\bibitem[\protect\citeauthoryear{Harary}{Harary}{1953}]{SignedGraph}
Harary, F. (1953).
\newblock On the notion of balance of a signed graph.
\newblock {\em Michigan Math. J.\/}~{\em 2\/}(2), 143--146.

\bibitem[\protect\citeauthoryear{Holland, Laskey, and Leinhardt}{Holland
  et~al.}{1983}]{holland1983stochastic}
Holland, P.~W., K.~B. Laskey, and S.~Leinhardt (1983).
\newblock Stochastic blockmodels: First steps.
\newblock {\em Social networks\/}~{\em 5\/}(2), 109--137.

\bibitem[\protect\citeauthoryear{Horn and Johnson}{Horn and
  Johnson}{1985}]{HornJohnson}
Horn, R. and C.~Johnson (1985).
\newblock {\em Matrix Analysis}.
\newblock Cambridge University Press.

\bibitem[\protect\citeauthoryear{Ji and Jin}{Ji and Jin}{2016}]{SCC-JiJin}
Ji, P. and J.~Jin (2016).
\newblock Coauthorship and citation networks for statisticians.
\newblock {\em Ann. Appl. Statist.\/}~{\em 10\/}(4), 1779--1812.

\bibitem[\protect\citeauthoryear{Jiang, Li, and Yao}{Jiang
  et~al.}{2023}]{Yao2021}
Jiang, B., J.~Li, and Q.~Yao (2023).
\newblock Autoregressive networks.
\newblock {\em J. Mach. Learn. Res.\/}~{\em 24\/}(227), 1--69.

\bibitem[\protect\citeauthoryear{Jin}{Jin}{2015}]{SCORE}
Jin, J. (2015).
\newblock Fast community detection by {SCORE}.
\newblock {\em Ann. Statist.\/}~{\em 43\/}(1), 57--89.

\bibitem[\protect\citeauthoryear{Jin, Ke, and Luo}{Jin et~al.}{2021}]{SP2019}
Jin, J., Z.~T. Ke, and S.~Luo (2021).
\newblock Optimal adaptivity of signed-polygon statistics for network testing.
\newblock {\em Ann. Statist.\/}~{\em 49\/}(6), 3408--3433.

\bibitem[\protect\citeauthoryear{Jin, Ke, and Luo}{Jin
  et~al.}{2023}]{Mixed-SCORE}
Jin, J., Z.~T. Ke, and S.~Luo (2023).
\newblock Mixed membership estimation for social networks.
\newblock {\em Journal of Econometrics\/}~{\em 239\/}(2), 105369.

\bibitem[\protect\citeauthoryear{Johnson and Reams}{Johnson and
  Reams}{2009}]{DADJohnson}
Johnson, C. and R.~Reams (2009).
\newblock Scaling of symmetric matrices by positive diagonal congruence.
\newblock {\em Linear Multilinear A\/}~{\em 57}, 123--140.

\bibitem[\protect\citeauthoryear{Karrer and Newman}{Karrer and
  Newman}{2011}]{DCBM}
Karrer, B. and M.~Newman (2011).
\newblock Stochastic blockmodels and community structure in networks.
\newblock {\em Phys. Rev. E\/}~{\em 83\/}(1), 016107.

\bibitem[\protect\citeauthoryear{Lee}{Lee}{2019}]{lee2019u}
Lee, A.~J. (2019).
\newblock {\em U-statistics: Theory and Practice}.
\newblock Routledge.

\bibitem[\protect\citeauthoryear{Li and Li}{Li and Li}{2018}]{TSTLi2018}
Li, Y. and H.~Li (2018).
\newblock Two-sample test of community memberships of weighted stochastic block
  models.
\newblock {\em arXiv:1811.12593\/}.

\bibitem[\protect\citeauthoryear{Liu, Choi, Xie, and Roeder}{Liu
  et~al.}{2018}]{Liu927}
Liu, F., D.~Choi, L.~Xie, and K.~Roeder (2018).
\newblock Global spectral clustering in dynamic networks.
\newblock {\em Proc. Nat. Acad. Sci.\/}~{\em 115\/}(5), 927--932.

\bibitem[\protect\citeauthoryear{Segerstolpe}{Segerstolpe}{2016}]{segerstolpe2016single}
Segerstolpe, {\AA}sa, {\it et al}. (2016).
\newblock Single-cell transcriptome profiling of human pancreatic islets in
  health and type 2 diabetes.
\newblock {\em Cell metabolism\/}~{\em 24\/}(4), 593--607.

\bibitem[\protect\citeauthoryear{Sinkhorn and Knopp}{Sinkhorn and
  Knopp}{1967}]{sinkhorn1967concerning}
Sinkhorn, R. and P.~Knopp (1967).
\newblock Concerning nonnegative matrices and doubly stochastic matrices.
\newblock {\em Pacific Journal of Mathematics\/}~{\em 21\/}(2), 343--348.

\bibitem[\protect\citeauthoryear{Stuart}{Stuart}{2019}]{RN262}
Stuart, T.~{\it et al}. (2019).
\newblock Comprehensive integration of single-cell data.
\newblock {\em Cell\/}~{\em 177\/}(7), 1888--+.

\bibitem[\protect\citeauthoryear{Tang, Athreya, Sussman, Lyzinski, and
  Priebe}{Tang et~al.}{2017}]{tang2017a}
Tang, M., A.~Athreya, D.~L. Sussman, V.~Lyzinski, and C.~E. Priebe (2017, 08).
\newblock A nonparametric two-sample hypothesis testing problem for random
  graphs.
\newblock {\em Bernoulli\/}~{\em 23\/}(3), 1599--1630.

\bibitem[\protect\citeauthoryear{Tang}{Tang}{2017}]{tang2017b}
Tang, M.~{\it et al}. (2017).
\newblock A semiparametric two-sample hypothesis testing problem for random
  graphs.
\newblock {\em J. Comput. Graph Statist.\/}~{\em 26}, 344--354.

\bibitem[\protect\citeauthoryear{Wang, Yu, and Rinaldo}{Wang
  et~al.}{2021}]{wang2018optimal}
Wang, D., Y.~Yu, and A.~Rinaldo (2021).
\newblock Optimal change point detection and localization in sparse dynamic
  networks.
\newblock {\em Ann. Statist.\/}~{\em 49\/}(1), 203--232.

\bibitem[\protect\citeauthoryear{Wang, Liang, and Ji}{Wang
  et~al.}{2020}]{DCMM-Ji}
Wang, Z., Y.~Liang, and P.~Ji (2020).
\newblock Spectral algorithms for community detection in directed networks.
\newblock {\em J. Mach. Learn. Res.\/}~{\em 21\/}(153), 1--45.

\bibitem[\protect\citeauthoryear{Yuan, Liu, Feng, and Shang}{Yuan
  et~al.}{2022}]{Yuan2018TestHyper}
Yuan, M., R.~Liu, Y.~Feng, and Z.~Shang (2022).
\newblock Testing community structure for hypergraphs.
\newblock {\em Ann. Statist.\/}~{\em 50\/}(1), 147--169.

\bibitem[\protect\citeauthoryear{Zhu, Lei, Devlin, and Roeder}{Zhu
  et~al.}{2017}]{zhu2017testing}
Zhu, L., J.~Lei, B.~Devlin, and K.~Roeder (2017).
\newblock Testing high-dimensional covariance matrices, with application to
  detecting schizophrenia risk genes.
\newblock {\em Ann. Appl. Statist.\/}~{\em 11\/}(3), 1810.

\end{thebibliography}
\addcontentsline{toc}{section}{References}

\spacingset{1.45}

\newpage

\setcounter{equation}{0}
 
\appendix
\section{Proof of Lemma \ref{lemma:comp}}
Notice the similarity in structure of $Q_n$, $C_n$ and $\widetilde{C}_n$, to show
Lemma \ref{lemma:comp}, it suffices to show that for any matrix $X\in\mathbb{R}^{n\times n}$ with $X_{ii} = 0, 1\leq i\leq n$ and $X_{ij} \in \left\lbrace -1, 0, 1\right\rbrace , 1\leq i, j\leq n$,
\begin{equation}
\label{equation:q}
\sum_{i_1, i_2, i_3, i_4 (dist)} X_{i_1 i_2}(X')_{i_2 i_3}X_{i_3 i_4} (X')_{i_4 i_1} = \text{tr}([XX']^2) - \text{tr}(XX' \circ XX')- \text{tr}(X'X \circ X'X) + 1_n'|X| 1_n,
\end{equation}
and the complexity for calculating the right hand side is as claimed.

We show \eqref{equation:q} first. By definition of $\text{tr}([XX']^2)$, we only need to show 
\[
\sum_{i_1, i_2, i_3, i_4 (not \ dist)} X_{i_1 i_2}(X')_{i_2 i_3}X_{i_3 i_4} (X')_{i_4 i_1} = 
\text{tr}(XX' \circ XX')+\text{tr}(X'X \circ X'X) - 1_n'|X| 1_n.
\]
We then count the non-zeros terms where $i_1, i_2, i_3, i_4$ are not distinct. By $X_{ii} = 0$, those terms could only have one of the forms in $\left\lbrace (i, j, i, k), (i, j, k, j), (i, j, i, j)\right\rbrace $, where $i, j, k$ are distinct numbers ranging from $1$ to $n$. Summation of $(i, j, i, k)$ type terms is 
\[
\sum_{i, j, k(dist)} X_{ij}^2 X_{ik}^2 = \sum_{i=1}^n \left( \sum_{j=1}^n X_{ij}^2\right)^2 - \sum_{i=1}^n \sum_{j=1}^n X_{ij}^4 = \text{tr}(XX'\circ XX') - 1_n'|X|1_n,
\]
and summation of $(i, j, k, j)$ type terms is $\text{tr}(X'X\circ X'X) - 1_n'|X|1_n$. Summation of $(i, j, i, j)$ type terms is 
\[
\sum_{i, j(dist)} X_{ij}^4 = \sum_{i, j(dist)} |X_{ij}| = 1_n'|X|1_n.
\]
Thus
\[
\sum_{i_1, i_2, i_3, i_4 (not \ dist)} X_{i_1 i_2}(X')_{i_2 i_3}X_{i_3 i_4} (X')_{i_4 i_1} = 
 \text{tr}(XX' \circ XX')+ \text{tr}(X'X \circ X'X) - 1_n'|X| 1_n,
\]
which proves \eqref{equation:q}. 

It remains to consider the computation cost. Using matrix product, the complexity of $1_n'|X| 1_n$ is $O(n\bar{d})$, and the complexity of $\text{tr}([XX']^2)$ and $\text{tr}(XX'\circ XX')$ is the same as complexity of calculating $XX'$. Below we show for any $ B\in\mathbb{R}^{n\times n}$, the complexity of calculating $BC$ is $O(n^2\bar{t})$ where $\bar t$ is the averaged degree of graph induced by $C$, then since the averaged degree for $X'$ is $O(\bar d)$ (when $X\in \{A,\widetilde A, A^*\}$), the complexity of calculating $XX'$ is $O(n^2\bar{d})$. For each $(i, j)$, we can see $BC(i, j) = \sum_{k=1}^n B_{ik} C_{kj}$ requires $O(\bar{t})$ summations, thus the computation cost for $BC$ is $O(n^2\bar{t})$. This completes proof of the first claim in Lemma. 

When using the adjacency list representation of $X\in\{A,\widetilde A, A^*\}$, we can create two dictionaries for each node $i$. The key sets of the two dictionaries are 
\[K^+_i=\{j: \exists \ k\  s.t. \ X_{ik}X'_{kj}=1\}\quad \text{and} \quad K^-_i=\{j: \exists \ k\  s.t. \ X_{ik}X'_{kj}=-1\}.\]
For each $j\in K^+_i$, $Dic^+_i(j)=\#\{k: X_{ik}X'_{kj}=1\}$ and for each $j\in K^-_i$, $Dic^-_i(j)=\#\{k: X_{ik}X'_{kj}=-1\}$. The dictionaries can be constructed by searching the in-neighbors of out-neighbors of each node ($j$ is an in-neighbor of $i$ if $X_{ji}\neq 0$ and is an out-neighbor if $X_{ij}\neq 0$). Overall, it requires $O(n \bar d d_{\max})$ complexity to construct $Dic^+_i$ and $Dic^-_i$ for $i=1,\cdots,n$. And it's easy to see that 
\begin{equation}
q(X) = \sum_{i=1}^n \sum_{j \in K_i^+\cup K_i^-} Dic^+_i(j) (Dic^+_i(j)-1) + Dic^-_i(j) (Dic^-_i(j)-1)-2Dic^+_i(j) Dic^-_i(j).
\end{equation}
Note that when $j \notin K^\pm_i$, $Dic_i^\pm(j)$ returns 0 at $O(1)$ complexity. This summation also has $O(n \bar d d_{\max})$ computation complexity, which completes the proof of Lemma.

\section{Proof of Lemma \ref{lemma:iden1}}
Consider the first part of the claim. Suppose 
\[
\Omega = \Theta \Pi P \Gamma' Z'
\] 
as in Model (\ref{model1a})-(\ref{model1c}) where $P$ is fully indecomposable. 
It is sufficient to show we can write 
\[
\Omega =  \widetilde{\Theta} \widetilde{\Pi} \widetilde{P} \widetilde{\Gamma}' \widetilde{Z},  
\] 
where $(\widetilde{\Theta}, \widetilde{\Pi}, \widetilde{P}, \widetilde{\Gamma}, \widetilde{Z})$ are as in  Model (\ref{model1a})-(\ref{model1c}) and $\widetilde{P}$ is doubly stochastic. 
By \cite{sinkhorn1967concerning} (see also \cite{DADJohnson}), there are  $K \times K$ diagonal matrices $D_1$ and $D_2$ with positive diagonal entries such that 
\[
D_1 P D_2 
\]  
is doubly stochastic.  At the same time, for each $1 \leq i \leq n$, 
there are $\tilde{\theta}_i > 0, \tilde{\zeta}_i > 0$ and weight vectors  $\tilde{\pi}_i$ and $\tilde{\gamma}_i$ (i.e., all the entries are non-negative, with a unit-$\ell_1$-norm) such that 
\[
\theta_i \pi_i' D_1^{-1} = \tilde{\theta}_i \tilde{\pi}_i',  \qquad   \zeta_i \gamma_i'  D_2^{-1} = \tilde{\zeta}_i \tilde{\gamma}_i'.  
\]  
Therefore, if we let 
\[
\widetilde{\Theta} = \diag(\tilde{\theta}_1, \ldots, \tilde{\theta}_n),  \qquad 
\widetilde{Z} =  \diag(\tilde{\zeta}_1, \ldots, \tilde{\zeta}_n), 
\] 
\[
\widetilde{\Pi} = [\pi_1, \pi_2, \ldots, \pi_n]', \qquad  \widetilde{\Gamma}  = [\gamma_1,\gamma_2, \ldots, \gamma_n]', 
\] 
and 
\[
\widetilde{P} = D_1 P D_2, 
\] 
then it is seen that 
\[
\Omega =  \widetilde{\Theta} \widetilde{\Pi} \widetilde{P} \widetilde{\Gamma}' \widetilde{Z}. 
\] 
This proves the claim.

We now consider the second part of the claim. Suppose we have 
\begin{equation} \label{identpf1} 
\Omega = \Theta \Pi P \Gamma' Z' =  \widetilde{\Theta}  \widetilde{\Pi}  \widetilde{P} \widetilde{\Gamma}'   
\widetilde{Z}',  
\end{equation} 
where both $(\Theta, \Pi, P, \Gamma, Z)$ and $(\widetilde{\Theta}, \widetilde{\Pi}, \widetilde{P}, \widetilde{\Gamma}, 
\widetilde{Z})$ satisfy the conditions of Lemma \ref{lemma:iden1}.  The goal is to show 
 $(\Theta, \Pi, P, \Gamma, Z) = (c_0\widetilde{\Theta}, \widetilde{\Pi}, \widetilde{P}, \widetilde{\Gamma}, 
c_0^{-1}\widetilde{Z})$ (once this is proved,  the identifiability follows from $\|\theta\|=\|\zeta\|$).  Note that to show this, 
  it  suffices to show that 
\begin{equation} \label{identpf1a} 
(\Theta, \Pi, P) = (c_0 \widetilde{\Theta}, \widetilde{\Pi}, \widetilde{P}); 
\end{equation}
the proof of $(\Gamma, Z) = (\widetilde{\Gamma},c_0^{-1}\widetilde{Z})$ is similar by symmetry. 

We now show (\ref{identpf1a}). 
We  show $\widetilde{P} = P$ first. By the conditions of the lemma, 
each of the $K$ community has at least one node which is both pure 
as a citer and as a citee. Without loss of generality, assume for each $1 \leq k \leq K$, node $k$ 
is pure in community $k$ both as a citer and as a citee. 
Comparing the $K \times K$ sub-matrix of  $\Theta \Pi P \Gamma' Z'$ and $\widetilde{\Theta}  \widetilde{\Pi}  \widetilde{P} \widetilde{\Gamma}'   \widetilde{Z}'$ consisting of the first $K$ rows and $K$ columns.    
It follows 
\[
\diag(\theta_1, \ldots, \theta_K) \cdot P \cdot \diag(\zeta_1, \ldots, \zeta_K) = \diag(\tilde{\theta}_1, \ldots, \tilde{\theta}_K) \cdot \widetilde{P}  \cdot \diag(\tilde{\zeta}_1, \ldots, \tilde{\zeta}_K). 
\] 
Since both $P$ and $\widetilde{P}$ are double stochastic and fully indecomposable, by Sinkhorn' theorem \citep{sinkhorn1967concerning}, there exists a constant $c_0 > 0$ such that 
\begin{equation} \label{identpf2a} 
P = \widetilde{P}, \qquad \theta_k  = c_0 \tilde{\theta}_k, \qquad \zeta_k = c_0^{-1} \tilde{\zeta}_k, \qquad 1 \leq k \leq K. 
\end{equation} 

Next, we show  $\Pi=\widetilde{\Pi}$. From (\ref{identpf1}) and (\ref{identpf2a}), we know $\Pi P \Gamma' =  \widetilde{\Pi}  P \widetilde{\Gamma}'$. Consider the first $K$ columns of this equation, we have $\Pi P I_K' =  \widetilde{\Pi}  P I_K'$. Since $P$ is non-singular, we immediately have $\Pi=\widetilde{\Pi}$ which completes the proof.

\section{Proof of Theorems \ref{thm:null}-\ref{thm:LB} for directed-DCMM}  

\subsection{Proof of Theorem \ref{thm:null}} 

Recall that $\Omega_{ij}  = \widetilde{\Omega}_{ij}$ for all $1\leq i, j\leq n$ under the null hypothesis, 
it follows from  definition  that 
\[
A_{ij} - \widetilde{A}_{ij} = (W_{ij} + {\Omega}_{ij}) - (\widetilde{W}_{ij} + \widetilde{\Omega}_{ij}) = W_{ij} - \widetilde{W}_{ij},
\]
which indicates
\begin{align*}
Q_n  = \sum_{i_1, i_2, i_3, i_4 (dist)} (W_{i_1 i_2} -  \widetilde{W}_{i_1 i_2}) (W_{i_2 i_3} - \widetilde{W}_{i_2 i_3}) (W_{i_3 i_4} - \widetilde{W}_{i_3 i_4}) (W_{i_4 i_1} - \widetilde{W}_{i_4 i_1}). 
\end{align*}
For distinct $i_1, i_2, i_3, i_4$, random variables $W_{i_1i_2}, W_{i_2i_3}, W_{i_3i_4}, W_{i_4i_1},$ $\widetilde{W}_{i_1i_2}, \widetilde{W}_{i_2i_3}, \widetilde{W}_{i_3i_4}, \widetilde{W}_{i_4i_1}$ are mutually independent.

We start with  deriving the mean and variance of $Q_n$.  
For the mean, it follows from that $W_{ij}$ and $\widetilde{W}_{ij}$ are mean zero for all $1\leq i\neq j \leq n$ and  independence that \begin{align*}
\mathbb{E}[Q_n] 
& = \sum_{i_1, i_2, i_3, i_4 (dist)} \mathbb{E}\big[W_{i_1 i_2} -  \widetilde{W}_{i_1 i_2}\big] \mathbb{E}\big[ W_{i_2 i_3} - \widetilde{W}_{i_2 i_3}\big] \mathbb{E}\big[ W_{i_3 i_4} - \widetilde{W}_{i_3 i_4}\big] \mathbb{E}\big[ W_{i_4 i_1} - \widetilde{W}_{i_4 i_1}\big] = 0. 
\end{align*}
Consider the variance. First we group the terms in $Q_n$ into uncorrelated groups. Notice that for each term in $Q_n$ indexed by $(i_1, i_2, i_3, i_4)$, there are $7$ other terms in $Q_n$ that are identical to it up to permutation. Topologically, these $8$ terms are the representation of the same quadrilateral with $4$ possible starting points and $2$ possible directions. Define $I_4(n) = \left\lbrace (i_1, i_2, i_3, i_4), (i_1, i_2, i_4, i_3), (i_1, i_3, i_2, i_4), 1\leq i_1<i_2<i_3<i_4\leq n \right\rbrace $, then each such $8$-term group can be represented with one unique element in $I_4(n)$.
Therefore, we can rewrite $Q_n$ as
\[
Q_n = 8\sum_{I_4(n)} (W_{i_1 i_2} -  \widetilde{W}_{i_1 i_2}) (W_{i_2 i_3} - \widetilde{W}_{i_2 i_3}) (W_{i_3 i_4} - \widetilde{W}_{i_3 i_4}) (W_{i_4 i_1} - \widetilde{W}_{i_4 i_1}),
\]
where the terms in the summation are now uncorrelated with each other since the underlying quadrilateral is different. It follows that \beq\label{temp-VarQ}
\mathrm{Var}(Q_n) = 64 \sum_{I_4(n)}\mathrm{Var}\bigg( (W_{i_1 i_2} -  \widetilde{W}_{i_1 i_2}) (W_{i_2 i_3} - \widetilde{W}_{i_2 i_3}) (W_{i_3 i_4} - \widetilde{W}_{i_3 i_4}) (W_{i_4 i_1} - \widetilde{W}_{i_4 i_1})\bigg). 
\eeq
For distinct $i_1, \cdots, i_4$ and by previous argument that $ (W_{i_1 i_2} -  \widetilde{W}_{i_1 i_2}) \cdots (W_{i_4 i_1} - \widetilde{W}_{i_4 i_1})$ has zero mean, we obtain
 \begin{align*}
& \quad \mathrm{Var}\bigg( (W_{i_1 i_2} -  \widetilde{W}_{i_1 i_2}) (W_{i_2 i_3} - \widetilde{W}_{i_2 i_3}) (W_{i_3 i_4} - \widetilde{W}_{i_3 i_4}) (W_{i_4 i_1} - \widetilde{W}_{i_4 i_1})\bigg) \\
& =  \mathbb{E}\big[ (W_{i_1 i_2} -  \widetilde{W}_{i_1 i_2})^2 (W_{i_2 i_3} - \widetilde{W}_{i_2 i_3})^2(W_{i_3 i_4} - \widetilde{W}_{i_3 i_4})^2 (W_{i_4 i_1} - \widetilde{W}_{i_4 i_1})^2\big] \\
& =  \mathbb{E}(W_{i_1 i_2} -  \widetilde{W}_{i_1 i_2})^2  \mathbb{E}(W_{i_2 i_3} - \widetilde{W}_{i_2 i_3})^2 \mathbb{E}(W_{i_3 i_4} - \widetilde{W}_{i_3 i_4})^2 \mathbb{E}(W_{i_4 i_1} - \widetilde{W}_{i_4 i_1})^2,
\end{align*}
 by the mutual  independence between $W_{i_1i_2}, W_{i_2i_3}, W_{i_3i_4}, W_{i_4i_1},$ $\widetilde{W}_{i_1i_2}, \widetilde{W}_{i_2i_3}, \widetilde{W}_{i_3i_4}, \widetilde{W}_{i_4i_1}$. 
Since $\mathrm{Var}(W_{ij}) = \mathrm{Var}(\widetilde{W}_{ij}) = \Omega_{ij}(1 - \Omega_{ij})$ and recall that $W_{ij}$ and $\widetilde{W}_{ij}$ are independent, we find 
\[
\mathbb{E}(W_{i_1 i_2} -  \widetilde{W}_{i_1 i_2})^2 = \mathbb{E}[W_{i_1 i_2}^2] + \mathbb{E}[ \widetilde{W}_{i_1 i_2}^2] = 2\Omega_{i_1i_2}(1 - \Omega_{i_1i_2}) = 2[1+o(1)]\cdot \Omega_{i_1i_2}.
\]
 The last equation is by $\Omega_{ij}=\theta_i\pi_i'P\pi_j\theta_j\leq \theta_i \theta_j \|\pi_i\|\|P\|\|\pi_j\|\leq C \theta_{\max}^2\goto0$. We further have 
\[
\mathbb{E}(W_{i_1 i_2} -  \widetilde{W}_{i_1 i_2})^2  \mathbb{E}(W_{i_2 i_3} - \widetilde{W}_{i_2 i_3})^2 \mathbb{E}(W_{i_3 i_4} - \widetilde{W}_{i_3 i_4})^2  \mathbb{E}(W_{i_4 i_1} - \widetilde{W}_{i_4 i_1})^2 = 16[1+o(1)]\cdot\Omega_{i_1i_2}\Omega_{i_2i_3}\Omega_{i_3i_4}\Omega_{i_4i_1}. 
\]
Plug it back into \eqref{temp-VarQ}, we obtain
 \begin{align*}
\mathrm{Var}(Q_n) & = 64\cdot 16[1+o(1)]  \cdot  \sum_{I_4(n)} \Omega_{i_1i_2}\Omega_{i_2i_3}\Omega_{i_3i_4}\Omega_{i_4i_1} \\
& = 128[1+o(1)]\cdot \sum_{i_1,i_2,i_3,i_4(dist)}\Omega_{i_1i_2}\Omega_{i_2i_3}\Omega_{i_3i_4}\Omega_{i_4i_1} \\
& = 128[1+o(1)]\cdot \mathbb{E}[C_n]. 
\end{align*}
Additionally, under the null hypothesis, it's not hard to see $\mathbb{E}[C_n] = \mathbb{E}[\widetilde{C}_n]$. Hence, \[
\mathrm{Var}(Q_n)  = 128[1+o(1)]\cdot \mathbb{E}[C_n] = 64[1+o(1)]\cdot (\mathbb{E}[C_n] + \mathbb{E}[\widetilde{C}_n]), 
\]
which completes the proof of the  first claim. 

Notice  that $(W_{ij} - \widetilde W_{ij})$ are independent with each other for $1\leq i < j\leq n$ and has zero mean.  
Using martingale central limit theorem (the proof is analogous to that of Theorem \ref{thm:null-d} thus omitted), we obtain 
\beq\label{normalityA}
\frac{Q_n}{\sqrt{\mathrm{Var}(Q_n)}} \stackrel{d}{\goto} N(0,1). 
\eeq

Consider the   second  claim. We prove it under both the null and alternative hypothesis. First, consider the mean. By definition, $\mathbb{E}[C_n] = \sum_{i_1,i_2, i_3,i_4 (dist)}\Omega_{i_1i_2}\Omega_{i_2i_3}\Omega_{i_3i_4}\Omega_{i_4i_1}$, so
\[
\mathbb{E}[C_n] = \tr(\Omega^4)-\sum_{\substack{\text{non-distinct}\\ i_1,i_2, i_3,i_4}}\Omega_{i_1i_2}\Omega_{i_2i_3}\Omega_{i_3i_4}\Omega_{i_4i_1},
\]
where by an analogy of \eqref{thm1-claim3-key2}, $\tr(\Omega^4) \asymp \lambda_1^4 \asymp \|\theta\|^8.$
It remains to control the remainder term. Note that $\Omega_{ij}=\theta_i\theta_j(\pi_i'P\pi_j)\leq C\theta_i\theta_j$, where the last inequality is from Condition (\ref{cond-P}). Hence,  
\begin{align*} 
&\sum_{\substack{\text{non-distinct}\\ i_1,i_2, i_3,i_4}}\Omega_{i_1i_2}\Omega_{i_2i_3}\Omega_{i_3i_4}\Omega_{i_4i_1}
\leq \sum_{\substack{\text{non-distinct}\\ i_1,i_2, i_3, i_4}}C\theta_{i_1}^2\theta_{i_2}^2\theta_{i_3}^2\theta_{i_4}^2\cr
&\leq \sum_{i_1,i_2,i_3}C\theta_{i_1}^4\theta_{i_2}^2\theta_{i_{3}}^2 =  C\|\theta\|^{4}\|\theta\|_4^4,
\end{align*}
which gives $\mathbb{E}[C_n] =  \tr(\Omega^4) + O(\|\theta\|^4\|\theta\|_4^4)$. 

Consider the variance of $C_n$. 
We decompose $(C_n-\mathbb{E}[C_n])$ as the sum of five terms:
\begin{align*}
X_1&= 4 \sum_{i_1, i_2, i_3, i_4 (dist)}W_{i_1i_2}\Omega_{i_2i_3}\Omega_{i_3i_4}\Omega_{i_4i_1}, \qquad
X_2 =  4\sum_{i_1, i_2, i_3, i_4 (dist)}
 W_{i_1i_2}W_{i_2i_3}\Omega_{i_3i_4}\Omega_{i_4i_1},\cr
X_3
&= 2 \sum_{i_1, i_2, i_3, i_4 (dist)} W_{i_1i_2}\Omega_{i_2i_3}W_{i_3i_4}\Omega_{i_4i_1} ,\qquad 
X_4
=4 \sum_{i_1, i_2, i_3, i_4 (dist)}\Omega_{i_1i_2}W_{i_2i_3}W_{i_3i_4}W_{i_4i_1},\cr
X_5 
& =  \sum_{i_1, i_2, i_3, i_4 (dist)} W_{i_1i_2}W_{i_2i_3}W_{i_3i_4}W_{i_4i_1}.
\end{align*}
By Cauchy-Schwarz inequality, $\mathrm{Var}(C_n) \leq 5\sum_{i=1}^5\mathrm{Var}(X_i)$ holds for random variables $X_1, X_2, \cdots, X_5$. It suffices to upper bound $\mathrm{Var}(X_i)$, for $i = 1, 2, \cdots, 5$. 

Consider $X_1$. Recall that
	\[
	X_1 = 4  \sum_{i_1, i_2, i_3, i_4 (dist)} W_{i_1i_2} \Omega_{i_2i_3}\Omega_{i_3i_4} \Omega_{i_4i_1} = 8 \sum_{i_1< i_2}\Bigl( \sum_{\substack{i_3,i_4\notin\{i_1,i_2\}\\ i_3\neq i_4}}\Omega_{i_2i_3} \Omega_{i_3i_4} \Omega_{i_4i_1} \Bigr) W_{i_1i_2}.
	\]
It is easily seen that $\mathbb{E}[X_1]=0$. Furthermore, we have
\beq  \label{LemmaA4-D1}
\mathrm{Var}(X_1) =  64 \sum_{i_1<i_2 }\Bigl( \sum_{\substack{i_3,i_4\notin\{i_1,i_2\}\\ i_3\neq i_4}}\Omega_{i_2i_3} \Omega_{i_3i_4} \Omega_{i_4i_1} \Bigr)^2\cdot \mathrm{Var}(W_{i_1i_2}). 
\eeq
By condition (\ref{cond-P}), 
\[
\Bigl| \sum_{\substack{i_3,i_4\notin\{i_1,i_2\}\\ i_3\neq i_4}}\Omega_{i_2i_3} \Omega_{i_3i_4} \Omega_{i_4i_1} \Bigr| \leq C\sum_{i_3,i_4}\theta_{i_1}\theta_{i_2}\theta_{i_3}^2\theta_{i_4}^2 \leq C \|\theta\|^4\cdot \theta_{i_1}\theta_{i_2}.
\]
We plug it into \eqref{LemmaA4-D1}  and use $\mathrm{Var}(W_{i_1i_2})\leq \Omega_{i_1i_2}\leq C\theta_{i_1}\theta_{i_2}$. It yields that 
\begin{eqnarray} \label{LemmaC4-D1-UB}
\mathrm{Var}(X_1) & \leq & C\sum_{i_1, i_2 (dist)}(\|\theta\|^4\theta_{i_1}\theta_{i_2})^2\cdot \theta_{i_1}\theta_{i_2} \leq   C \|\theta\|^8\|\theta\|_3^6. 
\end{eqnarray}

Consider $X_2$. Recall that
\[
X_2 = 4 \sum_{i_1, i_2, i_3, i_4 (dist)} \Omega_{i_3i_4}  \Omega_{i_4i_1} W_{i_1i_2} W_{i_2i_3}= 4 \sum_{i_1, i_2,i_3 (dist)}\Bigl(\sum_{i_4\notin\{i_1,i_2,i_3\}} \Omega_{i_3i_4}  \Omega_{i_4i_1} \Bigr)W_{i_1i_2} W_{i_2i_3}. 
\]
It is easy to see that $\mathbb{E}[X_2]=0$. We note that for $W_{k\ell}W_{\ell i}$ and $W_{k'\ell'}W_{\ell' i'}$ to be correlated, we must have either  $(k',\ell',i')=(k,\ell, i)$ or $(k',\ell',i')=(i,\ell, k)$; in other words, the two underlying paths $k$-$\ell$-$i$ and $k'$-$\ell'$-$i'$ have to be equal. We therefore  have
\begin{align*}
\mathrm{Var}(X_2) &\leq C\sum_{i_1, i_2, i_3 (dist)}\mathrm{Var}\Bigl[\Bigl(\sum_{i_4\notin\{i_1,i_2,i_3\}} {\Omega}_{i_3i_4}  {\Omega}_{i_4i_1} \Bigr)W_{i_1i_2} W_{i_2i_3}\Bigr]\cr
&\leq  C\sum_{i_1, i_2, i_3 (dist)}\Bigl(\sum_{i_4\notin\{i_1,i_2,i_3\}}  {\Omega}_{i_3i_4}  {\Omega}_{i_4i_1} \Bigr)^2\cdot\mathrm{Var}(W_{i_1i_2} W_{i_2i_3}). 
\end{align*}
By condition (\ref{cond-P}), we have
\[
\Bigl|\sum_{i_4\notin\{i_1,i_2,i_3\}}  {\Omega}_{i_3i_4}  {\Omega}_{i_4i_1}  \Bigr|\leq C\sum_{i_4}\theta_{i_1}\theta_{i_3}\theta_{i_4}^2 =  C \|\theta\|^2\cdot \theta_{i_1}\theta_{i_3}. 
\]
Combining the above gives
\begin{align*}
\mathrm{Var}(X_2)  & \leq C\sum_{i_1,i_2,i_3}(\|\theta\|^2\theta_{i_1}\theta_{i_3})^2\cdot \theta_{i_1}\theta_{i_2}^2\theta_{i_3}  = C \|\theta\|^6\|\theta\|_3^6. 
\end{align*}
Since $\|\theta\|\to \infty$, the right hand side is $o(\|\theta\|^8\|\theta\|_3^6)$. 

Then we consider $X_3$. It is easy to see that $\mathbb{E}[X_3]=0$. To calculate its variance, note that $W_{jk}W_{\ell i}$ and $W_{j'k'}W_{\ell' i'}$ are uncorrelated unless (i) $\{j',k'\}=\{j,k\}$ and $\{\ell',i'\}=\{\ell,i\}$ or (ii) $\{j',k'\}=\{\ell ,i\}$ and $\{\ell',i'\}=\{j,k\}$. We immediately have
\begin{align*}
\mathrm{Var}(X_3) &\leq C \sum_{i_1, i_2, i_3, i_4 (dist)}\mathrm{Var}\bigl(  {\Omega}_{i_2i_3}  {\Omega}_{i_1i_4} W_{i_1i_2} W_{i_3i_4} \bigr)\cr
&\leq C \sum_{i_1, i_2, i_3, i_4 (dist)}  {\Omega}^2_{i_2i_3} {\Omega}^2_{i_1i_4}\cdot \mathrm{Var}( W_{i_1i_2} W_{i_3i_4})\cr
&\leq C\sum_{i_1, i_2, i_3, i_4 } (\theta_{i_2}\theta_{i_3})^2(\theta_{i_1}\theta_{i_4})^2\cdot \theta_{i_1}\theta_{i_2}\theta_{i_3}\theta_{i_4} \leq C\|\theta\|_3^{12}. 
\end{align*}
Since $\|\theta\|_3^3\leq \theta_{\max}\|\theta\|^2=o(\|\theta\|^2)$, the right hand side is $o(\|\theta\|^8)$. 

For $X_4$, first recall that
\[
X_4 =  4 \sum_{i_1, i_2, i_3, i_4 (dist)}  W_{i_1i_2} W_{i_2i_3} W_{i_3i_4} \Omega_{i_4i_1},  
\]
which has mean 0.
Each index choice $(i,j,k,\ell)$ defines a undirected path $j$-$k$-$\ell$-$i$ in the complete graph of $n$ nodes. If the two paths $j$-$k$-$\ell$-$i$ and $j'$-$k'$-$\ell'$-$i'$ are not exactly overlapping, then $W_{jk}W_{k\ell}W_{\ell i}\cdot W_{j'k'} W_{k' \ell'} W_{\ell' i'}$ is mean-zero, thus $W_{jk}W_{k\ell}W_{\ell i}\Omega_{ij}$ and $W_{j'k'} W_{k' \ell'} W_{\ell' i'}\Omega_{i'j'}$ are uncorrelated. In the sum above, each unique path $j$-$k$-$\ell$-$i$ is counted twice as $(i,j,k,\ell)$ and $(j,i,\ell,k)$. We then immediately have 
	\begin{align*}
	\mathrm{Var}(X_4) & =  32 \sum_{i_1, i_2, i_3, i_4 (dist)} \mathrm{Var}\bigl(  W_{i_1i_2} W_{i_2i_3} W_{i_3i_4} \Omega_{i_4i_1} \bigr)\cr
	&=  32  \sum_{i_1, i_2, i_3, i_4 (dist)} \Omega_{i_4i_1}^2\cdot \mathrm{Var}\bigl(W_{i_1i_2} W_{i_2i_3} W_{i_3i_4}\bigr).
	\end{align*}
Moreover, $\mathrm{Var}( W_{i_1i_2} W_{i_2i_3} W_{i_3i_4})\leq \Omega_{i_1i_2} \Omega_{i_2i_3}\Omega_{i_3i_4}\leq C\theta_{i_1}\theta_{i_2}^2\theta_{i_3}^2\theta_{i_4}$. It follows that
	\begin{align*}
	\mathrm{Var}(X_4)& \leq C\sum_{i_1, i_2, i_3, i_4 (dist)} (\theta_{i_4}\theta_{i_1})^2\cdot \theta_{i_1}\theta_{i_2}^2\theta_{i_3}^2\theta_{i_4} \leq C  \|\theta\|^4\|\theta\|_3^6. 
	\end{align*}
Since  $\|\theta\|_3^3=o(\|\theta\|^2)$, the right hand side is $o(\|\theta\|^8)$. 


Finally, we consider $X_5$. Mimicking previous argument and it follows that \[
\mathrm{Var}(X_5) \leq C\sum_{i_1, i_2, i_3, i_4 (dist)} \Omega_{i_1i_2} \Omega_{i_2i_3}\Omega_{i_3i_4}\Omega_{i_4i_1} \leq C\sum_{i_1, i_2, i_3, i_4 (dist)}\theta_{i_1}^2 \theta^2_{i_2}\theta_{i_3}^2\theta_{i_4}^2 = C\|\theta\|^8.
\]
Combining above, we obtain $\mathrm{Var}(C_n) \leq C\|\theta\|^8 + C\|\theta\|^8\|\theta\|_3^6$, thus completes the proof of the variance part in the second claim. 

Consider the last  part in the second claim. By Markov's inequality, for any $\eps>0$, \[
\mathbb{P}\bigg(\bigg|\frac{{C}_n}{\mathbb{E}[C_n]}  - 1\bigg|\geq \eps\bigg) \leq \frac{1}{\eps^2}\mathbb{E}\bigg(\frac{{C}_n}{\mathbb{E}[C_n]}  - 1\bigg)^2 = \frac{1}{\eps^2\mathbb{E}[C_n]^2}\mathbb{E}\big({C}_n - \mathbb{E}[C_n]\big)^2.
\]
Here by the   first two parts in second claim of Theorem \ref{thm:null} 
that $\mathbb{E}[C_n]\asymp\|\theta\|^8$ and    $\mathbb{E}\big({C}_n - \mathbb{E}[C_n]\big)^2\leq C\|\theta\|^8\cdot \left[1+\|\theta\|_3^6 \right]$, the rightmost term is no greater than \[
 \frac{C\|\theta\|^8\cdot[1+\|\theta\|_3^6]}{\eps^2\|\theta\|^{16}} =  \frac{C}{\eps^2\|\theta\|^{8}} + \frac{C\|\theta\|_3^6}{\eps^2\|\theta\|^{8}} \leq \frac{C}{\eps^2\|\theta\|^{8}} + \frac{C\theta_{\max}^2\|\theta\|^4}{\eps^2\|\theta\|^{8}} = o(1),
\]
where the last two steps follows from $\theta_{\max}\goto0$ and $\|\theta\|\goto\infty$.
This proves $C_n/\mathbb{E}[C_n]\stackrel{p}{\goto}1$. Similarly, we obtain  $\widetilde{C}_n/\mathbb{E}[\widetilde{C}_n]\stackrel{p}{\goto}1$. 
Combining   (\ref{normalityA})  and Slutsky's theorem, we get $\psi_n\goto N(0,1)$ in law. \qed


\subsection{Proof of Theorem \ref{thm:alt}}

Introduce the vector $\bar\theta\in\mathbb{R}^n$ such that for $1\leq i\leq n$
\beq\label{def:bartheta}
\bar\theta_i = \theta_i + \tilde{\theta}_i.
\eeq
By $\Omega_{ij}\leq  C \theta_i\theta_j$, $\widetilde{\Omega}_{ij}\leq  C \tilde\theta_i\tilde\theta_j$, we obtain \beq\label{bartheta1}
|\Delta_{ij}| \leq \Omega_{ij} + \widetilde{\Omega}_{ij} \leq  C \bar\theta_i\bar\theta_j 
\eeq

Consider the mean  of $Q_n$. Recall that in the proof of  Theorem \ref{thm:null}, the random variables $A_{i_1i_2}$, $A_{i_2i_3}$, $A_{i_3i_4}$, $A_{i_4i_1}$, $\widetilde{A}_{i_1i_2}$, $\widetilde{A}_{i_2i_3}$, $\widetilde{A}_{i_3i_4}$, $\widetilde{A}_{i_4i_1}$ are  mutually  independent, it follows that 
\begin{align*}
\mathbb{E}[Q_n] & = \sum_{i_1, i_2, i_3, i_4 (dist)} \mathbb{E}[(A_{i_1 i_2} -  \widetilde{A}_{i_1 i_2}) (A_{i_2 i_3} - \widetilde{A}_{i_2 i_3}) (A_{i_3 i_4} - \widetilde{A}_{i_3 i_4}) (A_{i_4 i_1} - \widetilde{A}_{i_4 i_1})] \\
& =  \sum_{i_1, i_2, i_3, i_4 (dist)} \mathbb{E}[A_{i_1 i_2} -  \widetilde{A}_{i_1 i_2}] \mathbb{E}[A_{i_2 i_3} - \widetilde{A}_{i_2 i_3}] \mathbb{E}[A_{i_3 i_4} - \widetilde{A}_{i_3 i_4}] \mathbb{E}[A_{i_4 i_1} - \widetilde{A}_{i_4 i_1}]. 
\end{align*}
Together with $\mathbb{E}[A_{ij}] = \Omega_{ij}$, $\mathbb{E}[\widetilde{A}_{ij}] = \widetilde{\Omega}_{ij}$ and that $\Delta_{ij} = \Omega_{ij} - \widetilde{\Omega}_{ij}$, we obtain 
\begin{align*}
\mathbb{E}[Q_n] 
 = \sum_{i_1, i_2, i_3, i_4 (dist)} \Delta_{i_1i_2}\Delta_{i_2i_3}\Delta_{i_3i_4}\Delta_{i_4i_1} 
=  \tr(\Delta^4) - \sum_{\substack{i_1, i_2, i_3, i_4\\ \text{ non-distinct}}}\Delta_{i_1i_2}\Delta_{i_2i_3}\Delta_{i_3i_4}\Delta_{i_4i_1}. 
\end{align*}
The remaining part  of this section is to show, for sufficiently large $n$,  
\beq\label{Mean-Rem}
\sum_{\substack{i_1, i_2, i_3, i_4 \\ \text{non-distinct}} } \Delta_{i_1i_2}\Delta_{i_2i_3}\Delta_{i_3i_4}\Delta_{i_4i_1}=o\big(\|\bar\theta\|^2\cdot\tr(\Delta^2)\big).
\eeq
That $i_1, i_2, i_3, i_4$ are non-distinct implies that there are a pair of identical indices, thus 
\[
\begin{split}
   \Big|\sum_{\substack{i_1, i_2, i_3, i_4 \\ \text{non-distinct}} } \Delta_{i_1i_2}\Delta_{i_2i_3}\Delta_{i_3i_4}\Delta_{i_4i_1}\Big|&\leq 4 \sum_{i_1,i_2,i_3} |\Delta_{i_1 i_2}\Delta_{i_2 i_3}\Delta_{i_3 i_1}\Delta_{i_1 i_1}|+2\sum_{i_1,i_2,i_3} |\Delta_{i_1 i_2}^2\Delta_{i_2 i_3}^2| \\
   &\leq 2\sum_{i_1,i_2,i_3} (\Delta_{i_1 i_2}^2\Delta_{i_3 i_1}^2+\Delta_{i_2 i_3}^2\Delta_{i_1 i_1}^2+\Delta_{i_1 i_2}^2\Delta_{i_2 i_3}^2)\\
   &\leq C \sum_{i_1,i_2,i_3} (\Delta_{i_1 i_2}^2 \bar{\theta}_{i_3}^2 \bar{\theta}_{\max}^2+\Delta_{i_2 i_3}^2 \bar{\theta}_{i_1}^2 \bar{\theta}_{\max}^2+\Delta_{i_1 i_2}^2 \bar{\theta}_{i_3}^2 \bar{\theta}_{\max}^2)\\
   &= C\cdot \bar{\theta}_{\max}^2\|\bar\theta\|^2\tr(\Delta^2)=o(\|\bar\theta\|^2\cdot\tr(\Delta^2)),
\end{split}
\]
where the second line is by Cauchy-Schwarz inequality. 
Thus we complete the proof of \eqref{Mean-Rem}. Furthermore, by $\delta_1^2/(\lambda_1+\tilde\lambda_1)\to \infty$, we know $\|\bar\theta\|^2\asymp \|\theta\|^2+\|\tilde\theta\|^2\asymp \lambda_1 +\tilde\lambda_1=o(\delta_1^2)$. Therefore, $\|\bar\theta\|^2\cdot\tr(\Delta^2)=o(\delta_1^4)=o(\tr(\Delta^4))$.

Consider the  variance  of $Q_n$. Recall that 
\[
A_{ij} - \widetilde{A}_{ij} = (W_{ij} + {\Omega}_{ij}) - (\widetilde{W}_{ij} + \widetilde{\Omega}_{ij}) = (W_{ij} - \widetilde{W}_{ij}) + \Delta_{ij}, \mbox{ for  $i\neq j$},
\]
which indicates
\begin{align*}
Q_n & = \sum_{i_1, i_2, i_3, i_4 (dist)} (A_{i_1 i_2} -  \widetilde{A}_{i_1 i_2}) (A_{i_2 i_3} - \widetilde{A}_{i_2 i_3}) (A_{i_3 i_4} - \widetilde{A}_{i_3 i_4}) (A_{i_4 i_1} - \widetilde{A}_{i_4 i_1}) \\
& = \sum_{i_1, i_2, i_3, i_4 (dist)} (W_{i_1 i_2} -  \widetilde{W}_{i_1 i_2} + \Delta_{i_1i_2}) (W_{i_2 i_3} - \widetilde{W}_{i_2 i_3} + \Delta_{i_2i_3}) \cdots (W_{i_4 i_1} - \widetilde{W}_{i_4 i_1} + \Delta_{i_4i_1}). 
\end{align*}
By symmetry, we decompose $Q_n$ as the sum of six terms
\[
S_1 = \sum_{i_1, i_2, i_3, i_4 (dist)} (W_{i_1 i_2} -  \widetilde{W}_{i_1 i_2}) (W_{i_2 i_3} - \widetilde{W}_{i_2 i_3}) (W_{i_3 i_4} - \widetilde{W}_{i_3 i_4}) (W_{i_4 i_1} - \widetilde{W}_{i_4 i_1}),
\]
\[
S_2 = 4\sum_{i_1, i_2, i_3, i_4 (dist)} (W_{i_1 i_2} -  \widetilde{W}_{i_1 i_2}) (W_{i_2 i_3} - \widetilde{W}_{i_2 i_3}) (W_{i_3 i_4} - \widetilde{W}_{i_3 i_4}) \Delta_{i_4i_1},
\]
\[
S_3 = 4\sum_{i_1, i_2, i_3, i_4 (dist)} (W_{i_1 i_2} -  \widetilde{W}_{i_1 i_2}) (W_{i_2 i_3} - \widetilde{W}_{i_2 i_3}) \Delta_{i_3i_4} \Delta_{i_4i_1},
\]
\[
S_4 = 2\sum_{i_1, i_2, i_3, i_4 (dist)} (W_{i_1 i_2} -  \widetilde{W}_{i_1 i_2}) \Delta_{i_2i_3} (W_{i_3 i_4} - \widetilde{W}_{i_3 i_4}) \Delta_{i_4i_1},
\]
\[
S_5 = 4\sum_{i_1, i_2, i_3, i_4 (dist)} (W_{i_1 i_2} -  \widetilde{W}_{i_1 i_2}) \Delta_{i_2i_3} \Delta_{i_3 i_4}\Delta_{i_4i_1}, \quad \mbox{and} \quad S_6 = \sum_{i_1, i_2, i_3, i_4 (dist)} \Delta_{i_1 i_2} \Delta_{i_2i_3} \Delta_{i_3 i_4}\Delta_{i_4i_1}.
\]
Recall the basic inequality that $\mathrm{Var}(X_1+\cdots+X_6) \leq 6[\mathrm{Var}(X_1) + \cdots +  \mathrm{Var}(X_6)]$ for random variables $X_1, X_2, \cdots, X_6$. It suffices to control the variance of $S_1, \cdots, S_6$. 

Consider the variance of $S_1$.  Recall that $W_{i_1i_2}, W_{i_2i_3}, W_{i_3i_4}, W_{i_4i_1}, \widetilde{W}_{i_1i_2}, \widetilde{W}_{i_2i_3}, \widetilde{W}_{i_3i_4},\widetilde{W}_{i_4i_1}$ are mutually independent and 
\beq\label{VarW}
\mathbb{E}(W_{i_1 i_2} -  \widetilde{W}_{i_1 i_2})^2 = \mathbb{E}[W_{i_1 i_2}^2] + \mathbb{E}[ \widetilde{W}_{i_1 i_2}^2] = \Omega_{i_1i_2}(1 - \Omega_{i_1i_2}) +  \widetilde{\Omega}_{i_1i_2}(1 - \widetilde{\Omega}_{i_1i_2})\leq C\bar{\theta}_{i_1}\bar{\theta}_{i_2}.
\eeq
Mimicking the proof of Theorem \ref{thm:null}, we directly have \beq
\mathrm{Var}(S_1) \leq C \|\bar{\theta}\|^8.
\eeq

Consider the variance of $S_2$. Recall that 
\[
S_2 = 8\sum_{\substack{i_1, i_2, i_3, i_4 (dist) \\ i_1<i_4} } (W_{i_1 i_2} -  \widetilde{W}_{i_1 i_2}) (W_{i_2 i_3} - \widetilde{W}_{i_2 i_3}) (W_{i_3 i_4} - \widetilde{W}_{i_3 i_4}) \Delta_{i_4i_1},
\]
where the terms in the summation are  mean zero and uncorrelated with each other.  We obtain 
\begin{align*}
\mathrm{Var}(S_2) & = 64 \sum_{\substack{i_1, i_2, i_3, i_4 (dist) \\ i_1<i_4} }\mathbb{E}[(W_{i_1 i_2} -  \widetilde{W}_{i_1 i_2})^2 (W_{i_2 i_3} - \widetilde{W}_{i_2 i_3})^2 (W_{i_3 i_4} - \widetilde{W}_{i_3 i_4})^2 \Delta_{i_4i_1}^2] \\
&  \leq C\sum_{i_1, i_2, i_3, i_4 (dist)}(\bar{\theta}_{i_1}\bar{\theta}_{i_2})(\bar{\theta}_{i_2}\bar{\theta}_{i_3})(\bar{\theta}_{i_3}\bar{\theta}_{i_4})\Delta_{i_4i_1}^2  \leq C \sum_{i_1, i_2, i_3, i_4 } \bar\theta_{i_1}^3\bar\theta_{i_2}^2\bar\theta_{i_3}^2\bar\theta_{i_4}^3 = o(\|\bar\theta\|^8), 
\end{align*}
where we've used $\bar\theta_{\max}\goto0$ and so $\|\bar\theta\|^3_3 \leq\bar\theta_{\max}\|\bar\theta\|^2$ in the last inequality.   

Consider the variance of $S_3$. Rewrite it as \[
S_3 = 8\sum_{\substack{i_1, i_2, i_3 (dist) \\ i_1<i_3} } (W_{i_1 i_2} -  \widetilde{W}_{i_1 i_2}) (W_{i_2 i_3} - \widetilde{W}_{i_2 i_3}) \bigg(\sum_{i_4\notin\{i_1, i_2, i_3\}}\Delta_{i_3i_4}\Delta_{i_4i_1}\bigg). 
\]
By similar argument, 
\begin{align*}
\mathrm{Var}(S_3) & = 64\sum_{\substack{i_1, i_2, i_3 (dist) \\ i_1<i_3} } \mathbb{E}\bigg[(W_{i_1 i_2} -  \widetilde{W}_{i_1 i_2})^2 (W_{i_2 i_3} - \widetilde{W}_{i_2 i_3})^2\bigg(\sum_{i_4\notin\{i_1, i_2, i_3\}}\Delta_{i_3i_4}\Delta_{i_4i_1}\bigg)^2\bigg] \\
& \leq C \sum_{i_1, i_2, i_3}(\bar{\theta}_{i_1}\bar{\theta}_{i_2})(\bar{\theta}_{i_2}\bar{\theta}_{i_3})\bigg(\sum_{i_4\notin\{i_1, i_2, i_3\}}\Delta_{i_3i_4}\Delta_{i_4i_1}\bigg)^2. 
\end{align*}
Since $\bar\theta_{\max} = o(1)$, the above term is no more than
\begin{align*}
\sum_{i_1, i_2, i_3} \bar\theta_{i_2}^2\bigg(\sum_{i_4\notin\{i_1, i_2, i_3\}}\Delta_{i_3i_4}\Delta_{i_4i_1}\bigg)^2 
\leq 
\sum_{i_1, i_2, i_3}  \bar\theta_{i_2}^2\bigg(\sum_{i_4\notin\{i_1, i_2, i_3\}}\Delta_{i_3i_4}^2\bigg)\bigg(\sum_{i_4\notin\{i_1, i_2, i_3\}}\Delta_{i_4i_1}^2\bigg),
\end{align*}
where we've used the Cauchy-Schwarz inequality. Each term above are non-negative, so the sum is no more than 
\[
\sum_{i_1, i_2, i_3}  \bar\theta_{i_2}^2\bigg(\sum_{i_4}\Delta_{i_3i_4}^2\bigg)\bigg(\sum_{i_5}\Delta_{i_5i_1}^2\bigg) = \sum_{i_1, i_2, i_3, i_4, i_5} \bar\theta_{i_2}^2\Delta_{i_3i_4}^2\Delta_{i_5i_1}^2  =\|\bar\theta\|^2\cdot[\tr(\Delta^2)]^2.
\]
By Young's inequality that $ab \leq \frac{a^3}{3} + \frac{2b^{3/2}}{3}$, the variance of $S_3$ is upper bound as 
\beq
\mathrm{Var}(S_3) \leq C (\|\bar\theta\|^6 + [\tr(\Delta^2)]^3) \leq C(\|\bar\theta\|^8+ [\tr(\Delta^2)]^3). 
\eeq

Consider the variance of $S_4$.  Mimicking the argument in $S_2$, we find 
\begin{align*}
\mathrm{Var}(S_4) & = 16  \sum_{i_1, i_2, i_3, i_4 (dist)}\mathbb{E}[(W_{i_1 i_2} -  \widetilde{W}_{i_1 i_2})^2 \Delta_{i_2i_3}^2 (W_{i_3 i_4} - \widetilde{W}_{i_3 i_4})^2 \Delta_{i_4i_1}^2] \\
& \leq C\sum_{i_1, i_2, i_3, i_4 (dist)}(\bar{\theta}_{i_1}\bar{\theta}_{i_2})\Delta_{i_2i_3}^2(\bar{\theta}_{i_3}\bar{\theta}_{i_4})\Delta_{i_1i_4}^2 \leq C\sum_{i_1, i_2, i_3, i_4 } \bar\theta_{i_1}^3\bar\theta_{i_2}^3\bar\theta_{i_3}^3\bar\theta_{i_4}^3 = \|\bar\theta\|_3^{12}  = o(\|\bar\theta\|^8), 
\end{align*}
where we've used $\|\bar\theta\|^3_3=o(\|\bar{\theta}\|^2)$  in the last inequality.  

Consider the variance of $S_5$. Rewrite it as \[
S_5 = 8\sum_{i_1< i_2 } (W_{i_1 i_2} -  \widetilde{W}_{i_1 i_2}) \bigg(\sum_{i_3, i_4\notin\{i_1, i_2\}}\Delta_{i_2i_3}\Delta_{i_3i_4}\Delta_{i_4i_1}\bigg). 
\]
Again, the terms in the above summation are mean zero and uncorrelated with each other, which indicates 
\beq\label{S5A}
\begin{split}
\mathrm{Var}(S_5) & = 64 \sum_{i_1< i_2}\mathbb{E}\bigg[(W_{i_1 i_2} -  \widetilde{W}_{i_1 i_2})^2  \bigg(\sum_{i_3, i_4\notin\{i_1, i_2\}}\Delta_{i_2i_3}\Delta_{i_3i_4}\Delta_{i_4i_1}\bigg)^2\bigg] \\
& \leq C\sum_{i_1, i_2 (dist)} (\bar{\theta}_{i_1}\bar{\theta}_{i_2})  \bigg(\sum_{i_3, i_4\notin\{i_1, i_2\}}\Delta_{i_2i_3}\Delta_{i_3i_4}\Delta_{i_4i_1}\bigg)^2. 
\end{split}
\eeq
Using Cauchy-Schwarz inequality, we find 
\begin{align*} 
\bigg(\sum_{i_3, i_4\notin\{i_1, i_2\}}\Delta_{i_2i_3}\Delta_{i_3i_4}\Delta_{i_4i_1}\bigg)^2
&  \leq \bigg(\sum_{i_3, i_4\notin\{i_1, i_2\}}\Delta_{i_3i_4}^2 \bigg)\bigg( \sum_{i_3, i_4\notin\{i_1, i_2\}}\Delta_{i_2i_3}^2\Delta_{i_4i_1}^2\bigg). 
\end{align*}
Plug it into \eqref{S5A} and notice $\bar\theta_{i_1}\bar\theta_{i_2} \leq \bar\theta_{\max}^2  = o(1)$, we obtain 
\beq
\mathrm{Var}(S_5) \leq \sum_{i_1, i_2}\bigg(\sum_{i_3, i_4}\Delta_{i_3i_4}^2 \bigg)\bigg( \sum_{i_3, i_4}\Delta_{i_2i_3}^2\Delta_{i_4i_1}^2\bigg) = [\tr(\Delta^2)]^3. 
\eeq
Combining the upper bound on variance of $S_1,\cdots,S_5$  and notice that $S_6$ is non-stochastic (so the variance is $0$), we get 
\[
\mathrm{Var}(Q_n)\leq C[\mathrm{Var}(S_1)  + \cdots + \mathrm{Var}(S_6)] \leq C [\|\bar\theta\|^8 +  [\tr(\Delta^2)]^3 ]\leq C[\|\theta\|^8+ \|\tilde\theta\|^8+  [\tr(\Delta^2)]^3]. 
\]
 The last inequality is by $\|\bar\theta\|^8\leq 16(\|\theta\|^2+\|\tilde\theta\|^2)^4\leq 16\cdot 8(\|\theta\|^8+\|\tilde\theta\|^8)$. Using the fact that $\tr(\Omega^4)\asymp \|\theta\|^8$ and $\tr(\tilde\Omega^4)\asymp \|\tilde \theta\|^8$, we showed the first claim.

Consider the last claim. 
It suffices to show for any fixed constant $c > 0$, 
\[
\mathbb{P}_{H_1^{(n)}}\biggl(Q_n \leq cz_{\alpha} \cdot \sqrt{{C}_n + \widetilde{C}_n }   \biggr) \goto 0. 
\]
Fixing $0 < \eps < 1$, let $A_{\eps}$ be the event  $\big\{ ({C}_n + \widetilde{C}_n) \leq (1 + \eps)\cdot\mathbb{E} [  {C}_n + \widetilde{C}_n]  \big\}$.  
By the second claim of theorem \ref{thm:null},  over the event $A_{\eps}$, ${C}_n + \widetilde{C}_n \leq C(\|\theta\|^8 + \|\tilde\theta\|^8)$  and  $\mathbb{P}(A_{\eps}^c) = o(1)$.   Therefore,  
\begin{align*} 
& \mathbb{P}_{H_1^{(n)}}\biggl(Q_n \leq cz_{\alpha} \cdot \sqrt{{C}_n + \widetilde{C}_n }   \biggr)  \\
\leq & \mathbb{P}_{H_1^{(n)}}\biggl(Q_n \leq cz_{\alpha} \cdot \sqrt{{C}_n + \widetilde{C}_n } , A_{\eps}\biggr) + \mathbb{P}(A_{\eps}^c)  \\
\leq & \mathbb{P}_{H_1^{(n)}} \biggl(Q_n \leq Cz_{\alpha} \cdot (\|\theta\|^4 + \|\tilde\theta\|^4 )  \biggr) + o(1),  
\end{align*} 
where $C$ denotes a generic constant and by Chebyshev's inequality, the first term in the last line 
\begin{equation} \label{mypfA1}
\leq  [\mathbb{E}(Q_n) - C z_{\alpha} \cdot (\|\theta\|^4+\|\tilde\theta\|^4)]^{-2} \cdot \mathrm{Var}(Q_n). 
\end{equation} 
 Recall that $\mathbb{E}(Q_n)=  \tr(\Delta^4 )+o\big(\|\bar\theta\|^2\cdot\tr(\Delta^2)\big)$. Under the condition that $\delta_1^2/(\lambda_1+\widetilde{\lambda}_1)\to \infty$, $\|\bar\theta\|^2\cdot\tr(\Delta^2)\asymp(\|\theta\|^2+\|\tilde\theta\|^2)\cdot\tr(\Delta^2)\asymp (\lambda_1+\tilde\lambda_1)\delta_1^2=o(\delta_1^4)$, so  $\mathbb{E}(Q_n)\asymp  \tr(\Delta^4 )\asymp \delta_1^4$. Meanwhile, $ \|\theta\|^4 +  \|\tilde\theta\|^4\asymp \lambda_1^2+\widetilde{\lambda}_1^2=o(\delta_1^4)$, thus  
\beq\label{claim:mean}
\mathbb{E}[Q_n] - C  z_{\alpha}  ( \|\theta\|^4 +  \|\tilde\theta\|^4)   \geq  \frac{1}{2}\tr(\Delta^4 ). 
\eeq
Furthermore, notice $
\mathrm{Var}(Q_n) \leq   C[\|\theta\|^8+ \|\tilde\theta\|^8+  [\tr(\Delta^2)]^3]  \leq C[ \lambda_1^4 + \widetilde{\lambda}_1^4 + \delta_1^6]$,
 the right hand side of (\ref{mypfA1}) does not exceed 
\begin{equation} \label{mypfC2} 
C \times  \frac{\lambda_1^4 + \widetilde{\lambda}_1^4 + \delta_1^6 }{\big(\tr(\Delta^4)\big)^2}  \leq C \times  \frac{(\lambda_1 + \widetilde{\lambda}_1)^4 + \delta_1^6}{\delta_1^8} \goto 0, 
\end{equation} 
where  we've used the property that $\delta_1^2 \gg \lambda_1 + \widetilde{\lambda}_1 \asymp \|\bar\theta\|^2$  and that $\|\bar\theta\|\goto\infty$. Therefore $\psi\goto\infty$ in probability under the alternative hypothesis, and the Type II error goes to $0$.  

Under the null,  
\[
\frac{Q_n}{8\sqrt{({C}_n + \widetilde{C}_n)}} \stackrel{d}{\longrightarrow} N(0,1), 
\]
so the Type I error is 
\[
 \mathbb{P}_{H_0^{(n)}}\biggl(\frac{Q_n }{{8\sqrt{({C}_n + \widetilde{C}_n)}}  } \geq  z_{\alpha} \biggr)   = \alpha + o(1), 
\]
Combining above, the power of the IBM test goes to $1$ as $n\goto\infty.$


\subsection{Proof of Theorem \ref{thm:LF}-\ref{thm:LB}}

Notice that Theorem \ref{thm:LF} follows directly from Theorem \ref{thm:LB}, we thus only prove Theorem \ref{thm:LB}. 

Our first step is to construct $\widetilde{\Omega}(\sigma)\in\mathcal{M}_n(\beta_n, K+1, c_0)$ satisfying 
\beq\label{construct-key}
\widetilde{\Omega}_{ij}(\sigma) = \Omega_{ij} + \eps_n \sigma_i\sigma_j\theta_i\theta_j\Pi_{iK}\Pi_{jK},
\eeq
where $\eps_n$  is a diminishing sequence with its value to be specify and  $\sigma = (\sigma_1, ..., \sigma_n)$ is a binary vector (i.e.,  $\sigma_i\in\{-1, 1\}$)  for $i = 1, 2, ..., n$. 

For $\Omega = \Theta\Pi P\Pi'\Theta \in\mathcal{M}_n(\beta_n, K, c_0)$, we introduce $\check\Pi(\sigma)\in\mathbb{R}^{n, K+1}$ as follow
\begin{align*}
&\check{\Pi}_{i \ell}(\sigma)= \Pi_{i\ell}, \quad 1\leq i\leq n, 1\leq \ell\leq K-1, \quad\mbox{ and } \quad \check{\Pi}_{iK}(\sigma) = \frac{1 + \sigma_i}{2}  \Pi_{iK}, \quad \check{\Pi}_{i,K+1}(\sigma) = \frac{1 - \sigma_i}{2} \Pi_{iK}. 
\end{align*}
It's not hard to see $\check\Pi(\sigma)$ is a non-negative matrix with row sums equal to $1$. Therefore it is a valid membership matrix with $(K+1)$ communities. 
Rewrite \[
P = \left(\begin{matrix} P_0& \alpha\\\alpha^T & 1
\end{matrix}\right) \in\mathbb{R}^{K,K},
\]
we introduce   $\check P \in\mathbb{R}^{K+1, K+1}$ such that \[
\check{P} = \left(\begin{matrix} P_0 & \alpha & \alpha \\
\alpha^T & 1+\eps_n & 1-\eps_n \\
\alpha^T & 1-\eps_n & 1 + \eps_n 
\end{matrix}\right). 
\]
It should be noted that $\check P$ is not a valid probability matrix for DCMM with $(K+1)$ communities, as the last two diagonal elements of $\check P$ are not equal to $1$. 
For notation simplicity, we write $\check\Pi$ in short for $\check\Pi(\sigma)$ in the rest of the proof. 
Write the $i$-th row of $\check{\Pi}$ as $\check{\pi}_i'$, and denote the first $(K-1)$ entries of $\check{\pi}_i'$ by $\pi'_{i[K-1]}$. 
Let $\widetilde{\Omega}(\sigma) = \Theta \check{\Pi} \check{P}\check{\Pi}'\Theta$, and we have
\begin{align*}
\widetilde{\Omega}_{ij}(\sigma) = & \theta_i\theta_j \check{\pi}_i' \check{P}\check{\pi}_j = \theta_i\theta_j \left[  \pi_{i[K-1]}'P_0 \pi_{j[K-1]} + (1+\eps_n)(\check{\Pi}_{iK}\check{\Pi}_{jK}+\check{\Pi}_{i,K+1}\check{\Pi}_{j,K+1})\right. \\
& \left.+ (1-\eps_n)(\check{\Pi}_{iK}\check{\Pi}_{j,K+1}+\check{\Pi}_{i,K+1}\check{\Pi}_{jK}) + (\check{\Pi}_{iK}+\check{\Pi}_{i,K+1}) \pi_{j[K-1]}'\alpha
+ (\check{\Pi}_{jK}+\check{\Pi}_{j,K+1}) \pi_{i[K-1]}'\alpha\right] .
\end{align*}
Notice that $\check{\Pi}_{iK}+\check{\Pi}_{i,K+1}=\Pi_{iK}$ and that
\[
\Omega_{ij}  = \theta_i\theta_j \pi_i'P\pi_j= \theta_i\theta_j \left[ \pi_{i[K-1]}'P_0 \pi_{j[K-1]} + \Pi_{iK} \pi_{j[K-1]}'\alpha + \Pi_{jK} \pi_{i[K-1]}'\alpha+ \Pi_{iK}\Pi_{jK}\right] ,
\]
 we obtain
\[
\widetilde{\Omega}_{ij}(\sigma) = \Omega_{ij} + \eps_n\theta_i\theta_j (\check{\Pi}_{iK}-\check{\Pi}_{i,K+1})
(\check{\Pi}_{jK}-\check{\Pi}_{j,K+1}) =  \Omega_{ij}+ \eps_n\theta_i\theta_j \sigma_i\sigma_j \Pi_{iK}\Pi_{jK}.
\]
Therefore $\widetilde\Omega(\sigma) = \Theta \check\Pi\check P\check \Pi'\Theta$ satisfies 
\eqref{construct-key}. However, as we have mentioned before, $\check P$ is not a valid probability matrix for DCMM. We still need to find $\widetilde \Theta$, $\widetilde\Pi$ and $\widetilde P$ such that 
 $\widetilde\Omega(\sigma) = \widetilde\Theta \widetilde\Pi\widetilde P \widetilde \Pi'\widetilde\Theta \in\mathcal{M}_n(\beta_n, K+1, c_0)$.

Write $D = \diag(1, \dots, 1, \sqrt{1+\eps_n}, \sqrt{1+\eps_n})$, and let $\widetilde{P}=D^{-1} \check{P}D^{-1}$ (obviously,  $\|\tilde{P}\|_{\max}\leq \|P\|_{\max}\leq c_0^{-1}$). It's not hard to verify $\widetilde P \in\mathbb{R}^{K+1, K+1}$ is a valid probability matrix for DCMM with unit diagonals.
Introduce a diagonal matrix $G \in\mathbb{R}^{n,n}$ such that 
\[
G_{ii} = \sum_{k=1}^{K-1} \Pi_{ik} + \sqrt{1+\eps_n} \cdot\Pi_{iK}. 
\]
Also, introduce $\widetilde{\Pi}\in\mathbb{R}^{n, K+1}$ such that $\widetilde{\Pi} = G^{-1}\check\Pi D$. Then it's not hard to verify that $\widetilde{\Pi}$ is a valid membership matrix for DCMM with $(K+1)$ communities, as it's a non-negative matrix and all rows sum up to $1$. 
Combining above, we have
\[
\widetilde{\Omega} = \Theta \check{\Pi} \check{P} \check{\Pi}' \Theta = \Theta G   \widetilde{\Pi} \widetilde{P}\widetilde{\Pi}' G \Theta = \widetilde{\Theta}\widetilde{\Pi} \widetilde{P}\widetilde{\Pi}'\widetilde{\Theta},
\]
where $\widetilde\Theta = \Theta G \in\mathbb{R}^{n,n}$ is a diagonal matrix.
Recall that $\Omega \in \mathcal{M}_n(\beta_n,K, c_0)$. By $G_{ii} \in [1, \sqrt{1 + \eps_n}]$, we have $\tilde{\theta}_i =\theta_i \cdot G_{ii}  \in [\theta_i,\sqrt{1+\epsilon_n}\cdot \theta_i]$. Additionally, $\eps_n = o(1)$.  Therefore,  $\tilde{\theta}_{\max} \leq  \sqrt{1+\eps_n}\theta_{\max} \leq  \sqrt{1+\eps_n}K \beta_n  \leq (K+1)\beta_n$ and $\|\tilde{\theta}\|\geq \|\theta\|\geq \beta_n^{-1}$ for sufficiently large $n$.

Consider the spectral norm of $\widetilde\Omega(\sigma)$. Rewrite $\widetilde{\Omega}(\sigma)$ as 
\[
\widetilde{\Omega}(\sigma) = \Omega + \eps_n (\theta\circ \sigma\circ \pi^{(K)})\cdot (\theta\circ \sigma\circ \pi^{(K)})^T,
\]
where $\pi^{(K)}$ is the $K$-th column of $\Pi$. Then it's not hard to see the following upper bound
\beq\label{lb:spectral}
\|\widetilde{\Omega}(\sigma)-\Omega\| = \eps_n \|\theta\circ \pi^{(K)}\|^2 \leq \eps_n \|\theta\|^2.
\eeq
Recall that $\eps_n = o(1)$, we have \[
\|\widetilde{\Omega}(\sigma)\|\geq \|\Omega\| - \|\widetilde{\Omega}(\sigma)-\Omega\| \geq c_0K^{-1}\|{\theta}\|^2- \eps_n \|\theta\|^2, 
\]
where for sufficiently large $n$, the rightmost term is \[
(c_0/K - \eps_n)\|\theta\|^2 \geq (c_0/K - \eps_n)\|\tilde\theta\|^2/(1+\eps_n) \geq c_0 \|\tilde\theta\|^2/(K+1) .
\]
Similarly, we have 
\beq\label{tildeOmega-tmp}
\|\widetilde{\Omega}(\sigma)\|\leq \|\Omega\| + \|\widetilde{\Omega}(\sigma)-\Omega\| \leq c_0^{-1}K \|{\theta}\|^2+ \eps_n \|\theta\|^2 , 
\eeq
where for sufficiently large $n$ and $\eps_n\goto0$, the rightmost term is bounded by \[
(c_0^{-1}K + \eps_n)\|\tilde\theta\|^2 \leq c_0^{-1}(K+1)\|\tilde\theta\|^2. 
\]  Combining above, we conclude that  as long as $\eps_n=o(1)$, $\widetilde{\Omega} \in \mathcal{M}_n(\beta_n, K+1, c_0)$ and it satisfies \eqref{construct-key}.

Our next step is to show  for given $\{\beta_n\}_{n=1}^{\infty}$ and $\{\rho_n\}_{n=1}^{\infty}$, $\{\eps_n\}_{n=1}^{\infty}$ can be chosen such that  $(\Omega, \widetilde\Omega(\sigma))\in \mathcal{S}_n(\beta_n,\rho_n, K, K+1, c_0)$ . We've already shown  in the first step that $\widetilde\Omega(\sigma)\in\mathcal{M}_n(\beta_n, K+1, c_0)$  if $\eps_n=o(1)$.  It remains to show  $\{\eps_n\}_{n=1}^{\infty}$ can additionally satisfy\[
\frac{\|\Omega - \widetilde\Omega(\sigma)\|}{\sqrt{\|\Omega\| + \|\widetilde\Omega(\sigma)\|}} \geq  \sqrt{\rho_n},
\]
for given $\rho_n\goto 0$. 
Without loss of generality, we may assume 
\[
\|\theta\circ\pi^{(K)}\|^2\geq \|\theta\circ\pi^{(k)}\|^2, \qquad \mbox{for $1\leq k\leq K$},
\]
where $\pi^{(k)}$ is the $k$-th column of $\Pi$. 
As a result, 
\[
\|\theta\circ\pi^{(K)}\|^2\geq \frac{1}{K} \sum_{k=1}^K \|\theta \circ \pi^{(k)}\|^2 = \frac{1}{K}\sum_{i=1}^n\sum_{k=1}^K \theta_i^2\Pi_{i, k}^2.
\]
Noticing $\sum_{k=1}^K \Pi_{i,k} = 1$, we have $\sum_{k=1}^K\Pi_{i,k}^2 \geq 1/K$ by elementary algebra, which implies \[
\|\widetilde{\Omega}(\sigma)-\Omega\|  = \eps_n \|\theta\circ\pi^{(K)}\|^2\geq \frac{\eps_n}{K^2} \sum_{i=1}^n \theta_i^2 = \eps_n\cdot \frac{\|\theta\|^2}{K^2}. 
\]
Hence, 
\[
 \frac{\|\widetilde{\Omega}-\Omega\|}{\sqrt{\|\Omega\| + \|\widetilde\Omega\|}}\geq \frac{\eps_n\|\theta\|^2/K^2}{\sqrt{(2c_0^{-1}K+\eps_n) \|\theta\|^2}} \geq \eps_n \|\theta\| \cdot\frac{\sqrt{c_0}}{\sqrt{3}  K^{2.5}},
\]
where we've used $\|\Omega\| \leq c_0^{-1} K \|\theta\|^2$ and $\|\widetilde\Omega\|\leq (c_0^{-1}K + \eps_n)\|\theta\|^2$ (by our derivation, we actually have $\frac{\|\widetilde{\Omega}-\Omega\|}{\sqrt{\|\Omega\| + \|\widetilde\Omega\|}}\asymp \epsilon_n\|\theta\|$).
Therefore, for any given sequence $\rho_n=o(1)$,  we can find sequence $\eps_n$ such that  $\sqrt{\rho_n} = \eps_n \|\theta\| \cdot\frac{\sqrt{c_0}}{\sqrt{3} K^{2.5}}$.  Consequently, $\eps_n=o(1)$ and $(\Omega, \widetilde{\Omega}(\sigma)) \in\mathcal{S}_n(\beta_n, \rho_n, K, K+1, c_0)$.

Our last step is to construct $H_0^{(n)}$ and $H_{1,\sigma}^{(n)}$ as follows:
\[
H_0^{(n)}: A\sim \mathbb{P}_n , \widetilde{A}\sim \mathbb{P}_n, \qquad H_{1,\sigma}^{(n)}: A \sim \mathbb{P}_n, \widetilde{A} \sim \mathbb{Q}_n(\sigma),
\]
where $\mathbb{P}_n$ is the distribution of adjacency matrix indicated by $\Omega$, and $\mathbb{Q}_n(\sigma)$ is the distribution of adjacency matrix indicated by $\widetilde{\Omega}(\sigma)$. As shown in the second step, we have $(\Omega, \widetilde{\Omega}(\sigma)) \in\mathcal{S}_n(\beta_n,\rho_n, K, K+1, c_0)$ and $(\Omega, \Omega)\in \mathcal{S}_n^*(\beta_n, K, c_0)$. 

Write $\mathbb{Q}_n = \frac{1}{2^n}\sum_{\sigma \in \left\lbrace\pm 1 \right\rbrace ^n} \mathbb{Q}_n(\sigma)$. 
It suffices to show 
\begin{equation}
\int\left( \frac{d \mathbb{Q}_n }{d \mathbb{P}_n}\right)^2d\mathbb{P}_n = 1 + o(1), 
\quad \text{as} \quad n\goto\infty.
\end{equation}
Let $\sigma, \sigma'$ both be uniformly sampled from $\{1 ,-1\}^n$ independently. We re-write the $\chi^2$-distance as 
\begin{align*}
\int\left( \frac{d \mathbb{Q}_n }{d \mathbb{P}_n}\right)^2d\mathbb{P}_n = & \mathbb{E}_{\sigma, \sigma'}\left[ 
\prod_{1\leq i<j\leq n} \left( \frac{\widetilde{\Omega}_{ij}(\sigma)\widetilde{\Omega}_{ij}(\sigma')}{\Omega_{ij}} + \frac{(1-\widetilde{\Omega}_{ij}(\sigma))(1-\widetilde{\Omega}_{ij}(\sigma'))}{(1-\Omega_{ij})}
\right) \right] \\
= & \mathbb{E}_{\sigma, \sigma'}\left[ \prod_{1\leq i<j\leq n} \left(
1 + \frac{\Delta_{ij}(\sigma)\Delta_{ij}(\sigma')}{\Omega_{ij}(1-\Omega_{ij})}
\right) \right] \\
= & \mathbb{E}_{\sigma, \sigma'}\left[ \prod_{1\leq i<j\leq n} \left(1+
\frac{  \eps_n^2\theta_i^2\theta_j^2 \sigma_i\sigma_j\sigma_i'\sigma_j' \Pi_{iK}^2\Pi_{jK}^2     }{\Omega_{ij}(1-\Omega_{ij})}
\right) \right] .
\end{align*}
Note that  $\sigma \circ \sigma'$ can also be viewed as generated uniformly from $\{1 ,-1\}^n$ (thus replace $\sigma_i\times \sigma_i'$ by $\sigma_i$), and by $e^x\geq 1+x$, the above equation can be rewritten as 
\[
= \mathbb{E}_{\sigma}\left[ \prod_{1\leq i<j\leq n} \left(1+
\frac{  \eps_n^2\theta_i^2\theta_j^2 \sigma_i\sigma_j \Pi_{iK}^2\Pi_{jK}^2     }{\Omega_{ij}(1-\Omega_{ij})}
\right) \right] \leq 
\mathbb{E}_{\sigma}\left[ \exp\left\lbrace \sum_{1\leq i<j\leq n}
\frac{  \eps_n^2\theta_i^2\theta_j^2 \sigma_i\sigma_j \Pi_{iK}^2\Pi_{jK}^2     }{\Omega_{ij}(1-\Omega_{ij})}
\right\rbrace \right] .
\]
Introduce \[
S_{n} = \sum_{1\leq i < j\leq n}\frac{\eps_n^2  \theta_i^2\theta_j^2  \Pi_{iK}^2\Pi_{jK}^2     }{\Omega_{ij}(1-\Omega_{ij})}\times \sigma_i\sigma_j.
\]
Let $M^{(n)}_{ij}=\eps_n^2  \theta_i^2\theta_j^2  \Pi_{iK}^2\Pi_{jK}^2   \Omega_{ij}^{-1}(1-\Omega_{ij})^{-1}$, then $S_n=\sum_{1\leq i < j\leq n}M^{(n)}_{ij} \sigma_i\sigma_j$. By Proposition 8.13 in \cite{Foucart2013}, we have
\beq
\mathbb{P}(|S_n|\geq t)\leq 2 \exp\Big(-\min\Big\{\frac{3t^2}{32\|M^{(n)}\|^2_F},\frac{t}{8\|M^{(n)}\|}\Big\}\Big).
\eeq
Meanwhile,
\beq
\|M^{(n)}\|^2\leq \|M^{(n)}\|_F^2=\sum_{ i \neq j}\Big[\frac{\eps_n^2  \theta_i^2\theta_j^2  \Pi_{iK}^2\Pi_{jK}^2     }{\Omega_{ij}(1-\Omega_{ij})}\Big]^2.
\eeq
Notice $\Omega_{ij} = \theta_i\theta_j\pi_i'P\pi_j\geq \theta_i\theta_j\Pi_{iK}\Pi_{jK}$ and $0\leq \Pi_{iK}\leq 1$ for $1\leq i\leq n$, the above quantity is no more than \[
2\eps_n^4\sum_{1\leq i < j\leq n} \frac{\theta_i^2\theta_j^2}{(1-\Omega_{ij})^2} \asymp \eps_n^4 \|\theta\|^4\asymp\rho_n^2=o(1),
\]
where we've used $\Omega_{ij}\leq c_0^{-1}\theta_{\max}^2\goto0$ for $1\leq i, j \leq n$.
 
Since $\mathbb{P}(|S_n|\geq t)\times \exp(t)\to 0$ as $t\to \infty$ for large $n$, we can apply the tail-sum formula and get
\beq
\begin{split}
\mathbb{E}_{\sigma}(\exp(|S_n|))&=1+\int_0^{\infty}\exp(t)\mathbb{P}(|S_n|> t)dt\\
&\leq 1+ \int_0^{\infty}2\exp\Big(t-\frac{3t^2}{32\|M^{(n)}\|^2_F}\Big)dt +\int_0^{\infty}2\exp\Big(t-\frac{t}{8\|M^{(n)}\|}\Big)dt=1+o(1),
\end{split}
\eeq
where the last step is from $\|M^{(n)}\|\leq \|M^{(n)}\|_F=o(1)$.
\[1\leq \int\left( \frac{d \mathbb{Q}_n }{d \mathbb{P}_n}\right)^2d\mathbb{P}_n \leq \mathbb{E}_{\sigma}(\exp(S_n))\leq \mathbb{E}_{\sigma}(\exp(|S_n|))=1+o(1) \]
completes the proof. 

\section{Proof of Theorem \ref{thm:null-d}-\ref{thm:LB-d}}

\subsection{Proof of Theorem \ref{thm:null-d}} 

Recall that $\Omega = \widetilde\Omega$ under the null hypothesis, it follows that \[
A_{ij} - \widetilde A_{ij} = (W_{ij} + \Omega_{ij}) -  (\widetilde W_{ij} + \widetilde\Omega_{ij}) = W_{ij} - \widetilde W_{ij}. 
\]
Introduce 
\beq\label{def:O}
O = W - \widetilde W \quad \mbox{ and }\quad \Omega_{ij}^* = \Omega_{ij}(1-\Omega_{ij}), \mbox{ for $1\leq i, j\leq n$}.
\eeq
 It follows from basic probability and the independence between $W$ and $\widetilde W$ that
\beq\label{var-O}
\mathrm{Var}(O_{ij}) = \mathrm{Var}(W_{ij} - \widetilde W_{ij}) = \mathrm{Var}(W_{ij}) +\mathrm{Var}( \widetilde W_{ij}) = 2 \Omega_{ij}^*.
\eeq
Furthermore, we rewrite $Q_n$ as follows
\[
Q_n = \sum_{i_1, i_2, j_1, j_2 (dist)}  O_{i_1 j_1} O_{i_1 j_2} O_{i_2 j_1}  
O_{i_2 j_2} .
\]
Notice that  for distinct indices $i_1, i_2, j_1, j_2$,  the random variables $W_{i_1 j_1}, W_{i_2 j_1}, W_{i_1 j_2}, W_{i_2 j_2}$ and $\widetilde W_{i_1 j_1}, \widetilde W_{i_2 j_1}, \widetilde W_{i_1 j_2}, \widetilde W_{i_2 j_2}$  are mutually independent. Therefore, $\{O_{ij}\}_{i\neq j}$ are also mutually independent. 

%

We now consider the mean and variance part of $Q_n$. 
For the mean, it follows from independence that 
\[
\mathbb{E}[Q_n] = \sum_{i_1, i_2, j_1, j_2 (dist)}  \mathbb{E}[O_{i_1 j_1}] \mathbb{E}[O_{i_1 j_2}] \mathbb{E}[O_{i_2 j_1}] \mathbb{E}[O_{i_2 j_2}].
\]
Moreover, the above term equals to $0$ since $O_{ij}$ have zero means for $1\leq i, j \leq n$.
We then consider the variance. We first group the terms in $Q_n$ into uncorrelated groups. Notice that for each term in $Q_n$ indexed by $(i_1, i_2, j_1, j_2)$, there are $3$ other terms in $Q_n$ that are identical to it up to permutation, namely $(i_1, i_2, j_2, j_1)$, $(i_2, i_1, j_1, j_2)$ and $(i_2, i_1, j_2, j_1)$. 
 Define \beq\label{def:J4}
J_4(n)  = \{ (i_1, i_2, j_1, j_2) \qquad  \mbox{where } 1\leq i_1<i_2\leq n, 1\leq j_1<j_2\leq n \}.
\eeq
Then each such $4$-term group can be represented by one unique element in $J_4(n)$.

Therefore, we can rewrite $Q_n$ as
\[
Q_n = 4\sum_{J_4(n)} O_{i_1 j_1}O_{i_2 j_1}O_{i_1 j_2}O_{i_2 j_2},
\]
where the terms in the summation are now uncorrelated with each other since the underlying quadrilateral is different. It follows that \beq\label{temp-VarQ-d}
\mathrm{Var}(Q_n) = 16 \sum_{J_4(n)}\mathrm{Var}\big( O_{i_1 j_1}O_{i_2 j_1}O_{i_1 j_2}O_{i_2 j_2}\big). 
\eeq
For distinct $i_1, i_2, j_1, j_2$ and by previous argument that $O_{i_1 j_1} O_{i_1 j_2}O_{i_2 j_1} O_{i_2 j_2}$ has zero mean, we obtain
 \begin{align*}
& \quad \mathrm{Var}\big( O_{i_1 j_1}O_{i_2 j_1}O_{i_1 j_2}O_{i_2 j_2}\big) =  \mathbb{E}\big[ O_{i_1 j_1}^2 O_{i_1 j_2}^2O_{i_2 j_1}^2O_{i_2 j_2}^2\big] =  \mathbb{E}[O_{i_1 j_1}^2]  \mathbb{E}[O_{i_1 j_2}^2] \mathbb{E}[O_{i_2 j_1}^2] \mathbb{E}[O_{i_2j_2}^2],
\end{align*}
where we've again used the independence between $\{O_{ij}\}_{i\neq j}$. 
Recall \eqref{var-O} and $\Omega_{ij}^* = [1+o(1)]\Omega_{ij}$ since  $\Omega_{ij}=\theta_i\zeta_j(\pi_i'P\gamma_j)\leq \theta_i\zeta_j \|\pi_i\|\|P\|\|\gamma_j\|\leq C\theta_{\max}\zeta_{\max}\goto0$, we have 
\[
 \mathbb{E}[O_{i_1 j_1}^2]  \mathbb{E}[O_{i_1 j_2}^2] \mathbb{E}[O_{i_2 j_1}^2] \mathbb{E}[O_{i_2j_2}^2] = 16[1+o(1)]\cdot\Omega_{i_1j_1}\Omega_{i_1j_2}\Omega_{i_2j_1}\Omega_{i_2j_2}. 
\]
Plug it back into \eqref{temp-VarQ-d}, we obtain
 \begin{align*}
\mathrm{Var}(Q_n) & = 16\cdot 16[1+o(1)]  \cdot \sum_{J_4(n)} \Omega_{i_1j_1}\Omega_{i_1j_2}\Omega_{i_2j_1}\Omega_{i_2j_2} \\
& = 64[1+o(1)]\cdot \sum_{i_1,i_2,i_3,i_4(dist)}\Omega_{i_1j_1}\Omega_{i_1j_2}\Omega_{i_2j_1}\Omega_{i_2j_2} \\
& = 64[1+o(1)]\cdot \mathbb{E}[C_n]. 
\end{align*}
This  completes the proof of the mean and variance part in the  first claim. 

Next, we consider the  second claim.  
By definition, \[
\mathbb{E}[C_n] = \sum_{i_1, i_2, j_1, j_2(dist)} \Omega_{i_1j_1}\Omega_{i_1j_2}\Omega_{i_2j_1}\Omega_{i_2j_2}. 
\]
Moreover, it follows from basic linear algebra that 
\[
\text{tr}[(\Omega'\Omega)^2] = \sum_{i_1, i_2, j_1, j_2} \Omega_{i_1j_1}\Omega_{i_1j_2}\Omega_{i_2j_1}\Omega_{i_2j_2}.
\]
To show the first equation, notice \[
\text{tr}[(\Omega'\Omega)^2]  = \mathbb{E}[C_n]  + \sum_{\substack{i_1, i_2, j_1, j_2\\ \text{non-distinct}}} \Omega_{i_1j_1}\Omega_{i_1j_2}\Omega_{i_2j_1}\Omega_{i_2j_2} ,
\]
it suffices to show \beq\label{thm1-claim3-key}
\sum_{\substack{i_1, i_2, j_1, j_2\\ \text{non-distinct}}} \Omega_{i_1j_1}\Omega_{i_1j_2}\Omega_{i_2j_1}\Omega_{i_2j_2} \leq C[\|\theta\|^4\|\zeta\|_4^4 + \|\zeta\|^4\|\theta\|_4^4].
\eeq
That $i_1, i_2, j_1, j_2$ are not distinct implies that there must be an identical pair, so
\begin{align*}
\sum_{\substack{i_1, i_2, j_1, j_2\\ \text{non-distinct}}} \Omega_{i_1j_1}\Omega_{i_1j_2}\Omega_{i_2j_1}\Omega_{i_2j_2} & \leq \sum_{i_1, i_2, j_1} \Omega_{i_1j_1}^2 \Omega_{i_2j_1}^2  + 4  \sum_{i_1, i_2, j_1} \Omega_{i_1j_1}\Omega_{i_1i_2}\Omega_{i_2j_1}\Omega_{i_2i_2} \\
& \quad +  \sum_{i_1, j_1, j_2} \Omega_{i_1j_1}^2 \Omega_{i_1j_2}^2. 
\end{align*}
Notice $\Omega_{ij}\leq C  \theta_i\zeta_j$, we immediately have 
\[
\sum_{i_1, i_2, j_1} \Omega_{i_1j_1}^2 \Omega_{i_2j_1}^2 \leq C \sum_{i_1, i_2, j_1} \theta_{i_1}^2\theta_{i_2}^2 \zeta_{j_1}^4 =  C \|\theta\|^4\|\zeta\|_4^4,
\]
and \[
 \sum_{i_1, i_2, j_1} \Omega_{i_1j_1}\Omega_{i_1i_2}\Omega_{i_2j_1}\Omega_{i_2i_2} \leq C  \sum_{i_1, i_2, j_1} \theta_{i_1}^2\theta_{i_2}^2\zeta_{j_1}^2\zeta_{i_2}^2 = C  \|\theta\|^2\|\zeta\|^2\sum_{i_2}\theta_{i_2}^2\zeta_{i_2}^2\leq C \|\theta\|^2\|\zeta\|^2\|\theta\|_4^2\|\zeta\|_4^2,
\]
where we've used Cauchy-Schwarz inequality in the last equality. 
Similarly, we have $\sum_{i_1, j_1, j_2} \Omega_{i_1j_1}^2 \Omega_{i_1j_2}^2\leq \|\zeta\|^4\|\theta\|_4^4$.
Combining above, we conclude \begin{align*}
\sum_{\substack{i_1, i_2, j_1, j_2\\ \text{non-distinct}}} \Omega_{i_1j_1}\Omega_{i_1j_2}\Omega_{i_2j_1}\Omega_{i_2j_2} & \leq  C[\|\theta\|^4\|\zeta\|_4^4 +  \|\zeta\|^4\|\theta\|_4^4 + 4\|\theta\|^2\|\zeta\|^2\|\theta\|_4^2\|\zeta\|_4^2 ] \\
& \leq  C[\|\theta\|^4\|\zeta\|_4^4 +  \|\zeta\|^4\|\theta\|_4^4 + 2[\|\theta\|^4\|\zeta\|_4^4 +  \|\zeta\|^4\|\theta\|_4^4]] \\
& = 3 C [\|\theta\|^4\|\zeta\|_4^4 +  \|\zeta\|^4\|\theta\|_4^4],
\end{align*} 
where we've used basic inequality that $2xy\leq x^2 + y^2$ in the second last line.  This completes the proof of \eqref{thm1-claim3-key}.

Now, to complete the proof of the  mean part of the second claim, it remains to show 
\beq\label{thm1-claim3-key2}
\text{tr}[(\Omega'\Omega)^2] \asymp\|\theta\|^4\|\zeta\|^4=\|\theta\|^8, 
\qquad \mbox{ and} \qquad\|\theta\|^4\|\zeta\|_4^4 +  \|\zeta\|^4\|\theta\|_4^4 \ll \|\theta\|^4\|\zeta\|^4.
\eeq
For the first part of \eqref{thm1-claim3-key2}, since $\Omega$ is non-negative and $\Omega_{ij}\leq C\theta_i\zeta_j$, we have 
\[
\text{tr}[(\Omega'\Omega)^2] \leq C^4 \tr[(\theta \zeta' \zeta \theta')^2] = C^4\|\zeta\|^4\tr[(\theta\theta')^2] = C^4\|\theta\|^4\|\zeta\|^4. 
\]
At the same time, 
\begin{align*}
\tr[(\Omega'\Omega)^2] & = \tr[(P(\Gamma'Z^2 \Gamma) P'(\Pi' \Theta^2\Pi))^2] \\
&\geq \tr[(CI_K(\Gamma'Z^2 \Gamma) CI_K(\Pi' \Theta^2\Pi))^2] \\
& \geq C\lambda_{\min}(\Gamma'Z^2 \Gamma) \tr[(\Pi' \Theta^2\Pi)(\Gamma'Z^2 \Gamma)(\Pi' \Theta^2\Pi) ]\\
& \geq C\lambda_{\min}^2(\Gamma'Z^2 \Gamma)\tr[(\Pi' \Theta^2\Pi)^2]\\
&\geq  C\lambda_{\min}^2(\Gamma'Z^2 \Gamma) \lambda_{\min}^2(\Pi' \Theta^2\Pi),
\end{align*} 
where we've used $\tr(XY)\geq\lambda_{\min}(X)\tr(Y)$ and $\tr(Y)\geq \lambda_{\min}(Y)$ for symmetric matrices  $X,Y$  with non-negative eigenvalues. By (\ref{D-cond-P}), 
\[
 \lambda_{\min}^2(\Gamma'Z^2 \Gamma) \lambda_{\min}^2(\Pi' \Theta^2\Pi)\geq C \|\theta\|^4\|\zeta\|^4.
\]
From here, we have shown $\text{tr}[(\Omega'\Omega)^2]\asymp \|\theta\|^8$ since $\|\theta\|=\|\zeta\|$, which completes the first part of \eqref{thm1-claim3-key2}. 
For the second part of  \eqref{thm1-claim3-key2}, notice by $\theta_{\max}\goto0$, we have $\|\theta\|^4_4 = \sum \theta_i^4 \leq \theta_{\max}^2\sum \theta_i^2 \ll \|\theta\|^2$, which further implies $ \|\zeta\|^4\|\theta\|_4^4 \ll \|\theta\|^4\|\zeta\|^4.$ Similarly, $ \|\theta\|^4\|\zeta\|_4^4
\ll \|\theta\|^4\|\zeta\|^4$, which completes the second half of \eqref{thm1-claim3-key2}.

Consider the  variance part of the second claim. We decompose $(C_n - \mathbb{E}[C_n])$ as the sum of the following terms
\[
X_1 = 4\sum_{i_1, i_2, j_1, j_2 (dist)}W_{i_1j_1}\Omega_{i_1j_2}\Omega_{i_2j_1}\Omega_{i_2j_2}, \qquad X_2 = 2\sum_{i_1, i_2, j_1, j_2 (dist)}W_{i_1j_1}W_{i_1j_2}\Omega_{i_2j_1}\Omega_{i_2j_2},
\]
\[
X_3 = 2\sum_{i_1, i_2, j_1, j_2 (dist)}W_{i_1j_1}\Omega_{i_1j_2}\Omega_{i_2j_1}W_{i_2j_2}, \qquad X_4 = 2\sum_{i_1, i_2, j_1, j_2 (dist)}W_{i_1j_1}\Omega_{i_1j_2}W_{i_2j_1}\Omega_{i_2j_2},
\]
\[
X_5 = 4 \sum_{i_1, i_2, j_1, j_2 (dist)}\Omega_{i_1j_1}W_{i_1j_2}W_{i_2j_1}W_{i_2j_2}, \qquad X_6 = \sum_{i_1, i_2, j_1, j_2 (dist)}W_{i_1j_1}W_{i_1j_2}W_{i_2j_1}W_{i_2j_2}. 
\]
Using the inequality that $\mathrm{Var}(\sum_{i=1}^6X_i)\leq  6 \sum_{i=1}^6\mathrm{Var}(X_i)$ for random variables $X_1, ..., X_6$, it suffices to upper bound $\mathrm{Var}(X_i)$ for $i = 1, ..., 6.$ 

Consider $X_1$. Recall that
\[
X_1 =4\sum_{i_1, i_2, j_1, j_2 (dist)}W_{i_1j_1}\Omega_{i_1j_2}\Omega_{i_2j_1}\Omega_{i_2j_2} =  4  \sum_{i_1 \neq  j_1}\Bigl( \sum_{i_2, j_2\notin\{i_1, j_1\} }\Omega_{i_1j_2}\Omega_{i_2j_1}\Omega_{i_2j_2} \Bigr) W_{i_1j_1}.
\]
It is easily seen that $\mathbb{E}[X_1]=0$. Furthermore, we have
\beq  \label{LemmaC4-D1}
\mathrm{Var}(X_1) = 16 \sum_{i_1 \neq  j_1}\Bigl( \sum_{i_2, j_2\notin\{i_1, j_1\} }\Omega_{i_1j_2}\Omega_{i_2j_1}\Omega_{i_2j_2} \Bigr)^2 \mathrm{Var}(W_{i_1j_1}). 
\eeq
By that $\Omega_{ij}\leq C \theta_i\zeta_j$ for $1\leq i, j\leq n$, 
\[
\Bigl|\sum_{i_2, j_2\notin\{i_1, j_1\} }    \Omega_{i_1j_2}\Omega_{i_2j_1}\Omega_{i_2j_2}  \Bigr| \leq C\sum_{i_2,j_2} \theta_{i_1}\theta_{i_2}^2\zeta_{j_1}\zeta_{j_2}^2
 \leq C \|\theta\|^2\|\zeta\|^2\cdot \theta_{i_1}\zeta_{j_1}.
\]
We plug it into \eqref{LemmaC4-D1} and use $\mathrm{Var}(W_{i_1 j_1})\leq \Omega_{i_1 j_1}\leq C\theta_{i_1}\zeta_{j_1}$. It yields that 
\begin{eqnarray} 
\mathrm{Var}(X_1) & \leq & C\sum_{i_1, j_1 (dist)}(\|\theta\|^2\|\zeta\|^2\theta_{i_1} \zeta_{j_1})^2\cdot \theta_{i_1}\zeta_{j_1} \leq   C \|\theta\|^4\|\zeta\|^4\|\theta\|_3^3\|\zeta\|_3^3. 
\end{eqnarray}
By basic algebra and the fact that $\|\theta\|=\|\zeta\|$, $2 \|\theta\|^4\|\zeta\|^4\|\theta\|_3^3\|\zeta\|_3^3   \leq \|\theta\|^8 \cdot [ \|\theta\|_3^6 + \|\zeta\|_3^6 ]$.  

Consider $X_2$. Recall that
\[
X_2 = 2\sum_{i_1, i_2, j_1, j_2 (dist)}W_{i_1j_1}W_{i_1j_2}\Omega_{i_2j_1}\Omega_{i_2j_2} =2\sum_{i_1, j_1,j_2 (dist)}\Bigl(\sum_{i_2\notin\{i_1,j_1 , j_2\}} \Omega_{i_2j_1}  \Omega_{i_2j_2} \Bigr)W_{i_1j_1}W_{i_1j_2}. 
\]
It is easy to see that $\mathbb{E}[X_2]=0$. We then study its variance. We note that for $W_{\ell k}W_{\ell i}$ and $W_{\ell' k'}W_{\ell' i'}$ to be correlated, we must have that $(k',\ell',i')=(k,\ell, i)$ or $(k',\ell',i')=(i,\ell, k)$. 
 Therefore, 
\begin{align*}
\mathrm{Var}(X_2) &\leq C\sum_{i_1, j_1,j_2 (dist)}\mathrm{Var}\Bigl[\Bigl(\sum_{i_2\notin\{i_1,j_1 , j_2\}} \Omega_{i_2j_1}  \Omega_{i_2j_2} \Bigr)W_{i_1j_1}W_{i_1j_2}\Bigr]\cr
&\leq  C\sum_{i_1, j_1,j_2 (dist)}\Bigl(\sum_{i_2\notin\{i_1,j_1 , j_2\}} \Omega_{i_2j_1}  \Omega_{i_2j_2} \Bigr)^2 \cdot\mathrm{Var}(W_{i_1j_1}W_{i_1j_2}). 
\end{align*}
Notice that 
\[
\sum_{i_2\notin\{i_1,j_1 , j_2\}} \Omega_{i_2j_1}  \Omega_{i_2j_2} 
\leq C\sum_{i_2}\theta_{i_2}^2\zeta_{j_1}\zeta_{j_2} \leq C \|\theta\|^2\cdot \zeta_{j_1}\zeta_{j_2}. 
\]
Combining the above gives
\begin{align*}
\mathrm{Var}(X_2)  & \leq C\sum_{i_1,j_1,j_2}(\|\theta\|^2\zeta_{j_1}\zeta_{j_2})^2\cdot \theta_{i_1}^2\zeta_{j_1}\zeta_{j_2}  \leq C \|\theta\|^6\|\zeta\|_3^6  \ll C\|\theta\|^8\|\zeta\|_3^6, 
\end{align*}
as   $\|\theta\|=\|\zeta\|\to \infty$. 

Next, we consider $X_3$. Recall that  
\[
X_3 = \sum_{i_1, i_2, j_1, j_2 (dist)}W_{i_1j_1}\Omega_{i_1j_2}\Omega_{i_2j_1}W_{i_2j_2}.
\]
It's not hard to see $\mathbb{E}[X_3] = 0$ and $W_{i_1j_1}W_{i_2j_2}$ are uncorrelated with $W_{i_1'j_1'}W_{i_2'j_2'}$ unless (i) $(i_1, j_1) = (i_1', j_1')$ and $(i_2, j_2) = (i_2', j_2')$, or (ii) $(i_1, j_1) = (i_2', j_2')$ and $(i_2, j_2) = (i_1', j_1')$. Therefore, 
\begin{align*}
\mathrm{Var}(X_3) &  \leq C\sum_{i_1, i_2, j_1, j_2 (dist)}\Omega_{i_1j_2}^2\Omega_{i_2j_1}^2\mathrm{Var}(W_{i_1j_1}W_{i_2j_2})\\
&  \leq C\sum_{i_1, i_2, j_1, j_2 (dist)}(\theta_{i_1}^2\theta_{i_2}^2\zeta_{j_1}^2\zeta_{j_2}^2)\cdot(\theta_{i_1}\zeta_{j_1}\theta_{i_2}\zeta_{j_2})\\
& \leq C \|\theta\|_3^6 \|\zeta\|_3^6 \ll C \|\theta\|^8\|\zeta\|_3^6.
\end{align*}
The last line is by $\|\theta\|^2\gg\|\theta\|_3^3$ and $\|\theta\|\to \infty$. 
 
Next, we consider $X_4$. 
We can mimick the analysis of $X_2$ and derive 
\[
\mathrm{Var}(X_4)\leq  C \|\zeta\|^6\|\theta\|_3^6  \ll C\|\theta\|^8 \|\theta\|_3^6.
\]

Next, we consider $X_5$. It's not hard to see $\mathbb{E}[X_5] = 0$.   Mimicking the previous arguments,  
\begin{align*}
\mathrm{Var}(X_5) & \leq \sum_{i_1, j_1, i_2, j_2}\Omega_{i_1j_1}^2 \mathrm{Var}(W_{i_1j_2}W_{i_2j_1}W_{i_2j_2}) \\
& \leq  \sum_{i_1, j_1, i_2, j_2}\Omega_{i_1j_1}^2\Omega_{i_1j_2}\Omega_{i_2j_1}\Omega_{i_2j_2}\\
& \leq \sum_{i_1, j_1, i_2, j_2}\theta_{i_1}^3\theta_{i_2}^2\zeta_{j_1}^3 \zeta_{j_2}^2.
\end{align*}
At the same time, $\sum_{i_1, j_1, i_2, j_2}\theta_{i_1}^3\theta_{i_2}^2\zeta_{j_1}^3\zeta_{j_2}^2 = \|\theta\|^2\|\theta\|_3^3\|\zeta\|^2\|\zeta\|_3^3.$ By basic algebra,  
\[
\|\theta\|^2\|\theta\|_3^3\|\zeta\|^2\|\zeta\|_3^3\ll \|\theta\|^4\|\zeta\|^4=\|\theta\|^8.
\]
Finally we consider $X_6$. Mimicking previous argument and it follows from direct calculation that 
\[
\mathrm{Var}(X_6) \leq C\sum_{i_1, i_2, i_3, i_4 (dist)} \Omega_{i_1j_1} \Omega_{i_1j_2}\Omega_{i_2j_1}\Omega_{i_2j_2} \leq C\sum_{i_1, i_2, i_3, i_4 (dist)}\theta_{i_1}^2 \theta^2_{i_2}\zeta_{j_1}^2\zeta_{j_2}^2 \leq C\|\theta\|^4\|\zeta\|^4=C\|\theta\|^8.
\]
Combining above, we obtain  $\mathrm{Var}(C_n) \leq C \|\theta\|^8\cdot [1+\|\zeta\|_3^6 + \|\theta\|_3^6]$.  This completes the proof of the variance part (i.e.,  the second part) in the second claim. 
 
We now consider the  third part in the second claim.
By Chebyshev's inequality, for any $\eps>0$, \beq\label{tmp:cheb}
\mathbb{P}\bigg(\bigg|\frac{{C}_n}{\mathbb{E}[C_n]}  - 1\bigg|\geq \eps\bigg) \leq \frac{1}{\eps^2}\mathbb{E}\bigg(\frac{{C}_n}{\mathbb{E}[C_n]}  - 1\bigg)^2 = \frac{1}{\eps^2\mathbb{E}[C_n]^2}\mathbb{E}\big({C}_n - \mathbb{E}[C_n]\big)^2.
\eeq
Here, by  the mean and variance part in the second claim, we have  $\mathbb{E}[C_n]\asymp\|\theta\|^8$ and    \[
\mathbb{E}\big({C}_n - \mathbb{E}[C_n]\big)^2\leq C\|\theta\|^8 \cdot \left[1+ \|\zeta\|^4\|\theta\|_3^6 + \|\theta\|^4\|\zeta\|_3^6  \right].
\]
Therefore,  the rightmost term of \eqref{tmp:cheb} is no greater than
\[C\times \frac{\|\theta\|^8 \cdot \left[1+ \|\theta\|_3^6 + \|\zeta\|_3^6  \right]}{\|\theta\|^{16}}\ll C\times \frac{1+\|\theta\|^4+\|\zeta\|^4}{\|\theta\|^8}\to 0,
\]
where the last two steps follows from that $\|\theta\|^3_3\leq\theta_{\max}\|\theta\|^2$,  $\|\zeta\|^3_3\leq\zeta_{\max}\|\zeta\|^2$ and $\|\theta\|\to \infty$. 
This proves $C_n/\mathbb{E}[C_n]\stackrel{p}{\goto}1$. Similarly, we obtain  $\widetilde{C}_n/\mathbb{E}[\widetilde{C}_n]\stackrel{p}{\goto}1$. Once we can show the normality of $\frac{Q_n}{SD(Q_n)}$, then combined with Slutsky's theorem, we get $\psi_n\goto N(0,1/2)$ in law. 

It now remains to prove the  the normality of $\frac{Q_n}{SD(Q_n)}$. Recall \eqref{def:J4} and we notice that under the null 
\[
O = A - \widetilde A = W - \widetilde W.
\]
We know
\[
\frac{Q_n}{SD(Q_n)}=
 \frac{4\sum_{J_4(n)}O_{i_1j_1}O_{i_1j_2}O_{i_2j_1}O_{i_2j_2}}{\sqrt{256\sum_{J_4(n)} \Omega_{i_1j_1}^*\Omega_{i_1j_2}^*\Omega_{i_2j_1}^*\Omega_{i_2j_2}^* }}= \frac{\sum_{J_4(n)}O_{i_1j_1}O_{i_1j_2}O_{i_2j_1}O_{i_2j_2}}{4\sqrt{\sum_{J_4(n)} \Omega_{i_1j_1}^*\Omega_{i_1j_2}^*\Omega_{i_2j_1}^*\Omega_{i_2j_2}^*}}.
\] 
For $1\leq m\leq n$, define 
\[
S_{n,m}= \frac{\sum_{J_4(m)} O_{i_1j_1}O_{i_1j_2}O_{i_2j_1}O_{i_2j_2}}{4\sqrt{\sum_{J_4(n)}\Omega_{i_1j_1}^*\Omega_{i_1j_2}^*\Omega_{i_2j_1}^*\Omega_{i_2j_2}^* }} 
\]
and the $\sigma$-algebra $\mathcal{F}_{n,m} = \sigma(\{O_{ij}\}_{1\leq i, j\leq m})$. It is seen that $\mathbb{E}[S_{n,m}|\mathcal{F}_{n,m-1}]=S_{n,m-1}$. Letting $X_{n,m}=S_{n,m}-S_{n,m-1}$, we conclude that $\{X_{n,m}\}_{m=1}^n$ is a martingale difference sequence relative to the filtration $\big\{\mathcal{F}_{n,m}\big\}_{m=1}^n$.  The normality claim follows from the central limit theorem in \cite{hall2014martingale}, we thus only need to check its requirements:
\begin{enumerate}
\item[(a)]  $\sum_{m=1}^n \mathbb{E}(X_{n,m}^2|\mathcal{F}_{n,m-1})\overset{p}{\to}1$.  
\item[(b)] $\sum_{m=1}^n \mathbb{E}(X_{n,m}^21_{\{|X_{n,m}|>\epsilon\}}|\mathcal{F}_{n,m-1})\overset{p}{\to}0$, for any $\epsilon>0.$
\end{enumerate}
Note here that our analysis is related to literature on U-statistics (e.g., \cite{lee2019u}).  However, almost all existing works on $U$-statistics  assume that the variables are {\it identically distributed}.  In our problem, $O_{ij}$'s are not identically distributed and {\it the variances of different $O_{ij}$'s can be at different magnitudes} (which makes the problem even more challenging). For this reason, we use the results in \cite{hall2014martingale} for our proofs instead of the existing results on U-statistics.  

To check (a)-(b), we give an alternative expression of $X_{n,m}$. 
Write for short \[
M_n=16\sum_{J_4(n)}\Omega_{i_1j_1}^*\Omega_{i_1j_2}^*\Omega_{i_2j_1}^*\Omega_{i_2j_2}^*.
\] 
Introduce \beq\label{thm-Y}
\alpha_{(m-1)j_1j_2} = \sum_{1\leq i_1 \leq m - 1} O_{i_1j_1}O_{i_1j_2}, \quad \mbox{ and }\quad \beta_{(m-1)i_1i_2} = \sum_{1\leq j_1 \leq m - 1} O_{i_1j_1}O_{i_2j_1},
\eeq
we rewrite $X_{n,m}$ as
\beq \label{thm-X}
X_{n,m} = \frac{1}{\sqrt{M_n}}\bigg[ \sum_{ 1\leq j_1 < j_2\leq m} \alpha_{(m-1)j_1j_2}O_{mj_1}O_{mj_2}  + \sum_{1\leq i_1< i_2 \leq m   }  \beta_{(m-1)i_1i_2}O_{i_1m}O_{i_2m}   \bigg].
\eeq
Conditioning on $\mathcal{F}_{n,m-1}$, $O_{mj_1}O_{mj_2}$ and $O_{i_1m}O_{i_2m}$ are mutually uncorrelated and $\alpha_{(m-1)j_1j_2}, \beta_{(m-1)i_1i_2}$ are constants for $1\leq i_1, i_2, j_1, j_2 \leq m - 1$. Hence, $\mathbb{E}[X_{n,m}] = 0$ and 
\begin{align*}  
\mathbb{E}(X^2_{n,m}|\mathcal{F}_{n,m-1}) & = \frac{1}{M_n} \bigg[\sum_{j_1 < j_2< m}\alpha_{(m-1)j_1j_2}^2
\mathrm{Var}(O_{mj_1}O_{mj_2}) + \sum_{i_1 < i_2< m} \beta_{(m-1)i_1i_2}^2  \mathrm{Var}(O_{i_1m}O_{i_2m})   \bigg]  \\
& = \frac{4}{M_n} \bigg[\sum_{j_1 < j_2< m}\alpha_{(m-1)j_1j_2}^2
\Omega^*_{mj_1}\Omega^*_{mj_2} + \sum_{i_1 < i_2 < m} \beta_{(m-1)i_1i_2}^2  \Omega^*_{i_1m}\Omega^*_{i_2m}   \bigg].
\end{align*}

We now check (a). 
In the definition \eqref{thm-Y}, the terms in the sum are (unconditionally) mutually uncorrelated. As a result, $\mathbb{E}[\alpha^2_{(m-1)j_1j_2}]=4\sum_{i_1<m, i_1 \notin\{j_1,j_2\}}\Omega^*_{i_1j_1}\Omega^*_{i_1j_2}$ and $\mathbb{E}[\beta^2_{(m-1)i_1i_2}] = 4\sum_{j_1<m, j_1 \notin\{i_1,i_2\}}\Omega^*_{i_1j_1}\Omega^*_{i_2j_1}$. It follows that 
\begin{align*}  
 \mathbb{E}\Bigl[ \sum_{m=1}^n \mathbb{E}(X^2_{n,m}|\mathcal{F}_{n,m-1})  \Bigr]  
 =& 
 \frac{16}{M_n} \sum_{m=1}^n\bigg[ \sum_{j_1 < j_2< m}\sum_{i_1<m, i_1 \notin\{j_1,j_2\}}\Omega^*_{i_1j_1}\Omega^*_{i_1j_2}
\Omega^*_{mj_1}\Omega^*_{mj_2} \\
& 
\qquad +
\sum_{i_1 < i_2< m}\sum_{j_1<m, j_1 \notin\{i_1,i_2\}}\Omega^*_{i_1j_1}\Omega^*_{i_2j_1}\Omega^*_{i_1m}\Omega^*_{i_2m} 
 \bigg]. 
\end{align*}
which indicates
\beq\label{thm-check(a)-mean}
 \mathbb{E}\Bigl[ \sum_{m=1}^n \mathbb{E}(X^2_{n,m}|\mathcal{F}_{n,m-1})  \Bigr]  = \frac{16}{M_n}\sum_{J_4(n)}\Omega_{i_1j_1}^*\Omega_{i_1j_2}^*\Omega_{i_2j_1}^*\Omega_{i_2j_2}^* = 1.
\eeq
We then study the variance of $\sum_{m=1}^n \mathbb{E}(X_{n,m}^2|\mathcal{F}_{n,m-1})$. By \eqref{thm-Y}, $\alpha^2_{(m-1)j_1j_2}=\sum_{i_1} O^2_{i_1j_1}O^2_{i_1j_2} + \sum_{i_1\neq i_1'}O_{i_1j_1}O_{i_1j_2}O_{i_1'j_1}O_{i_1'j_2}$. Similarly,
$\beta^2_{(m-1)i_1i_2} = \sum_{j_1} O^2_{i_1j_1}O^2_{i_2j_1} + \sum_{j_1\neq j_1'}O_{i_1j_1}O_{i_2j_1}O_{i_1j_1'}O_{i_2j_1'}$
We then have a decomposition
\beq \label{thm-X2-decompose}
\sum_{m=1}^n \mathbb{E}(X_{n,m}^2|\mathcal{F}_{n,m-1}) = I_a + I_b + I_c + I_d, 
\eeq
where
\begin{align*}
I_a &=  \frac{4}{M_n}\sum_{m=1}^n\sum_{j_1 < j_2< m}\sum_{i_1<m, i_1 \notin\{j_1,j_2\}}
 O^2_{i_1j_1}O^2_{i_1j_2} \Omega^*_{mj_1}\Omega^*_{mj_2}.
\cr
I_b & = \frac{4}{M_n}\sum_{m=1}^n\sum_{j_1<j_2 <m}\sum_{\substack{i_1\neq i_1', i_1,i_1' < m\\ i_1', i_1' \notin \{j_1,j_2\}}}O_{i_1j_1}O_{i_1j_2}O_{i_1'j_1}O_{i_1'j_2}\Omega^*_{mj_1}\Omega^*_{mj_2}. \\
I_c & = \frac{4}{M_n}\sum_{m=1}^n\sum_{i_1 < i_2< m}\sum_{j_1<m, j_1 \notin\{i_1,i_2\}}
 O^2_{i_1j_1}O^2_{i_2j_1} \Omega^*_{i_1m}\Omega^*_{i_2m}. \\
I_d & = \frac{4}{M_n}\sum_{m=1}^n\sum_{i_1<i_2 <m}\sum_{\substack{j_1\neq j_1', j_1,j_1' < m\\ j_1',j_1' \notin \{i_1,i_2\}}}O_{i_1j_1}O_{i_2j_1}O_{i_1j_1'}O_{i_2j_1'} \Omega^*_{i_1m}\Omega^*_{i_2m}. 
\end{align*}
By $\mathrm{Var}(I_a + I_b + I_c + I_d) \leq 4 [\mathrm{Var}(I_a) + ... + \mathrm{Var}(I_d)]$, we only need to bound $\mathrm{Var}(I_a), ..., \mathrm{Var}(I_d)$ separately.

For $I_a$, we  rewrite
\begin{align*}
I_a & =  \frac{4}{M_n}\sum_{i_1=1}^n\sum_{1\leq j_1 < j_2\leq n} 
 O^2_{i_1j_1}O^2_{i_1j_2} \sum_{m > \max\{i_1, j_1, j_2\}}\Omega^*_{mj_1}\Omega^*_{mj_2} \\
 & \equiv  \frac{4}{M_n}\sum_{i_1=1}^n\sum_{1\leq j_1 < j_2\leq n} 
 O^2_{i_1j_1}O^2_{i_1j_2} \cdot b_{i_1j_1j_2}.
\end{align*}
The terms correspond to different $i_1$ are independent of each other. We now fix $i_1$ and calculate the covariance between $O^2_{i_1j_1} O^2_{i_1j_2}$ and $O^2_{i_1j_1'}  O^2_{i_1j_2'}$. There are three cases. Case (i): $(j_1,j_2)=(j_1',j_2')$. In this case, 
\[
\mathrm{Var}(O^2_{i_1j_1}O^2_{i_1j_2})\leq \mathbb{E}[O^4_{i_1j_1}O^4_{i_1j_2}]\leq \mathbb{E}[O^2_{i_1j_1}O^2_{i_1j_2}] = 4\Omega^*_{i_1j_1}\Omega^*_{i_1j_2}.
\] 
Case (ii): $j_1=j_1'$ but $j_2\neq j_2'$. In this case, we have
\[
\mathrm{Cov}(O^2_{i_1j_1}O^2_{i_1j_2}, O^2_{i_1j_1}O^2_{i_1j_2'})  = \text{Var}(O_{i_1j_1}^2)\mathbb{E}(O_{i_1j_2}^2)\mathbb{E}(O_{i_1j_2'}^2) \leq 8  \Omega^*_{i_1j_1}\Omega^*_{i_1j_2}\Omega^*_{i_1j_2'}.
\] 
Case (iii): $(j_1, j_2)\cap (j_1',j_2')=\emptyset$. The two terms are independent, and their covariance is zero. Combining the above gives
\begin{align*}
\mathrm{Var}&(I_a) \leq \frac{16 C}{M_n^2}\sum_{i_1=1}^n\biggl( \sum_{1\leq j_1 < j_2 \leq n}b^2_{i_1j_1j_2}\Omega^*_{i_1j_1}\Omega^*_{i_1j_2}
 + \sum_{j_1, j_2, j_2'} b_{i_1j_1j_2}b_{i_1j_1j_2'}\Omega^*_{i_1j_1}\Omega^*_{i_1j_2}\Omega^*_{i_1j_2'} \biggr)
\end{align*}
We now bound the right hand side. Recall that $\Omega^*_{ij}\leq C\theta_i\zeta_j$, we have $b_{i_1j_1j_2} \leq C \sum_{m}\theta_m^2 \zeta_{j_1}\zeta_{j_2}\leq C\|\theta\|^2\zeta_{j_1}\zeta_{j_2}$. 
As a result,
\begin{align*}
\mathrm{Var}(I_a) &\leq \frac{C}{M^2_n}\Bigl[\sum_{i_1, j_1, j_2} \|\theta\|^4 \theta_{i_1}^2 \zeta_{j_1}^3\zeta_{j_2}^3
+ \sum_{i_1,j_1, j_2,j_2'}
\|\theta\|^4\theta_{i_1}^3\zeta^3_{j_1}\zeta^2_{j_2}\zeta^2_{j_2'}\Bigr] \cr
 &\leq \frac{C}{M^2_n}(\|\theta\|^6\|\zeta\|_3^6 + \|\theta\|^4\|\zeta\|^4\|\zeta\|_3^3\|\theta\|_3^3 ). 
\end{align*}
Notice $\Omega^*_{ij}=\Omega_{ij}(1-\Omega_{ij})\geq c\Omega_{ij}$ (in our setting, all $\Omega_{ij}$'s are bounded away from $1$). As a result, we have \[
M_n\gtrsim \sum_{i_1, i_2, j_1, j_2}\Omega_{i_1j_1}\Omega_{i_1j_2}\Omega_{i_2j_1}\Omega_{i_2j_2}\gtrsim \mathbb{E}[C_n].
\]
By the  second claim, $\mathbb{E}[C_n]\asymp \|\theta\|^4\|\zeta\|^4$. Combining the above gives  
\beq \label{thm-check(a)-var1}
\mathrm{Var}(I_a) \lesssim  \frac{\|\theta\|^6\|\zeta\|_3^6 + \|\theta\|^4\|\zeta\|^4\|\zeta\|_3^3\|\theta\|_3^3 }{\|\theta\|^8\|\zeta\|^8} = \frac{\|\zeta\|_3^6}{\|\theta\|^2\|\zeta\|^8} + \frac{\|\theta\|_3^3\|\zeta\|_3^3}{\|\theta\|^4\|\zeta\|^4} = o(1),
\eeq
where we've used $\|\theta\|\goto\infty, \|\zeta\|\goto\infty$, $\|\theta\|_3^3\ll \|\theta\|^2$ and $\|\zeta\|^3_3\leq \|\zeta\|^2.$

For $I_b$, we can write
\[
I_b =  \frac{2}{M_n}\sum_{i_1,i_1',j_1, j_2 (dist)}c_{i_1j_1j_2j_2'}Y_{i_1j_1j_2j_2'}, \qquad Y_{i_1i_1'j_1j_2} \equiv O_{i_1j_1}O_{i_1j_2}O_{i_1'j_1}O_{i_1'j_2}, 
\]
where $c_{i_1j_1j_2j_2'}=\sum_{m>\max\{i_1j_1j_2j_2'\}}\Omega^*_{mj_1}\Omega^*_{mj_2}$. Similar to the bound for $b_{i_1j_1j_2}$, we can obtain $c_{i_1j_1j_2j_2'}\leq C\|\theta\|^2\zeta_{j_1}\zeta_{j_2}$. 
Note that $I_b$ has a mean zero, it follows that $\mathrm{Var}(I_b)=\mathbb{E}[I_b^2]$. For $(i_1,i_1',j_1, j_2)$ and $(\tilde i_1,\tilde i_1',\tilde j_1, \tilde j_2)$,  $\mathbb{E}[Y_{i_1i_1'j_1 j_2}Y_{\tilde i_1\tilde i_1'\tilde j_1 \tilde j_2}]\neq 0$ if and only if the underlying directed quadrilateral are the same.  Hence,
\begin{align*}
\mathrm{Var}(I_b) &\leq  \frac{C}{M_n^2} \sum_{ i_1,i_1',j_1, j_2} c_{i_1j_1j_2j_2'}^2 \mathbb{E}[Y^2_{i_1j_1j_2j_2'}]
\cr
&\lesssim\frac{1}{\|\theta\|^{8}\|\zeta\|^8}\sum_{ i_1,i_1',j_1, j_2} (\|\theta\|^4 \zeta^2_{j_1}\zeta_{j_2}^2)\Omega^*_{i_1j_1}\Omega^*_{i_1j_2}\Omega^*_{i_1'j_1}\Omega^*_{i_1'j_2}   \cr
&\lesssim \frac{1}{\|\theta\|^{8}\|\zeta\|^8}\sum_{ i_1,i_1',j_1, j_2} \|\theta\|^4 \theta^2_{i_1}\theta^2_{i_2} \zeta^4_{j_1}\zeta^4_{j_2} \cr
&\leq O(\|\zeta\|_4^8/\|\zeta\|^8).
\end{align*}
As a result,
\beq \label{thm-check(a)-var2}
\sqrt{\mathrm{Var}(I_b)} \leq \frac{C\sum_i\zeta_i^4}{(\sum_i\zeta_i^2)^2}\leq \frac{C\zeta^2_{\max}}{\sum_i\zeta_i^2}= o(1). 
\eeq
The analysis of $\mathrm{Var}(I_c)$ is similar to that of $\mathrm{Var}(I_a)$, and the analysis of $\mathrm{Var}(I_d)$ is similar to that of $\mathrm{Var}(I_b)$. We omit the detail and conclude that 
\beq \label{thm-check(a)-var3}
\mathrm{Var}(I_c) = o(1), \qquad \mbox{ and }\qquad \mathrm{Var}(I_d) = o(1).
\eeq
We plug \eqref{thm-check(a)-var1}, \eqref{thm-check(a)-var2} and \eqref{thm-check(a)-var3} into \eqref{thm-X2-decompose} and find that the variance of $\sum_{m=1}^n \mathbb{E}(X_{n,m}^2|\mathcal{F}_{n,m-1})$ is $o(1)$. Combining it with \eqref{thm-check(a)-mean}, we conclude that this random variable converges to $1$ in probability. This gives (a). 

We now check (b). By the Cauchy-Schwarz inequality and the Chebyshev's inequality, 
\begin{align*}
&\sum_{m=1}^n \mathbb{E}(X_{n,m}^21_{\{|X_{n,m}|>\epsilon\}}|\mathcal{F}_{n,m-1}) \\
\leq & \sum_{m=1}^n \sqrt{\mathbb{E}(X_{n,m}^4|\mathcal{F}_{n,m-1})}\sqrt{\mathbb{P}\big(|X_{n,m}\big|\geq \epsilon|\mathcal{F}_{n,m-1}\big)}\\
\leq& \epsilon^{-2}\sum_{m=1}^n \mathbb{E}(X_{n,m}^4|\mathcal{F}_{n,m-1}).
\end{align*}
Therefore, it suffices to show that the right hand side converges in probability to $0$. Note that the right hand is a nonnegative random variable. We only need to prove that its mean is vanishing, i.e., 
\beq \label{thm-check(b)}
\mathbb{E}\bigg[\sum_{m=1}^n\mathbb{E}\big[ X_{n,m}^4\big|\mathcal{F}_{n,m-1}\big]\bigg] = o(1). 
\eeq
We use the expression of $X_{n,m}$ in \eqref{thm-X}. Conditioning on $\mathcal{F}_{n,m-1}$, the $\alpha_{(m-1)j_1j_2}$'s and $\beta_{(m-1)i_1i_2}$'s are non-stochastic. 
By basic inequality that $(x+y)^4\leq 8(x^4 + y^4)$, we can bound $\mathbb{E}[X^4_{n,m}|\mathcal{F}_{n,m-1}] $ by the following terms
\[
\frac{8}{M_n^2}\mathbb{E}\bigg[\bigg(\sum_{ 1\leq j_1 < j_2\leq m} \alpha_{(m-1)j_1j_2}O_{mj_1}O_{mj_2}\bigg)^4 + \bigg( \sum_{1\leq i_1< i_2 \leq m   }  \beta_{(m-1)i_1i_2}O_{i_1m}O_{i_2m} \bigg)^4\bigg|\mathcal{F}_{n,m-1}\bigg].
\]
It suffices to control above two terms. Since the analysis is similar, we only provide the proof for the first one. 
It follows that 
\begin{align*}
&\frac{8}{M_n^2}\sum_{m=1}^n\mathbb{E}\bigg[\bigg(\sum_{ 1\leq j_1 < j_2\leq m} \alpha_{(m-1)j_1j_2}O_{mj_1}O_{mj_2}\bigg)^4\bigg|\mathcal{F}_{n,m-1}\bigg] \\
= & 
\frac{8}{M_n^2}\bigg\{\sum_{m=1}^n\sum_{j_1 < j_2 \leq m}\alpha^4_{(m-1)j_1j_2}\mathbb{E}[O^4_{mj_1}O^4_{mj_2}] + C  \sum_{j_1 < m \leq n}\sum^{m-1}_{\substack{j_2\neq j_2' < m\\ j_2,j_2'\notin\{j_1\}}}\alpha^2_{(m-1)j_1j_2}\alpha^2_{(m-1)j_1j_2'}\mathbb{E}[O^4_{mj_1}O^2_{mj_2}O^2_{mj_2'}]\cr
& + C  \sum_{m=1}^n\sum_{\substack{j_1, j_2, j_1', j_2 < m \\ distinct} }\alpha^2_{(m-1)j_1j_2}\alpha^2_{(m-1)j_1'j_2'}\mathbb{E}[O^2_{mj_1}O^2_{mj_2}O^2_{mj_1'}O^2_{mj_2'}]\bigg\}. 
\end{align*}
We shall use the independence across entries of $O$ and the fact that $\mathbb{E}[O^4_{ij}]\leq \mathbb{E}[O^2_{ij}]\leq \Omega_{ij}\leq C\theta_i\zeta_j$. 
Next, we claim that 
\beq \label{alpha_bound_1}
\mathbb{E}[\alpha_{(m-1)j_1j_2}^4] \lesssim  \|\theta\|^2\zeta_{j_1}\zeta_{j_2} +  \|\theta\|^4 \zeta_{j_1}^2\zeta_{j_2}^2, \quad
\mathbb{E}[\alpha_{(m-1)j_1j_2}^2\alpha_{(m-1)j_1j_2'}^2] \lesssim  \|\theta\|_3^3\zeta_{j_1}\zeta_{j_2}\zeta_{j_2'} + \|\theta\|^4 \zeta_{j_1}^2\zeta_{j_2}\zeta_{j_2'},
\eeq
\beq\label{alpha_bound_2}
\quad \mathbb{E}[\alpha_{(m-1)j_1j_2}^2\alpha_{(m-1)j_1'j_2'}^2] \lesssim \|\theta\|^4 \zeta_{j_1}\zeta_{j_1'}\zeta_{j_2}\zeta_{j_2'}. 
\eeq
The proofs are similar, so we only show the first claim of  \eqref{alpha_bound_1}. By direct calculation, 
\beq \label{alpha_bound_1_tmp1}
\mathbb{E}[\alpha_{(m-1)j_1j_2}^4] = \sum_{i_1 < m}\mathbb{E}[ O_{i_1j_1}^4O_{i_1j_2}^4] + C  \sum_{i_1 \neq i_1' < m} \mathbb{E}[ O_{i_1j_1}^2O_{i_1j_2}^2 O_{i_1'j_1}^2O_{i_1'j_2}^2].
\eeq
By independence and that $j_1, j_2$ are distinct, $\mathbb{E}[ O_{i_1j_1}^4O_{i_1j_2}^4] \leq \mathbb{E}[ O_{i_1j_1}^4]\mathbb{E}[O_{i_1j_2}^4] \leq C\theta_{i_1}^2\zeta_{j_1}\zeta_{j_2}.$ Similarly, 
$ \mathbb{E}[ O_{i_1j_1}^2O_{i_1j_2}^2 O_{i_1'j_1}^2O_{i_1'j_2}^2]\leq C\theta_{i_1}^2\theta_{i_1'}^2\zeta_{j_1}^2\zeta_{j_2}^2.$ Plug into \eqref{alpha_bound_1_tmp1}, we obtain 
\[
\mathbb{E}[\alpha_{(m-1)j_1j_2}^4] \lesssim \|\theta\|^2\zeta_{j_1}\zeta_{j_2} + \|\theta\|^4\zeta_{j_1}^2\zeta_{j_2}^2.
\]
Last, in proving \eqref{thm-check(a)-var1}, we have seen that $M_n\gtrsim \|\theta\|^4\|\zeta\|^4$. Combining the above, we find that
\begin{align*}
& \mathbb{E}\bigg[ \frac{8}{M_n^2}\sum_{m=1}^n\mathbb{E}\bigg[\bigg(\sum_{ 1\leq j_1 < j_2\leq m} \alpha_{(m-1)j_1j_2}O_{mj_1}O_{mj_2}\bigg)^4\bigg|\mathcal{F}_{n,m-1}\bigg]\bigg]  \\
\lesssim &\frac{1}{M_n^2}\bigg\{ \sum_{m=1}^n\sum_{j_1\neq j_2 \leq m}(\|\theta\|^2\zeta_{j_1}\zeta_{j_2} +  \|\theta\|^4 \zeta_{j_1}^2\zeta_{j_2}^2)\Omega^*_{mj_1}\Omega^*_{mj_2} \\
&+ C  \sum_{j_1<m\leq n}\sum^{m-1}_{\substack{j_2\neq j_2' < m\\ j_2,j_2'\notin\{j_1\}}}( \|\theta\|_3^3\zeta_{j_1}\zeta_{j_2}\zeta_{j_2'} +\|\theta\|^4 \zeta_{j_1}^2\zeta_{j_2}\zeta_{j_2'})\Omega^*_{mj_1}\Omega^*_{mj_2}\Omega^*_{mj_2'}\cr
& + C\sum_{m=1}^n  \sum_{\substack{j_1, j_2, j_1', j_2 < m \\ distinct} } \|\theta\|^4 \zeta_{j_1}\zeta_{j_1'}\zeta_{j_2}\zeta_{j_2'}\Omega^*_{mj_1}\Omega^*_{mj_2}\Omega^*_{mj_1'}\Omega^*_{mj_2'}\bigg\} \\
& \leq \frac{C}{\|\theta\|^8\|\zeta\|^8}\bigg[\big(\|\theta\|^4\|\zeta\|^4 + \|\theta\|^6 \|\zeta\|_3^6\big) + \big(\|\theta\|_3^6\|\zeta\|^6 + \|\theta\|^4\|\theta\|_3^3\|\zeta\|^4\|\zeta\|_3^3\big) + \|\theta\|^4\|\theta\|_4^4\|\zeta\|^8\bigg],
\end{align*}
where the last line $\goto0$ as $\|\theta\|\goto\infty$, $\|\zeta\|\goto\infty$ and $\|\theta\|_3^3 + \|\theta\|_4^4 \ll \|\theta\|^2$ following arguments before.
As a result,
\[
\mathbb{E}\bigg[ \frac{1}{M_n^2}\sum_{m=1}^n\mathbb{E}\bigg[\bigg(\sum_{ 1\leq j_1 < j_2\leq m} \alpha_{(m-1)j_1j_2}O_{mj_1}O_{mj_2}\bigg)^4\bigg|\mathcal{F}_{n,m-1}\bigg]\bigg] =o(1). 
\]
Following a similar argument, we can show 
\[
\mathbb{E}\bigg[ 
\frac{1}{M_n^2}\sum_{m=1}^n \mathbb{E}\bigg[ \bigg( \sum_{1\leq i_1< i_2 \leq m   }  \beta_{(m-1)i_1i_2}O_{i_1m}O_{i_2m} \bigg)^4\bigg|\mathcal{F}_{n,m-1}\bigg]\bigg]=o(1). 
\]
Combining above gives \eqref{thm-check(b)}. We have proved (b). \qed


\subsection{Proof of Theorem \ref{thm:alt-d}}

Introduce \beq\label{def:barOmega}
\bar\Omega_{ij} = \Omega_{ij} + \widetilde\Omega_{ij}.
\eeq
It's not hard to see $\mathrm{Var}(O_{ij}) \leq \bar\Omega_{ij} $ for $1\leq i, j \leq n.$
Introduce vectors $\bar\theta, \bar\zeta\in\mathbb{R}^n$ such that $\bar\theta_{i} = \theta_i + \tilde\theta_i$ and $\bar\zeta_{i} = \zeta_i + \tilde\zeta_i$ for $1\leq i\leq n.$ 
 By (\ref{D-cond-theta}), we have $\|\bar\theta\|\to \infty$ and $\|\bar\zeta\|\to \infty$. 
By $\Omega_{ij} \leq C\theta_i\zeta_j$ and $\widetilde\Omega_{ij}\leq C\tilde\theta_i\tilde\zeta_j$ for $1\leq i, j\leq n$, we obtain \beq\label{bound-delta}
|\Delta_{ij}| = |\Omega_{ij} - \widetilde\Omega_{ij}|\leq \bar\Omega_{ij}  \leq C\bar\theta_i\bar\zeta_j.
\eeq
Notice the above bound is rather crude, but it is enough when requiring $\|\theta\|=\|\zeta\|$ and $\|\tilde\theta\|=\|\tilde\zeta\|$. 

Consider the mean part of the first claim. We have 
\begin{align*}
\mathbb{E}[Q_n] & = \sum_{i_1, i_2, j_1, j_2 (dist)} \mathbb{E}(A_{i_1j_1} - \widetilde A_{i_1j_1})(A_{i_1j_2} - \widetilde A_{i_1j_2})(A_{i_2j_1} - \widetilde A_{i_2j_1})(A_{i_2j_2} - \widetilde A_{i_2j_2})
\\
& =  \sum_{i_1, i_2, j_1, j_2 (dist)} \mathbb{E}[A_{i_1j_1} - \widetilde A_{i_1j_1}] \mathbb{E}[A_{i_1j_2} - \widetilde A_{i_1j_2}] \mathbb{E}[A_{i_2j_1} - \widetilde A_{i_2j_1}] \mathbb{E}[A_{i_2j_2} - \widetilde A_{i_2j_2}]. 
\end{align*}
Together with  $\mathbb{E}[A_{ij}] = \Omega_{ij}$, $\mathbb{E}[\widetilde A_{ij}] = \widetilde\Omega_{ij}$  and $\Delta_{ij} = \Omega_{ij} - \widetilde\Omega_{ij}$, we obtain \[
\mathbb{E}[Q_n] = \sum_{i_1, i_2, j_1, j_2(dist)} \Delta_{i_1j_1}\Delta_{i_1j_2}\Delta_{i_2j_1}\Delta_{i_2j_2} = \tr(\Delta'\Delta)^2 - \sum_{\substack{ i_1, i_2, j_1, j_2 \\ \text{non-distinct}}} \Delta_{i_1j_1}\Delta_{i_1j_2}\Delta_{i_2j_1}\Delta_{i_2j_2}. 
\]
Note that $\|\bar\theta\|^2\asymp\|\theta\|^2+\|\tilde\theta\|^2=\|\zeta\|^2+\|\tilde\zeta\|^2\asymp \|\bar\zeta\|^2$.   It then remains to show 
\beq\label{mean-ndist}
\sum_{\substack{ i_1, i_2, j_1, j_2 \\ \text{non-distinct}}} \Delta_{i_1j_1}\Delta_{i_1j_2}\Delta_{i_2j_1}\Delta_{i_2j_2} = o\big( (\|\bar\theta\|^2 + \|\bar\zeta\|^2)\cdot\tr(\Delta'\Delta)\big).
\eeq
That $i_1, i_2, j_1, j_2$ are non-distinct implies that there are a  pair of identical indices. Therefore, 
\[
\begin{split}
   &\Big|\sum_{\substack{i_1, i_2, j_1, j_2 \\ \text{non-distinct}} } \Delta_{i_1j_1}\Delta_{i_1j_2}\Delta_{i_2j_1}\Delta_{i_2j_2}\Big|\\
   &\leq 4 \sum_{i_1,i_2,j_1} |\Delta_{i_1j_1}\Delta_{i_1i_2}\Delta_{i_2j_1}\Delta_{i_2i_2}|+\sum_{i_1,j_1,j_2} |\Delta_{i_1 j_1}^2\Delta_{i_1 j_2}^2| +\sum_{i_1,i_2,j_1} |\Delta_{i_1 j_1}^2\Delta_{i_2 j_1}^2|\\
   &\leq \sum_{i_1,i_2,j_1} (\Delta_{i_1 j_1}^2\Delta_{i_2 i_2}^2+\Delta_{i_1 i_2}^2\Delta_{i_2 j_1}^2)+\sum_{i_1,j_1,j_2} \Delta_{i_1 j_1}^2\Delta_{i_1 j_2}^2 +\sum_{i_1,i_2,j_1} \Delta_{i_1 j_1}^2\Delta_{i_2 j_1}^2\\
   &\leq C \sum_{i_1,i_2,j_1} (\Delta_{i_1 j_1}^2 \bar{\theta}_{i_2}^2 \bar{\zeta}_{\max}^2+\Delta_{i_1 i_2}^2 \bar{\theta}_{\max}^2 \bar{\zeta}_{j_1}^2)+C\sum_{i_1,j_1,j_2} \Delta_{i_1 j_1}^2\bar{\theta}_{\max}^2\bar{\zeta}_{j_2}^2 +C\sum_{i_1,i_2,j_1} \Delta_{i_1 j_1}^2\bar{\theta}_{i_2}^2\bar{\zeta}_{\max}^2
   \\
   &= C\cdot(\bar{\zeta}_{\max}^2\|\bar\theta\|^2+ \bar{\theta}_{\max}^2\|\bar\zeta\|^2)\cdot\tr(\Delta'\Delta)=o((\|\bar\theta\|^2+\|\bar\zeta\|^2)\cdot\tr(\Delta'\Delta)). 
\end{split}
\]
This completes \eqref{mean-ndist}. Furthermore, by $\delta_1^2/(\lambda_1+\tilde\lambda_1)\to \infty$, we know $\|\bar\theta\|^2\asymp \|\theta\|^2+\|\tilde\theta\|^2\asymp \lambda_1 +\tilde\lambda_1=o(\delta_1^2)$. Therefore, $(\|\bar\theta\|^2+\|\bar\zeta\|^2)\cdot\tr(\Delta'\Delta)=o(\delta_1^4)=o(\tr([\Delta'\Delta]^2))$.

We then consider the  variance part of the first claim. 
It suffices to show 
\[
\mathrm{Var}(Q_n) \leq C \big(\|\bar\theta\|^4\|\bar\zeta\|^4 + [\tr(\Delta'\Delta)]^3\big),
\]
Since $\|\bar\theta\|^4\|\bar\zeta\|^4\asymp\|\bar\theta\|^8\asymp\|\theta\|^8+\|\tilde{\theta}\|^8$. 
Recall \eqref{def:O} and that $\Delta = \Omega - \widetilde\Omega$, we first decompose \[
A_{ij} - \widetilde A_{ij} = (W_{ij} + \Omega_{ij}) - (\widetilde W_{ij} + \widetilde\Omega_{ij}) = O_{ij} + \Delta_{ij}. 
\]
We then decompose $Q_n-\mathbb{E}(Q_n)$  as the sum of the following terms 
\[
S_1 = \sum_{i_1, i_2, j_1, j_2 (dist)} O_{i_1j_1}O_{i_1j_2}O_{i_2j_1}O_{i_2j_2}, \quad S_2 = 4 \sum_{i_1, i_2, j_1, j_2 (dist)} O_{i_1j_1}O_{i_1j_2}O_{i_2j_1}\Delta_{i_2j_2}, 
\]
\[
S_3 = 2 \sum_{i_1, i_2, j_1, j_2 (dist)} O_{i_1j_1}O_{i_1j_2}\Delta_{i_2j_1}\Delta_{i_2j_2}, \quad S_4 =  2\sum_{i_1, i_2, j_1, j_2 (dist)} O_{i_1j_1}\Delta_{i_1j_2}O_{i_2j_1}\Delta_{i_2j_2}, 
\]
\[
S_5 = 2\sum_{i_1, i_2, j_1, j_2 (dist)} O_{i_1j_1}\Delta_{i_1j_2}\Delta_{i_2j_1}O_{i_2j_2}, \quad S_6 =4 \sum_{i_1, i_2, j_1, j_2 (dist)} O_{i_1j_1}\Delta_{i_1j_2}\Delta_{i_2j_1}\Delta_{i_2j_2}.
\]
Recall the basic probability that $\mathrm{Var}(S_1 + ... + S_6)\leq 6[\mathrm{Var}(S_1) + ... + \mathrm{Var}(S_6) ].$ It suffices to control $\mathrm{Var}(S_1), ..., \mathrm{Var}(S_6)$.  

Consider $\mathrm{Var}(S_1)$. Mimicking the proof of Theorem \ref{thm:null-d}, we have 
\beq\label{VarS1}
\mathrm{Var}(S_1) = C \sum_{i_1, i_2, j_1, j_2} \bar\Omega_{i_1j_1}\bar\Omega_{i_1j_2}\bar\Omega_{i_2j_1}\bar\Omega_{i_2j_2}\leq C\sum_{i_1, i_2, j_1, j_2} \bar{\theta}_{i_1}^2\bar{\theta}_{i_2}^2\bar{\zeta}_{j_1}^2\bar{\zeta}_{j_2}^2  = C\|\bar\theta\|^4\|\bar\zeta\|^4.
\eeq

Consider $\mathrm{Var}(S_2)$.  
It's not hard to see the terms in the summation are uncorrelated with each other. Hence, 
\[
\mathrm{Var}(S_2)  =16 \sum_{\substack{i_1, i_2, j_1, j_2 (dist)}} \mathrm{Var}(O_{i_1j_1}O_{i_1j_2}O_{i_2j_1})\cdot\Delta_{i_2j_2}^2 \leq 16 \sum_{\substack{i_1, i_2, j_1, j_2 (dist)}} \bar\Omega_{i_1j_1}\bar\Omega_{i_1j_2}\bar\Omega_{i_2j_1}\Delta_{i_2j_2}^2.  
\]
Here we've used the independence between $O_{i_1j_1}, O_{i_1j_2}, O_{i_2j_1}$, so \[
\mathrm{Var}(O_{i_1j_1}O_{i_1j_2}O_{i_2j_1})\leq \mathbb{E}[O_{i_1j_1}^2O_{i_1j_2}^2O_{i_2j_1}^2]\leq \mathbb{E}[O_{i_1j_1}^2]\mathbb{E}[O_{i_1j_2}^2]\mathbb{E}[O_{i_2j_1}^2] \leq \bar\Omega_{i_1j_1}\bar\Omega_{i_1j_2}\bar\Omega_{i_2j_1}. 
\]
By \eqref{bound-delta}, we control 
\beq\label{VarS2}
\mathrm{Var}(S_2) \leq 16 \sum_{\substack{i_1, i_2, j_1, j_2 (dist)}} \bar\Omega_{i_1j_1}\bar\Omega_{i_1j_2}\bar\Omega_{i_2j_1}\Delta_{i_2j_2}^2 \lesssim \|\bar\theta\|^2\|\bar\zeta\|^2\|\bar\theta\|_3^3\|\bar\zeta\|_3^3 = o(\|\bar\theta\|^4\|\bar\zeta\|^4). 
\eeq

Consider $\mathrm{Var}(S_3)$. Rewrite it as 
\[
S_3 = 4 \sum_{\substack{i_1, j_1, j_2} (dist) \\ j_1<j_2}   O_{i_1 j_1}O_{i_1 j_2} \bigg(\sum_{i_2\notin\{i_1, j_1, j_2\}}\Delta_{i_2j_1}\Delta_{i_2j_2}\bigg). 
\]
The terms in the above summation are mean zero and uncorrelated with each other, which indicates 
\begin{align*}
\mathrm{Var}(S_3) & =  16 \sum_{\substack{i_1, j_1, j_2 (dist) \\ i_2<j_2} } \mathbb{E}\bigg[O_{i_1 j_1}^2O_{i_1 j_2}^2  \bigg(\sum_{i_2\notin\{i_1, j_1, j_2\}}\Delta_{i_2j_1}\Delta_{i_2j_2}\bigg)^2\bigg] \\
& \leq C\sum_{i_1, j_1, j_2} \bar{\Omega}_{i_1j_1}\bar{\Omega}_{i_1j_2}\bigg(\sum_{i_2}\Delta_{i_2j_1}\Delta_{i_2j_2}\bigg)^2. 
\end{align*}
Combining with  \eqref{bound-delta} and   $\bar\zeta_{\max} = o(1)$, the above term is no more than
\begin{align*}
\sum_{i_1, j_1, j_2} \bar\theta_{i_1}^2\bigg(\sum_{i_2}\Delta_{i_2j_1}\Delta_{i_2j_2}\bigg)^2 
\leq 
\sum_{i_1, j_1, j_2} \bar\theta_{i_1}^2\bigg(\sum_{i_2}\Delta_{i_2j_1}^2\bigg)\bigg(\sum_{i_2}\Delta_{i_2j_2}^2\bigg),
\end{align*}
where we've used the Cauchy-Schwarz inequality. Each term above are non-negative, so the sum is no more than 
\[
 \sum_{i_1, i_2, i_3, j_1, j_2} \bar\theta_{i_1}^2\Delta_{i_2j_1}^2\Delta_{i_3j_2}^2  =\|\bar\theta\|^2\cdot[\tr(\Delta'\Delta)]^2.
\]
By Young's inequality that $ab \leq \frac{a^3}{3} + \frac{2b^{3/2}}{3}$, the variance of $S_3$ is upper bounded by 
\beq\label{VarS3}
\mathrm{Var}(S_3)\leq C[\|\bar\theta\|^6 + [\tr(\Delta'\Delta)]^3\leq C[\|\bar\theta\|^8 + [\tr(\Delta'\Delta)]^3] \asymp [\|\bar\theta\|^4\|\bar\zeta\|^4 + [\tr(\Delta'\Delta)]^3.
\eeq
By symmetry, we know 
\beq
\mathrm{Var}(S_4)\leq C\|\bar\zeta\|^6 + [\tr(\Delta'\Delta)]^3 \lesssim [\|\bar\theta\|^4\|\bar\zeta\|^4 + [\tr(\Delta'\Delta)]^3.
\eeq

Consider the variance of $S_5$.  We find 
\begin{align*}
\mathrm{Var}(S_5) & =  C \sum_{i_1, i_2, j_1,  j_2 (dist)}\mathbb{E}[O_{i_1 j_1}^2 \Delta_{i_1j_2}^2\Delta_{i_2 j_1}^2 O_{i_2j_2}^2] 
 \leq C\sum_{i_1, i_2, j_1, j_2 (dist)}\bar{\Omega}_{i_1j_1}\Delta_{i_1j_2}^2\Delta_{i_2j_1}^2 \bar\Omega_{i_2j_2}.  
\end{align*} 
By \eqref{bound-delta}, the above term is no more than 
\beq\label{VarS4}
C\sum_{i_1, i_2, j_1, j_2 } \bar\theta_{i_1}^3\bar\theta_{i_2}^3\bar\zeta_{j_1}^3\bar\zeta_{j_2}^3 = C \|\bar\theta\|_3^{6}\|\bar\zeta\|_3^6  = o(\|\bar\theta\|^4\|\bar\zeta\|^4),
\eeq
where we've used $\|\bar\theta\|^3_3 \leq\bar\theta_{\max}\|\bar\theta\|^2 =o(\|\bar\theta\|^2)$ and $\|\bar\zeta\|^3_3=o(\|\bar\zeta\|^2)$ in the last inequality.

Consider the variance of  $S_6$.  Rewrite it as 
\[
S_6 = 4 \sum_{i_1\neq  j_1}  O_{i_1 i_2} \bigg(\sum_{i_2, j_2\notin\{i_1, j_1 \}}\Delta_{i_1j_2}\Delta_{i_2j_1}\Delta_{i_2j_2}\bigg). 
\]
It's then not hard to see the terms in the above summation are mean zero and uncorrelated with each other, which indicates 
\begin{align*}
\mathrm{Var}(S_6) & = 16  \sum_{i_1 \neq   j_1}\mathbb{E}\bigg[ O_{i_1j_1}^2  \bigg(\sum_{i_2, j_2\notin\{i_1, j_1 \}}\Delta_{i_1j_2}\Delta_{i_2j_1}\Delta_{i_2j_2}\bigg)^2\bigg]  \leq C \sum_{i_1 \neq  j_1}\bar\Omega_{i_1j_1}    \bigg(\sum_{i_2, j_2\notin\{i_1, j_1 \}}\Delta_{i_1j_2}\Delta_{i_2j_1}\Delta_{i_2j_2}\bigg)^2, 
\end{align*}
where by \eqref{bound-delta}, the above term is no more than
\beq\label{S5temp}
C\sum_{i_1, j_1 (dist)}(\bar\theta_{i_1}\bar\zeta_{j_1}  )\bigg(\sum_{i_2, j_2 \notin\{i_1, j_1 \}}\Delta_{i_1j_2}\Delta_{i_2j_1}\Delta_{i_2j_2}\bigg)^2. 
\eeq
Using Cauchy-Schwarz inequality, we find 
\begin{align*} 
\bigg(\sum_{i_2, j_2 \notin\{i_1, j_1 \}} \Delta_{i_1j_2}\Delta_{i_2j_1}\Delta_{i_2j_2}\bigg)^2
&  \leq\bigg(\sum_{i_2, j_2 }\Delta_{i_1j_2}^2\Delta_{i_2j_1}^2\bigg)\bigg(\sum_{i_2, j_2 }\Delta_{i_2j_2}^2\bigg). 
\end{align*}
Plug it into \eqref{S5temp} and recall $\bar\theta_{\max} = o(1), \bar\zeta_{\max} = o(1)$, we obtain 
\beq \label{VarS5}
\mathrm{Var}(S_6) \leq C\sum_{i_1, j_1 (dist)}\bigg(\sum_{i_2, j_2 }\Delta_{i_1j_2}^2\Delta_{i_2j_1}^2\bigg)\bigg(\sum_{i_2, j_2 }\Delta_{i_2j_2}^2\bigg)\leq [\tr(\Delta'\Delta)]^3. 
\eeq
Combining \eqref{VarS1}-\eqref{VarS5}, we get 
\[
\mathrm{Var}(Q_n)\leq 6[\mathrm{Var}(S_1)  + \cdots + \mathrm{Var}(S_6)] \leq C [\|\bar\theta\|^4\|\bar\zeta\|^4 +  [\tr(\Delta'\Delta)]^3 ]\asymp C [\|\theta\|^8+\|\tilde\theta\|^8 +  [\tr(\Delta'\Delta)]^3 ]. 
\]
Using the fact that $\tr([\Omega\Omega']^2)\asymp \|\theta\|^8$ and $\tr([\tilde\Omega\tilde\Omega']^2)\asymp \|\tilde \theta\|^8$, we showed the first claim.

Consider the last claim. 
To show $\psi_n\goto\infty$ in probability, 
it suffices to show for any $c > 0$, \[
 \mathbb{P}_{H_1^{(n)}}\biggl(Q_n \leq c z_{\alpha} \cdot \sqrt{{C}_n + \widetilde{C}_n }   \biggr)  \goto0.
\]
fixing $0 < \eps < 1$, let $A_{\eps}$ be the event  $\big\{ ({C}_n + \widetilde{C}_n) \leq (1 + \eps)\cdot\mathbb{E} [  {C}_n + \widetilde{C}_n  ]         \big\}$, 
which occurs with probability tending to 1 by the second claim in Theorem \ref{thm:null-d}.  
Over the event $A_{\eps}$,  ${C}_n + \widetilde{C}_n \leq C(\|\theta\|^8 + \|\tilde\theta\|^8)$ for a sufficiently large constant $C>0$. 
Therefore, 
\begin{align*} 
& \mathbb{P}_{H_1^{(n)}}\biggl(Q_n \leq cz_{\alpha} \cdot \sqrt{{C}_n + \widetilde{C}_n }   \biggr)  \\
\leq & \mathbb{P}_{H_1^{(n)}}\biggl(Q_n \leq cz_{\alpha} \cdot \sqrt{{C}_n + \widetilde{C}_n } , A_{\eps}\biggr) + \mathbb{P}(A_{\eps}^c)  \\
\leq & \mathbb{P}_{H_1^{(n)}} \biggl(Q_n \leq Cz_{\alpha} \cdot (\|\theta\|^4 + \|\tilde\theta\|^4)  \biggr) + o(1),  
\end{align*} 
where $C$ is a generic constant and by Chebyshev's inequality, the first term in the last line 
\begin{equation} \label{mypfC1}
\leq  [\mathbb{E}(Q_n) - C z_{\alpha} \cdot (\|\theta\|^4+\|\tilde\theta\|^4)]^{-2} \cdot \mathrm{Var}(Q_n). 
\end{equation} 
Recall that $\mathbb{E}(Q_n)=  \tr[(\Delta'\Delta )^2]+o\big(\|\bar\theta\|^2\cdot\tr(\Delta'\Delta)\big)$. Under the condition that $\delta_1^2/(\lambda_1+\widetilde{\lambda}_1)\to \infty$, 
$\|\bar\theta\|^2\cdot\tr(\Delta'\Delta)\asymp(\|\theta\|^2+\|\tilde\theta\|^2)\cdot\tr(\Delta'\Delta)\asymp (\lambda_1+\tilde\lambda_1)\delta_1^2=o(\delta_1^4)$ where the second last equation is from the second claim of Theorem \ref{thm:null-d}: $\lambda_1^4\asymp \tr[(\Omega'\Omega)^2]\asymp\|\theta\|^8$. Consequently,  $\mathbb{E}(Q_n)\asymp  \tr[(\Delta'\Delta )^2]\asymp \delta_1^4$. Meanwhile, $ \|\theta\|^4 +  \|\tilde\theta\|^4\asymp \lambda_1^2+\widetilde{\lambda}_1^2=o(\delta_1^4)$, thus 
\beq
\mathbb{E}[Q_n] - C  z_{\alpha}  ( \|\theta\|^2\|\zeta\|^2 +  \|\tilde\theta\|^2\|\tilde\zeta\|^2)   \geq  \frac{1}{2}
\tr[(\Delta'\Delta )^2]. 
\eeq
Furthermore, notice 
$\mathrm{Var}(Q_n) \leq C[ \|\theta\|^8+\|\tilde{\theta}\|^8 + [ \tr(\Delta'\Delta) ]^3] \asymp [ \lambda_1^4 + \widetilde{\lambda}_1^4 + \delta_1^6]$,
 the right hand side of (\ref{mypfC1}) does not exceed 
\begin{equation} 
C  \times \frac{\lambda_1^4 + \widetilde{\lambda}_1^4 + \delta_1^6 }{\big[\tr[(\Delta'\Delta )^2]}\big]^2  \leq C \times \frac{(\lambda_1 + \widetilde{\lambda}_1)^4 + \delta_1^6}{\delta_1^8} \goto 0, 
\end{equation} 
where  we've used the condition that $\delta_1^2 \gg \lambda_1 + \widetilde{\lambda}_1 \asymp   \|\theta\|^2 +  \|\tilde\theta\|^2\to \infty$.
Therefore, $\psi_n\goto\infty$ in probability and so the Type II error $\goto 0$. 

Under the null,  
\[
\frac{Q_n}{\sqrt{32({C}_n + \widetilde{C}_n)}} \stackrel{d}{\longrightarrow} N(0,1), 
\]
so the Type I error is 
\[
 \mathbb{P}_{H_0^{(n)}}\biggl(\frac{Q_n }{{\sqrt{32({C}_n + \widetilde{C}_n)}}  } \geq  z_{\alpha} \biggr)   = \alpha + o(1),
\] 
where we recall that  $z_{\alpha}$ is the $\alpha$-upper quantile of standard normal. Combining above, we completes the proof that the power of IBM test goes to $1$ as $n\goto \infty$.

\subsection{Proof of Theorem \ref{thm:LB-d}}
\label{proof:lb-distance}
 
Our first step is to construct $\widetilde{\Omega}(\sigma)\in\mathcal{M}_n^{dir}(\beta_n, K+1, {\bf c})$ satisfying
\beq\label{construct-key2}
\widetilde{\Omega}_{ij}(\sigma) = \Omega_{ij} + \eps_n \sigma_i\sigma_j\theta_i\zeta_j\Pi_{iK}\Gamma_{jK},
\eeq
where ${\bf c} = (c_0, c_1, c_2)$ and $\eps_n$ is a diminishing sequence with its value to be specify and  $\sigma = (\sigma_1, ..., \sigma_n)$ is a binary vector i.e. $\sigma_i\in\{-1, 1\}$ for $i = 1, 2, ..., n$. 

For $\Omega = \Theta\Pi P\Gamma'Z \in\mathcal{M}_{n}^{dir}(\beta_n, K, \bf{c})$, we introduce $\check\Pi(\sigma),\check\Gamma(\sigma)\in\mathbb{R}^{n, K+1}$ as follow
\begin{align*}
&\check{\Pi}_{i \ell}(\sigma)= \Pi_{i\ell}, \quad 1\leq i\leq n, 1\leq \ell\leq K-1, \quad\mbox{ and } \quad \check{\Pi}_{iK}(\sigma) = \frac{1 + \sigma_i}{2}  \Pi_{iK}, \quad \check{\Pi}_{i,K+1}(\sigma) = \frac{1 - \sigma_i}{2} \Pi_{iK};\\
&\check{\Gamma}_{i \ell}(\sigma)= \Gamma_{i\ell}, \quad 1\leq i\leq n, 1\leq \ell\leq K-1, \quad\mbox{ and } \quad \check{\Gamma}_{iK}(\sigma) = \frac{1 + \sigma_i}{2}  \Gamma_{iK}, \quad \check{\Gamma}_{i,K+1}(\sigma) = \frac{1 - \sigma_i}{2} \Gamma_{iK}.
\end{align*}
It's not hard to see $\check\Pi(\sigma)$ is a non-negative matrix with row sums equal to $1$. Rewrite \[
P = \left(\begin{matrix} P_0& \alpha\\\beta^T & 1
\end{matrix}\right) \in\mathbb{R}^{K,K},
\]
we introduce   $\check P \in\mathbb{R}^{K+1, K+1}$ such that \[
\check{P} = \left(\begin{matrix} P_0 & \alpha & \alpha \\
\beta^T & 1+\eps_n & 1-\eps_n \\
\beta^T & 1-\eps_n & 1 + \eps_n 
\end{matrix}\right). 
\]
For notation simplicity, we write $\check\Pi$ in short for $\check\Pi(\sigma)$ in the rest of the proof. 
Write the $i$-th row of $\check{\Pi},\check{\Gamma}$ as $\check{\pi}_i',\check{\gamma}_i'$, and denote the first $(K-1)$ entries of $\check{\pi}_i',\check{\gamma}_i'$ by $\pi'_{i[K-1]},\gamma'_{i[K-1]}$. 
Let $\widetilde{\Omega}(\sigma) = \Theta \check{\Pi} \check{P}\check{\Gamma}'Z$, and we have
\begin{align*}
\widetilde{\Omega}_{ij}(\sigma) = & \theta_i\zeta_j \check{\pi}_i' \check{P}\check{\gamma}_j = \theta_i\zeta_j \left[  \pi_{i[K-1]}'P_0 \gamma_{j[K-1]} + (1+\eps_n)(\check{\Pi}_{iK}\check{\Gamma}_{jK}+\check{\Pi}_{i,K+1}\check{\Gamma}_{j,K+1})\right. \\
& \left.+ (1-\eps_n)(\check{\Pi}_{iK}\check{\Gamma}_{j,K+1}+\check{\Pi}_{i,K+1}\check{\Gamma}_{jK}) + (\check{\Pi}_{iK}+\check{\Pi}_{i,K+1}) \gamma_{j[K-1]}'\beta
+ (\check{\Gamma}_{jK}+\check{\Gamma}_{j,K+1}) \pi_{i[K-1]}'\alpha\right] .
\end{align*}
Notice that $\check{\Pi}_{iK}+\check{\Pi}_{i,K+1}=\Pi_{iK},\check{\Gamma}_{jK}+\check{\Gamma}_{j,K+1}=\gamma_{jK}$ and that
\[
\Omega_{ij}  = \theta_i\zeta_j \pi_i'P\gamma_j= \theta_i\zeta_j \left[ \pi_{i[K-1]}'P_0 \gamma_{j[K-1]} + \Pi_{iK} \gamma_{j[K-1]}'\beta + \Gamma_{jK} \pi_{i[K-1]}'\alpha+ \Pi_{iK}\Gamma_{jK}\right] ,
\]
 we obtain
\[
\widetilde{\Omega}_{ij}(\sigma) = \Omega_{ij} + \eps_n\theta_i\zeta_j (\check{\Pi}_{iK}-\check{\Pi}_{i,K+1})
(\check{\Gamma}_{jK}-\check{\Gamma}_{j,K+1}) = \Omega_{ij}+\eps_n\theta_i\zeta_j \sigma_i\sigma_j \Pi_{iK}\Gamma_{jK}.
\]
Therefore $\widetilde\Omega(\sigma) = \Theta \check\Pi\check P\check \Gamma'Z$ satisfies \eqref{construct-key2}. 

Now we want to find a decomposition $\widetilde\Omega(\sigma) = \widetilde\Theta \widetilde\Pi\widetilde P \widetilde \Gamma'\widetilde Z \in\mathcal{M}_n^{dir}(\beta_n, K+1, {\bf c})$. For a fixed $\sigma$. One of the following four cases must occur:
\begin{equation}
    \begin{split}
        (\romannumeral 1),\quad  \|\theta \circ \check\pi^{(K+1)}\|\geq \|\theta \circ \check\pi^{(K)}\| \quad \text{and} \quad  \|\zeta \circ \check\gamma^{(K+1)}\|\geq \|\zeta \circ \check\gamma^{(K)}\|\\
        (\romannumeral 2),\quad  \|\theta \circ \check\pi^{(K+1)}\|\leq \|\theta \circ \check\pi^{(K)}\| \quad \text{and} \quad  \|\zeta \circ \check\gamma^{(K+1)}\|\leq \|\zeta \circ \check\gamma^{(K)}\|\\
        (\romannumeral 3),\quad  \|\theta \circ \check\pi^{(K+1)}\|\geq \|\theta \circ \check\pi^{(K)}\| \quad \text{and} \quad  \|\zeta \circ \check\gamma^{(K+1)}\|\leq \|\zeta \circ \check\gamma^{(K)}\|\\
        (\romannumeral 4),\quad  \|\theta \circ \check\pi^{(K+1)}\|\leq \|\theta \circ \check\pi^{(K)}\| \quad \text{and} \quad  \|\zeta \circ \check\gamma^{(K+1)}\|\geq \|\zeta \circ \check\gamma^{(K)}\|\\
    \end{split}
\end{equation}

If (\romannumeral 1) or (\romannumeral 2) occurs. Notice that the second case can be converted to the first case by permuting the $K$-th and $(K+1)$-th rows and columns of $\check{P}$ (also permute the last two columns of $\check\Pi$ and $\check\Gamma$ so that $\widetilde{\Omega}(\sigma)$ is unchanged). Thus we only discuss (i).

For case (\romannumeral 1). Write $D = \diag(1, \dots, 1, \sqrt{1+\eps_n}, \sqrt{1+\eps_n})$, and let $\widetilde{P}=D^{-1} \check{P}D^{-1}$ (obviously, $\|\tilde{P}\|_{\max}\leq \|P\|_{\max}\leq (K+1)c_1^{-1}$), then $\widetilde P_{K,K}=\widetilde P_{K+1, K+1}=1$.
Introduce two diagonal matrices $G,H\in\mathbb{R}^{n,n}$ such that \[
G_{ii} = \sum_{k=1}^{K-1} \Pi_{ik} + \sqrt{1+\eps_n} \cdot\Pi_{iK},\quad H_{ii} = \sum_{k=1}^{K-1} \Gamma_{ik} + \sqrt{1+\eps_n} \cdot\Gamma_{iK}. 
\]
Also, introduce $\widetilde{\Pi},\widetilde{\Gamma}\in\mathbb{R}^{n, K+1}$ such that $\widetilde{\Pi} = G^{-1}\check\Pi D$ and $\widetilde{\Gamma} = H^{-1}\check\Gamma D$. Then it's not hard to verify that $\widetilde{\Pi}$ and $\widetilde{\Gamma}$ are valid membership matrices with $(K+1)$ communities, as they're non-negative and all rows sum up to $1.$
Combining above, we have
\[
\widetilde{\Omega} = \Theta \check{\Pi} \check{P} \check{\Gamma}' Z = \Theta G  \widetilde{\Pi} \widetilde{P}\widetilde{\Gamma}'H Z = \widetilde{\Theta}\widetilde{\Pi} \widetilde{P}\widetilde{\Gamma}'\widetilde{Z},
\]
where $\widetilde\Theta = \Theta G$ and $\widetilde{Z}=HZ$ are diagonal matrices.
Recall that $\Omega \in \mathcal{M}_n^{dir}(\beta_n,K, \bf{c})$. By $G_{ii},H_{ii} \in [1, \sqrt{1 + \eps_n}]$, we have $\tilde{\theta}_i \in [\theta_i,\sqrt{1+\epsilon_n}\cdot \theta_i]$ and $\tilde{\zeta}_i \in [\zeta_i,\sqrt{1+\epsilon_n}\cdot \zeta_i]$. Additionally, $\eps_n = o(1)$, therefore,  $\tilde{\theta}_{\max} \leq  \sqrt{1+\eps_n}\theta_{\max} \leq  \sqrt{1+\eps_n}K \beta_n  \leq (K+1)\beta_n$ and $\|\tilde{\theta}\|\geq \|\theta\|\geq (K+1)^{-1}\beta_n^{-1}$ for sufficiently large $n$. The same argument holds for $\zeta$.

Consider the spectral norm of $\widetilde\Omega(\sigma)$. Rewrite $\widetilde{\Omega}(\sigma)$ as 
\[
\widetilde{\Omega}(\sigma) = \Omega + \eps_n (\theta\circ \sigma\circ \pi^{(K)})\cdot (\zeta\circ \sigma\circ \gamma^{(K)})^T,
\]
where $\pi^{(K)}$ and $\gamma^{(K)}$ are the $K$-th column of $\Pi$ and $\Gamma$. Then
\beq
\|\widetilde{\Omega}(\sigma)-\Omega\| = \eps_n \|\theta\circ \pi^{(K)}\|\|\zeta\circ \gamma^{(K)}\| \leq \eps_n \|\theta\|\|\zeta\|.
\eeq
Recall that $\eps_n = o(1)$, we have \[
\|\widetilde{\Omega}(\sigma)\|\geq \|\Omega\| - \|\widetilde{\Omega}(\sigma)-\Omega\| \geq c_0K^{-1}\|\theta\|\|\zeta\|- \eps_n \|\theta\|\|\zeta\|, 
\]
where for sufficiently large $n$, the rightmost term is \[
(c_0/K - \eps_n)\|\theta\|\|\zeta\| \geq (c_0/K - \eps_n)\|\tilde\theta\|\|\tilde\zeta\|/(1+\eps_n) \geq c_0\|\tilde\theta\|\|\tilde\zeta\|/(K+1) .
\]
Similarly, we have 
\beq
\|\widetilde{\Omega}(\sigma)\|\leq \|\Omega\| + \|\widetilde{\Omega}(\sigma)-\Omega\| \leq c_0^{-1}K \|\theta\|\|\zeta\|+ \eps_n \|\theta\|\|\zeta\| , 
\eeq
where for sufficiently large $n$ and $\eps_n\goto0$, the rightmost term is bounded by \[
(c_0^{-1}K + \eps_n)\|\tilde\theta\|\|\tilde\zeta\| \leq c_0^{-1}(K+1)\|\tilde\theta\|\|\tilde\zeta\|. 
\]  
Lastly, we know 
\[\|\tilde{\theta}\circ \tilde{\pi}^{(K+1)}\|=\sqrt{1+\eps_n}\cdot \|\theta \circ \check\pi^{(K+1)}\|\geq \sqrt{(1+\eps_n)/2}\cdot \|\theta \circ \pi^{(K)}\|\geq c_2 (K+1)^{-1}2^{-(K+1)/2}\|\tilde{\theta}\|\]
by assumption (\romannumeral 1), $ \|\theta \circ \check\pi^{(K)}\|^2+ \|\theta \circ \check\pi^{(K+1)}\|^2= \|\theta \circ \pi^{(K)}\|^2$ and $\|\theta \circ \pi^{(K)}\|\geq c_2K^{-1}2^{-K/2}\|\theta\|$. One can show the same argument for $\zeta$. 

If (\romannumeral 3) or (\romannumeral 4) occurs. We only discuss (\romannumeral 3) due to symmetry. Permute the $K$-th and $(K+1)$-th column of $\check P$ and $\check \Gamma$ so that after permutation (\romannumeral 1) holds and $\widetilde\Omega(\sigma)$ remains unchanged. Write $D = \diag(1, \dots, 1, \sqrt{1-\eps_n}, \sqrt{1-\eps_n})$, and let $\widetilde{P}=D^{-1} \check{P}D^{-1}$ (obviously, $\|\tilde{P}\|_{\max}\leq \max\{\frac{1+\eps_n}{1-\eps_n}, \frac{\|P\|_{\max}}{\sqrt{1-\eps_n}}\}\leq (K+1) c_1^{-1}$ for sufficiently large $n$), then $\widetilde P_{K,K}=\widetilde P_{K+1, K+1}=1$.
Introduce two diagonal matrices $G,H\in\mathbb{R}^{n,n}$ such that \[
G_{ii} = \sum_{k=1}^{K-1} \Pi_{ik} + \sqrt{1-\eps_n} \cdot\Pi_{iK},\quad H_{ii} = \sum_{k=1}^{K-1} \Gamma_{ik} + \sqrt{1-\eps_n} \cdot\Gamma_{iK}. 
\]
Also, introduce $\widetilde{\Pi},\widetilde{\Gamma}\in\mathbb{R}^{n, K+1}$ such that $\widetilde{\Pi} = G^{-1}\check\Pi D$ and $\widetilde{\Gamma} = H^{-1}\check\Gamma D$. Then it's not hard to verify that $\widetilde{\Pi}$ and $\widetilde{\Gamma}$ are valid membership matrices with $(K+1)$ communities, as they're non-negative and all rows sum up to $1.$
Combining above, we have
\[
\widetilde{\Omega} = \Theta \check{\Pi} \check{P} \check{\Gamma}' Z = \Theta G  \widetilde{\Pi} \widetilde{P}\widetilde{\Gamma}'H Z = \widetilde{\Theta}\widetilde{\Pi} \widetilde{P}\widetilde{\Gamma}'\widetilde{Z},
\]
where $\widetilde\Theta = \Theta G$ and $\widetilde{Z}=HZ$ are diagonal matrices.
Recall that $\Omega \in \mathcal{M}_n^{dir}(\beta_n,K, {\bf c})$. By $G_{ii},H_{ii} \in [\sqrt{1 - \eps_n},1]$, we have $\tilde{\theta}_i \in [\sqrt{1-\epsilon_n}\cdot \theta_i,\theta_i]$ and $\tilde{\zeta}_i \in [\sqrt{1-\epsilon_n}\cdot \zeta_i,\zeta_i]$. Additionally, $\eps_n = o(1)$, therefore,  $\tilde{\theta}_{\max} \leq  \theta_{\max} \leq  (K+1)\beta_n$ and $\|\tilde{\theta}\|\geq \sqrt{1-\eps_n}\cdot\|\theta\|\geq (K+1)^{-1}\beta_n^{-1}$ for sufficiently large $n$. The same argument holds for $\zeta$.

Consider the spectral norm of $\widetilde\Omega(\sigma)$. Then
\beq
\|\widetilde{\Omega}(\sigma)-\Omega\| = \eps_n \|\theta\circ \pi^{(K)}\|\|\zeta\circ \gamma^{(K)}\| \leq \eps_n \|\theta\|\|\zeta\|.
\eeq
Recall that $\eps_n = o(1)$, we have \[
\|\widetilde{\Omega}(\sigma)\|\geq \|\Omega\| - \|\widetilde{\Omega}(\sigma)-\Omega\| \geq c_0K^{-1}\|\theta\|\|\zeta\|- \eps_n \|\theta\|\|\zeta\|, 
\]
where for sufficiently large $n$, the rightmost term is \[
(c_0/K - \eps_n)\|\theta\|\|\zeta\| \geq (c_0/K - \eps_n)\|\tilde\theta\|\|\tilde\zeta\| \geq c_0\|\tilde\theta\|\|\tilde\zeta\|/(K+1) .
\]
Similarly, we have 
\beq
\|\widetilde{\Omega}(\sigma)\|\leq \|\Omega\| + \|\widetilde{\Omega}(\sigma)-\Omega\| \leq c_0^{-1}K \|\theta\|\|\zeta\|+ \eps_n \|\theta\|\|\zeta\| , 
\eeq
where for sufficiently large $n$ and $\eps_n\goto0$, the rightmost term is bounded by \[
(c_0^{-1}K + \eps_n)\|\tilde\theta\|\|\tilde\zeta\|/(1-\eps_n) \leq c_0^{-1}(K+1)\|\tilde\theta\|\|\tilde\zeta\|. 
\]  
Lastly, we know 
\[\|\tilde{\theta}\circ \tilde{\pi}^{(K+1)}\|=\sqrt{1-\eps_n}\cdot \|\theta \circ \check\pi^{(K+1)}\|\geq \sqrt{(1-\eps_n)/2}\cdot \|\theta \circ \pi^{(K)}\|\geq c_2 (K+1)^{-1}2^{-(K+1)/2}\|\tilde{\theta}\|\]
by assumption (\romannumeral 1), $ \|\theta \circ \check\pi^{(K)}\|^2+ \|\theta \circ \check\pi^{(K+1)}\|^2= \|\theta \circ \pi^{(K)}\|^2$ and $\|\theta \circ \pi^{(K)}\|\geq c_2 K^{-1}2^{-K/2}\|\theta\|$. One can show the same argument for $\zeta$. Combining above, we conclude that as long as $\eps_n=o(1)$, $\widetilde{\Omega} \in \mathcal{M}_n^{dir}(\beta_n, K+1, \bf{c})$ and it satisfies \eqref{construct-key2}.

Our next step is to show for given $\{\beta_n\}_{n=1}^{\infty}$ and $\{\rho_n\}_{n=1}^{\infty}$, $\{\eps_n\}_{n=1}^{\infty}$ can be chosen such that $(\Omega, \widetilde\Omega(\sigma))\in\mathcal{S}^{dir}_n(\beta_n, \rho_n, K, K+1, {\bf c})$. We've already shown in the first step that $\widetilde\Omega(\sigma)\in\mathcal{M}_n^{dir}(\beta_n, K+1, {\bf c})$ if $\eps_n=o(1)$. It remains to show $\{\eps_n\}_{n=1}^{\infty}$ can additionally satisfy
\[
\frac{\|\Omega - \widetilde\Omega(\sigma)\|}{\sqrt{\|\Omega\| + \|\widetilde\Omega(\sigma)\|}} \geq \sqrt{\rho_n},
\]
for given $\rho_n\goto0$.  Recall that $\|\theta\circ\pi^{(K)}\|\geq c_2 K^{-1}2^{-K/2} \|\theta\|$ and $\|\zeta\circ\gamma^{(K)}\|\geq c_2 K^{-1}2^{-K/2} \|\zeta\|$. Hence, 
\[
 \frac{\|\widetilde{\Omega}-\Omega\|}{\sqrt{\|\Omega\| + \|\widetilde\Omega\|}}\geq \frac{\eps_n c_2^2 \|\theta\|\|\zeta\|K^{-2}2^{-K}}{\sqrt{(2c_0^{-1}K+\eps_n) \|\theta\|\|\zeta\|}} \geq \eps_n (\|\theta\|\|\zeta\|)^{1/2} \cdot\frac{c_2^2\sqrt{c_0}}{\sqrt{3} K^{2.5}2^K},
\]
where we've used $\|\Omega\| \leq c_0^{-1} K\|\theta\|\|\zeta\|$ and $\|\widetilde\Omega\|\leq (c_0^{-1}K + \eps_n)\|\theta\|\|\zeta\|$.
Therefore, for any given sequence $\rho_n=o(1)$, we can find sequence $\eps_n$ such that  $\sqrt{\rho_n} = \eps_n(\|\theta\|\|\zeta\|)^{1/2} \cdot\frac{c_2^2\sqrt{c_0}}{\sqrt{3} K^{2.5}2^K}$. Consequently, $\eps_n=o(1)$ and  $(\Omega, \widetilde{\Omega}(\sigma)) \in\mathcal{S}^{dir}_n(\beta_n, \rho_n, K, K+1, {\bf c}).$ 

Our last step is to construct $H_0^{(n)}$ and $H_{1,\sigma}^{(n)}$ as follows:
\[
H_0^{(n)}: A\sim \mathbb{P}_n , \widetilde{A}\sim \mathbb{P}_n, \qquad H_{1,\sigma}^{(n)}: A \sim \mathbb{P}_n, \widetilde{A} \sim \mathbb{Q}_n(\sigma),
\]
where $\mathbb{P}_n$ is the distribution of adjacency matrix indicated by $\Omega$, and $\mathbb{Q}_n(\sigma)$ is the distribution of adjacency matrix indicated by $\widetilde{\Omega}(\sigma)$. As shown in the second step, we have $(\Omega, \widetilde{\Omega}(\sigma)) \in\mathcal{S}_n^{dir}(\beta_n,\rho_n, K, K+1, \bf{c})$ and $(\Omega, \Omega)\in \mathcal{S}_n^{dir*}(\beta_n, K, \bf{c})$. 

Write $\mathbb{Q}_n = \frac{1}{2^n}\sum_{\sigma \in \left\lbrace\pm 1 \right\rbrace ^n} \mathbb{Q}_n(\sigma)$,
 it suffices to show 
\begin{equation}
\int\left( \frac{d \mathbb{Q}_n }{d \mathbb{P}_n}\right)^2d\mathbb{P}_n = 1 + o(1)
, \quad \text{as} \quad n\goto\infty.
\end{equation}
Let $\sigma, \sigma'$ both be uniformly sampled from $\{1 ,-1\}^n$ independently. We re-write the $\chi^2$-distance as 
\begin{align*}
\int\left( \frac{d \mathbb{Q}_n }{d \mathbb{P}_n}\right)^2d\mathbb{P}_n = & \mathbb{E}_{\sigma, \sigma'}\left[ 
\prod_{i\neq j} \left( \frac{\widetilde{\Omega}_{ij}(\sigma)\widetilde{\Omega}_{ij}(\sigma')}{\Omega_{ij}} + \frac{(1-\widetilde{\Omega}_{ij}(\sigma))(1-\widetilde{\Omega}_{ij}(\sigma'))}{(1-\Omega_{ij})}
\right) \right] \\
= & \mathbb{E}_{\sigma, \sigma'}\left[ \prod_{i\neq j} \left(
1 + \frac{\Delta_{ij}(\sigma)\Delta_{ij}(\sigma')}{\Omega_{ij}(1-\Omega_{ij})}
\right) \right] \\
= & \mathbb{E}_{\sigma, \sigma'}\left[ \prod_{i\neq j} \left(1+
\frac{  \eps_n^2\theta_i^2\zeta_j^2 \sigma_i\sigma_j\sigma_i'\sigma_j' \Pi_{iK}^2\Gamma_{jK}^2     }{\Omega_{ij}(1-\Omega_{ij})}
\right) \right] .
\end{align*}
Note that $\sigma \circ \sigma'$ can also be viewed as generated uniformly from $\{1 ,-1\}^n$ (thus replace $\sigma_i\times \sigma_i'$ by $\sigma_i$), and by $e^x\geq 1+x$, the above equation can be rewritten as 
\[
= \mathbb{E}_{\sigma}\left[ \prod_{i\neq j} \left(1+
\frac{  \eps_n^2\theta_i^2\zeta_j^2 \sigma_i\sigma_j \Pi_{iK}^2\Gamma_{jK}^2     }{\Omega_{ij}(1-\Omega_{ij})}
\right) \right] \leq 
\mathbb{E}_{\sigma}\left[ \exp\left\lbrace \sum_{i\neq j}
\frac{  \eps_n^2\theta_i^2\zeta_j^2 \sigma_i\sigma_j \Pi_{iK}^2\Gamma_{jK}^2     }{\Omega_{ij}(1-\Omega_{ij})}
\right\rbrace \right] .
\]
Introduce \[
S_{n} = \sum_{1\leq i < j\leq n}\Big[\frac{\eps_n^2  \theta_i^2\zeta_j^2  \Pi_{iK}^2\Gamma_{jK}^2     }{\Omega_{ij}(1-\Omega_{ij})}+\frac{\eps_n^2  \theta_j^2\zeta_i^2  \Pi_{jK}^2\Gamma_{iK}^2     }{\Omega_{ji}(1-\Omega_{ji})}\Big]\times \sigma_i\sigma_j.
\]
Let $M^{(n)}_{ij}=\eps_n^2  \theta_i^2\zeta_j^2  \Pi_{iK}^2\Gamma_{jK}^2   \Omega_{ij}^{-1}(1-\Omega_{ij})^{-1}+\eps_n^2  \theta_j^2\zeta_i^2  \Pi_{jK}^2\Gamma_{iK}^2   \Omega_{ji}^{-1}(1-\Omega_{ji})^{-1}$, then $S_n=\sum_{1\leq i < j\leq n}M^{(n)}_{ij} \sigma_i\sigma_j$. By Proposition 8.13 in \cite{Foucart2013}, we have
\beq
\mathbb{P}(|S_n|\geq t)\leq 2 \exp\Big(-\min\Big\{\frac{3t^2}{32\|M^{(n)}\|^2_F},\frac{t}{8\|M^{(n)}\|}\Big\}\Big).
\eeq
Meanwhile,
\beq
\begin{split}
\|M^{(n)}\|^2\leq \|M^{(n)}\|_F^2=&2\sum_{ i < j}\Big[\frac{\eps_n^2  \theta_i^2\zeta_j^2  \Pi_{iK}^2\Gamma_{jK}^2     }{\Omega_{ij}(1-\Omega_{ij})}+\frac{\eps_n^2  \theta_j^2\zeta_i^2  \Pi_{jK}^2\Gamma_{iK}^2     }{\Omega_{ji}(1-\Omega_{ji})}\Big]^2\\
\leq & 4 \sum_{ i \neq j}\Big[\frac{\eps_n^2  \theta_i^2\zeta_j^2  \Pi_{iK}^2\Gamma_{jK}^2     }{\Omega_{ij}(1-\Omega_{ij})}\Big]^2.
\end{split}
\eeq
Notice $\Omega_{ij} = \theta_i\zeta_j\pi_i'P\gamma_j\geq \theta_i\zeta_j\Pi_{iK}\Gamma_{jK}$ and $0\leq \Pi_{iK},\Gamma_{iK}\leq 1$ for $1\leq i\leq n$, the above quantity is no more than \[
4\eps_n^4\sum_{i\neq j} \frac{\theta_i^2\zeta_j^2}{(1-\Omega_{ij})^2} \asymp \eps_n^4 \|\theta\|^2\|\zeta\|^2\asymp\rho_n^2=o(1),
\]
where we've used $\Omega_{ij}\leq \|P\|_{\max}\theta_{\max}\zeta_{\max}\goto0$ for $1\leq i, j \leq n$.  
Since $\mathbb{P}(|S_n|\geq t)\times \exp(t)\to 0$ as $t\to \infty$ for large $n$, we can apply the tail-sum formula and get
\beq
\begin{split}
\mathbb{E}_{\sigma}(\exp(|S_n|))&=1+\int_0^{\infty}\exp(t)\mathbb{P}(|S_n|> t)dt\\
&\leq 1+ \int_0^{\infty}2\exp\Big(t-\frac{3t^2}{32\|M^{(n)}\|^2_F}\Big)dt +\int_0^{\infty}2\exp\Big(t-\frac{t}{8\|M^{(n)}\|}\Big)dt=1+o(1),
\end{split}
\eeq
where the last step is from $\|M^{(n)}\|\leq \|M^{(n)}\|_F=o(1)$.
\[1\leq \int\left( \frac{d \mathbb{Q}_n }{d \mathbb{P}_n}\right)^2d\mathbb{P}_n \leq \mathbb{E}_{\sigma}(\exp(S_n))\leq \mathbb{E}_{\sigma}(\exp(|S_n|))=1+o(1) \]
completes the proof.

\section{A higher-order IBM test statistic} \label{supp:high-order}

The main paper focuses on the IBM test with $m=2$. It is also interesting to consider the higher-order IBM test statistics. 
In this section, we present and study the IBM test with $m=3$. We focus on the case where $A$ and $\widetilde{A}$ are symmetrical, and the asymmetrical case can be studied similarly. Same as before, write $A^* = A - \widetilde{A}$. Define
\beq \label{IBM(m=3)}
Q_n:= U_n^{(3)} = \sum_{i_1, i_2, i_3 (dist),\,  i_4, i_5, i_6 (dist)} A_{i_1 i_2}^*  A_{i_2 i_3}^*  A_{i_3 i_4}^*  A_{i_4 i_5}^*  
A_{i_5 i_6}^*  A_{i_6 i_1}^*, 
\eeq 
and 
\begin{align} \label{ICC(m=3)}
C_n&:= \sum_{i_1, i_2, i_3 (dist), \, i_4, i_5, i_6 (dist)} A_{i_1 i_2}  A_{i_2 i_3}  A_{i_3 i_4}  A_{i_4 i_5}  
A_{i_5 i_6}  A_{i_6 i_1},\cr
\widetilde{C}_n&:= \sum_{i_1, i_2, i_3 (dist),\,  i_4, i_5, i_6 (dist)} \widetilde{A}_{i_1 i_2} \widetilde{A}_{i_2 i_3} \widetilde{A}_{i_3 i_4}  \widetilde{A}_{i_4 i_5}  
\widetilde{A}_{i_5 i_6}  \widetilde{A}_{i_6 i_1}. 
\end{align}
Here, $Q_n$ is the Interlacing Balance Measure (IBM) statistic for $m=3$, and $C_n$ and $\widetilde{C}_n$ are the Interlacing Cycle Count (ICC) statistics for $m=3$. To save notations, we still denote them by $(Q_n, C_n, \widetilde{C}_n)$, but we keep in mind that they are different from $(Q_n, C_n, \widetilde{C}_n)$ in the main paper. We propose the following test statistic:
\beq \label{TestStat(m=3)}
\phi_n =(1/\sqrt{384})\cdot Q_n/(C_n+\widetilde{C}_n)^{1/2}. 
\eeq
Under some regularity conditions, $\phi_n\to N(0,1)$ under the null hypothesis $H_0^{(n)}:\Omega=\widetilde{\Omega}$. We will evaluate the numerical performance of this test in Section~\ref{supp:IBM-higher-order}.   

In what follows, we address two questions. The first is how to compute $\phi_n$ efficiently using matrix operations. In Section~\ref{supp(m=3)-comp}, we derive a lemma, which is related to computation and serves as an analogy of Lemma~\ref{lemma:comp} for $m=3$.
The second question is why a constant $1/\sqrt{384}$ appears in \eqref{TestStat(m=3)}. To answer this question, in Section~\ref{supp(m=3)-var}, we show that $384(C_n+\widetilde{C}_n)$ is a good  estimator of the variance of $Q_n$ under the null hypothesis.

\subsection{Computation of the test statistic when $m=3$} \label{supp(m=3)-comp}

To compute $\phi_n$, we introduce a matrix function: For any $n\times n$ symmetric matrix $X$, define 
\begin{align} \label{supp(m=3)-q(X)}
q(X)&= \mathrm{trace}(X^6)-6\cdot \mathrm{trace}(X^2\circ X^4) +3\cdot {\bf 1}_n' (X\circ X)^3 {\bf 1}_n\cr
& + 6\cdot \mathrm{trace}((X\circ X \circ X) X^3) +4\cdot \mathrm{trace}(X^2\circ X^2\circ X^2)\cr
&-12 \cdot \mathrm{trace}((X\circ X)^2\circ X^2) + 4\cdot {\bf 1}_n' (X\circ X \circ X \circ X \circ X \circ X) {\bf 1}_n. 
\end{align}
In the main paper, we introduced a function $q(X)$ for computation of $\psi_n$ ($m=2$). 
To save notation, we still call the above function $q(X)$, however, we keep in mind that it is different from $q(X)$ in the main paper. The following lemma is proved in Section~\ref{supp(m=3)-proofs}:
\begin{lemma} \label{lem(m=3):compu}
Let $(Q_n, C_n, \widetilde{C}_n)$ be as defined in \eqref{IBM(m=3)}-\eqref{ICC(m=3)}. It holds that $Q_n=q(A-\widetilde{A})$, $C_n=q(A)$, and $\widetilde{C}_n=q(\widetilde{A})$. 
\end{lemma}
Using Lemma~\ref{lem(m=3):compu}, we propose to compute $\phi_n$ by 
\beq \label{supp(m=3)-phi-compu}
\phi_n = (1/\sqrt{384})\cdot q(A-\widetilde{A})/[q(A)+q(\widetilde{A})]^{1/2}. 
\eeq
From \eqref{supp(m=3)-q(X)}, the function $q(\cdot)$ only involves matrix operations such as trace, matrix multiplication, and Hadamard product. 
The complexity is the same as the one for $m=2$.

\subsection{The variance estimator when $m=3$} \label{supp(m=3)-var}
A key step for designing a valid test statistic based on $Q_n$ is to estimate its variance under the null hypothesis. We show that a good variance estimator is 
\beq \label{supp(m=3)-var-estimator}
\widehat{\mathrm{Var}(Q_n)} = 384(C_n+\widetilde{C}_n)
\eeq

We now justify \eqref{supp(m=3)-var-estimator}. Write $W^*=W-\widetilde{W}$ and define
\begin{align*} \label{supp(m=3)-S}
S_n&: = \sum_{i_1, i_2, i_3 (dist),\, i_4, i_5, i_6 (dist)} W_{i_1 i_2}^*  W_{i_2 i_3}^*  W_{i_3 i_4}^*  W_{i_4 i_5}^*  
W_{i_5 i_6}^*  W_{i_6 i_1}^*, \cr
S_n^*&: =  \sum_{\substack{i_1, i_2, i_3, i_4, i_5, i_6 (dist)}} W_{i_1 i_2}^*  W_{i_2 i_3}^*  W_{i_3 i_4}^*  W_{i_4 i_5}^*  
W_{i_5 i_6}^*  W_{i_6 i_1}^*. 
\end{align*}
By our model, when $H_0^{(n)}$ holds, $Q_n=S_n$. We introduce $S_n^*$ as a proxy to $S_n$. The following lemma shows that their variances are close to each other. 
\begin{lemma} \label{lem(m=3):distinct}
Under the conditions of Theorem~\ref{thm:null}, $\mathrm{Var}(S_n^*)/\mathrm{Var}(S_n)\to 1$, as $n\to\infty$. 
\end{lemma}

\noindent
By Lemma~\ref{lem(m=3):distinct}, it suffices to calculate $\mathrm{Var}(S_n^*)$. We note that for different $(i_1, i_2, i_3, i_4, i_5, i_6)$ and $(i'_1, i'_2, i'_3, i'_4, i'_5, i'_6)$, their corresponding summands in $S_n^*$ may be the same. Here is an example: When $(i_1, i_2, i_3, i_4, i_5, i_6)=(1,2,3,4,5,6)$ and $(i'_1, i'_2, i'_3, i'_4, i'_5, i'_6)=(3,4,5,6,1,2)$, 
\[
W_{12}^*  W_{23}^*  W_{34}^*  W_{45}^*  
W_{56}^*  W_{61}^* = W_{34}^*  W_{45}^*  
W_{56}^*  W_{61}^*W_{12}^*  W_{23}^*. 
\]
Such cases are easily identified, as $(1,2,3,4,5,6)$ and $(3, 4, 5, 6, 1, 2)$ correspond to the same 6-cycle in a compete graph.  
Let $I_6(n)$ be the collection of all 6-cycles in a complete graph with $n$ nodes. 
Each element in $I_6(n)$ is associated with 12 distinct $(i_1, i_2, i_3, i_4, i_5, i_6)$ (we can describe this 6-cycle by starting from any one of the six indices and using either clockwise or counterclockwise direction; this gives $6\times 2=12$). 
Therefore, we can re-write
\beq \label{Sproxy-new}
S_n^* = 12 \sum_{(i_1, i_2, i_3, i_4, i_5, i_6)\in I_6(n)} W^*_{i_1i_2} W^*_{i_2i_3} W^*_{i_3i_4} W^*_{i_4i_5} W^*_{i_5i_6} W^*_{i_6i_1},
\eeq
where in the summation here, we only pick one out of the 12 distinct $(i_1, i_2, i_3, i_4, i_5, i_6)$ associated with the same 6-cycle.
Now, any two distinct summands in \eqref{Sproxy-new} are uncorrelated. 
 It follows that
\[
\mathrm{Var}(S_n^*) = 12^2\sum_{(i_1, i_2, i_3, i_4, i_5, i_6)\in I_6(n)} \mathrm{Var}(W^*_{i_1i_2} W^*_{i_2i_3} W^*_{i_3i_4} W^*_{i_4i_5} W^*_{i_5i_6} W^*_{i_6i_1}).
\]
In the proof of Theorem~\ref{thm:null}, we have shown that $\{W^*_{ij}\}_{1\leq i<j\leq n}$ are independent mean-zero variables, and under the null hypothesis, $\mathrm{Var}(W^*_{ij})=2\Omega_{ij}(1-\Omega_{ij})= 2\Omega_{ij}[1+o(1)]$.  
Hence, 
\begin{align*}  
\mathrm{Var}(S_n^*) & = 12^2 \sum_{(i_1, i_2, i_3, i_4, i_5, i_6)\in I_6(n)} 2^6[1+o(1)] \cdot \Omega_{i_1i_2} \Omega_{i_2i_3} \Omega_{i_3i_4} \Omega_{i_4i_5} \Omega_{i_5i_6} \Omega_{i_6i_1}\cr
& = 12^2\cdot (1/12) \sum_{i_1, i_2, i_3, i_4, i_5, i_6 (dist)} 2^6[1+o(1)] \cdot \Omega_{i_1i_2} \Omega_{i_2i_3} \Omega_{i_3i_4} \Omega_{i_4i_5} \Omega_{i_5i_6} \Omega_{i_6i_1}\cr
& = [1+o(1)]\cdot 768 \sum_{i_1, i_2, i_3, i_4, i_5, i_6 (dist)} \Omega_{i_1i_2} \Omega_{i_2i_3} \Omega_{i_3i_4} \Omega_{i_4i_5} \Omega_{i_5i_6} \Omega_{i_6i_1}\cr
& = [1+o(1)]\cdot 768 \, \mathbb{E}[C_n] \cr
& = [1+o(1)]\cdot 384\, ( \mathbb{E}[C_n] +\mathbb{E}[\widetilde{C}_n] ), 
\end{align*}
where the second line is because the summation is changed back to be over all $(i_1, i_2, i_3, i_4, i_5, i_6)$, which results in the factor of  $(1/12)$, and the last two lines are because $\mathbb{E}[C_n] = \mathbb{E}[\widetilde{C}_n] = \sum_{i_1, i_2, i_3, i_4, i_5, i_6 (dist)} \Omega_{i_1i_2} \Omega_{i_2i_3} \Omega_{i_3i_4} \Omega_{i_4i_5} \Omega_{i_5i_6} \Omega_{i_6i_1}$. 
In addition, the $o(1)$ term can be moved out of the summation, because we can bound the aggregated effect of $\{|\mathrm{Var}(W^*_{ij})-2\Omega_{ij}|\}_{1\leq i<j\leq n}$ (similarly as in the proof of Theorem~\ref{thm:null}). Combining the above with Lemma~\ref{lem(m=3):distinct} gives
\beq \label{supp(m=3)-Var}
\mathrm{Var}(Q_n)=\mathrm{Var}(S_n)\sim 384( \mathbb{E}[C_n] +\mathbb{E}[\widetilde{C}_n] ), \qquad\mbox{under }H_0^{(n)}. 
\eeq
This justifies the variance estimator in \eqref{supp(m=3)-var-estimator}.

\subsection{Proofs of the supplementary lemmas for $m=3$} \label{supp(m=3)-proofs}

\subsubsection{Proof of Lemma~\ref{lem(m=3):compu}}
For any symmetric matrix $X$ and six indices $(i_1, i_2, i_3, i_4, i_5, i_6)$, write for short $h_{i_1i_2i_3i_4i_5i_6}(X):= X_{i_1i_2}X_{i_2i_3}X_{i_3i_4}X_{i_4i_5}X_{i_5i_6}X_{i_6i_1}$. The goal of this lemma is showing 
\beq \label{lem-m3compu-0}
\sum_{i_1, i_3, i_5 (dist),\, i_2, i_4, i_6 (dist) }h_{i_1i_2i_3i_4i_5i_6}(X)=q(X), \qquad\mbox{for any symmetric matrix $X$}. 
\eeq
For any $1\leq j\neq k\leq 6$, let
$
T_{jk}=\bigl\{
(i_1, i_2, i_3, i_4, i_5, i_6): \; i_j=i_k \bigr\}$. 
Write 
\beq \label{lem-m3compu-1add}
T=T_{13}\cup T_{15}\cup T_{35}\cup T_{24}\cup T_{26}\cup T_{46}.
\eeq
It is easy to see that the left hand side of \eqref{lem-m3compu-0} is equal to $\mathrm{trace}(X^6)-\zeta(X)$, where 
\beq \label{lem-m3compu-1}
\zeta(X):= \sum_{(i_1,i_2,i_3,i_4,i_5,i_6)\in T}h_{i_1i_2i_3i_4i_5i_6}(X). 
\eeq
The proof reduces to simplifying the expression of $\zeta(X)$. For notational simplicity, we omit $(i_1,i_2,i_3,i_4,i_5,i_6)$ in describing the sum; e.g., we write $\zeta(X)=\sum_{T}h_{i_1i_2i_3i_4i_5i_6}(X)$. 

We now study $\zeta(X)$. Since $T_{jk}$'s have intersections with each other, 
\[
\zeta(X)\;\; \neq\;\; \Bigl(\sum_{T_{13}}+\sum_{T_{15}}+\sum_{T_{35}}+ \sum_{T_{24}}+\sum_{T_{26}}+\sum_{T_{46}}\Bigr) h_{i_1i_2i_3i_4i_5i_6}(X). 
\]
To proceed with the calculations, we introduce a few notations. 
Define an index set $\mathbb{I}=\{(1,3), (1,5), (3,5), (2,4), (2,6), (4,6)\}$, so that we can re-express $T$ as $T=\cup_{(j,k)\in \mathbb{I}}\, T_{jk}$. For each $1\leq \ell\leq 6$, let $S_{\ell}$ be the collection of size-$\ell$ subsets of $\mathbb{I}$. Then, $S_1=\mathbb{I}$, and $S_2$ has ${6\choose 2}=15$ elements, and each element is a size-2 subset, such as $\{(1,3), (1,5)\}$, $\{(1,3), (3,5)\}$, and so on. For an element $I\in S_2$, we use $T(I)$ to denote the intersection of all $T_{jk}$'s with $(j,k)\in I$. For example, when $I=\{(1,3), (1,5)\}$, we have $T(I)=T_{13}\cap T_{15}$, consisting of all $(i_1, i_2, i_3, i_4, i_5, i_6)$ such that $i_1=i_3$ and $i_1=i_5$. Similarly, for any $I\in S_3$, we can define $T(I)$ in the same way; e.g., when $I=\{(1,3), (1,5), (2,4)\}$, $T(I)=T_{13}\cap T_{15}\cap T_{24}$.  Define
\beq \label{lem-m3compu-2}
P_{\ell}(X):=\sum_{I\in S_{\ell}}\, \sum_{ (i_1, i_2, i_3, i_4, i_5, i_6)\in T(I)} h_{i_1i_2i_3i_4i_5i_6}(X), \qquad 1\leq \ell\leq 6. 
\eeq
For example, 
\begin{align*}
P_1(X) &= \Bigl(\sum_{T_{13}}+\sum_{T_{15}}+\ldots+\sum_{T_{46}}\Bigr) h_{i_1i_2i_3i_4i_5i_6}(X),\cr
P_2(X) &=  \Bigl(\sum_{T_{13}\cap T_{15}}+\sum_{T_{13}\cap T_{35}}+\ldots + \sum_{T_{24}\cap T_{46}}\Bigr) h_{i_1i_2i_3i_4i_5i_6}(X),\cr
& \quad\vdots\cr
P_6 (X) &= \sum_{T_{13}\cap T_{15}\cap T_{35}\cap T_{24}\cap T_{26}\cap T_{46}} h_{i_1i_2i_3i_4i_5i_6}(X). 
\end{align*}
Using the inclusion-exclusion principle for set unions, we have
\beq \label{lem-m3compu-3}
\zeta(X) = P_1(X)-P_2(X)+P_3(X)-P_4(X)+P_5(X)-P_6(X). 
\eeq
It remains to compute $P_{\ell}(X)$ for $1\leq \ell\leq 6$. Consider $P_1(X)$. It is easy to see that 
\begin{align*}
\sum_{T_{13}}& h_{i_1i_2i_3i_4i_5i_6}(X)=\sum_{i_1, i_2, i_4, i_5, i_6} X_{i_1i_2}X_{i_2i_1}X_{i_1i_4}X_{i_4i_5}X_{i_5i_6}X_{i_6i_1}\cr
&=\sum_{i_1} (X^2)_{i_1i_1} (X^4)_{i_1i_1} = \mathrm{trace}(X^2\circ X^4). 
\end{align*}
By symmetry, $\sum_{T_{jk}}h_{i_1i_2i_3i_4i_5i_6}(X)$ is the same for all $(j,k)\in \mathbb{I}$. It follows that
\beq \label{lem-m3compu-P1}
P_1(X) = 6\cdot \mathrm{trace}(X^2\circ X^4). 
\eeq
Consider $P_2(X)$. There are three different cases. Here we describe a geometric perspective that help clarify things. Let us think $i_j$ as the vertexes of a six pointed star. Each element in $P_2(X)$ can be thought as connecting two pair of vertexes on the six pointed star and the connected vertexes share the same value. There are only three distinctive shapes that can be formed by two edges - two consecutive edges of a equilateral triangle,  a cross sign, and two paralleled lines. In total, they compose all 15 elements in $S_2$.

\begin{itemize}
\item Case 1: $T_{13}\cap T_{15}$. In this case, the formed shape is two consecutive edges of an equilateral triangle. There are 6 such elements in $S_2$, since each vertex of the star can define one. These 6 elements are equivalent under rotation. 

\item Case 2:  $T_{13}\cap T_{24}$. In this case, the formed shape is a cross. There are 6 such elements, since each short edge of the star can define one.

\item Case 3: $T_{13}\cap T_{46}$. In this case, the formed shape is two paralleled lines, and apparently there are 3 such elements.
\end{itemize}
According to the above analysis, 
\[
P_2(X) = 6\sum_{T_{13}\cap T_{15}} h_{i_1i_2i_3i_4i_5i_6}(X) + 6\sum_{T_{13}\cap T_{24}} h_{i_1i_2i_3i_4i_5i_6}(X) + 3\sum_{T_{13}\cap T_{46}} h_{i_1i_2i_3i_4i_5i_6}(X). 
\]
By direct calculations,
\begin{align*}
\sum_{T_{13}\cap T_{15}} h_{i_1\cdots i_6}(X) &= \sum_{i_1, i_2, i_4, i_6}X^2_{i_1i_2}X^2_{i_1i_4}X^2_{i_1i_6} = \sum_{i_1}\bigl[(X^2)_{i_1i_1}\bigr]^3 = \mathrm{trace}(X^2\circ X^2\circ X^2),\cr
\sum_{T_{13}\cap T_{24}} h_{i_1\cdots i_6}(X) &= \sum_{i_1, i_2, i_5, i_6}X^3_{i_1i_2}X_{i_2i_5}X_{i_5i_6}X_{i_6i_1}  = \sum_{i_1, i_2}X^3_{i_1i_2}(X^3)_{i_2i_1} = \mathrm{trace}((X\circ X\circ X) X^3), \cr
\sum_{T_{13}\cap T_{46}} h_{i_1\cdots i_6}(X) &= \sum_{i_1, i_2, i_4, i_5}X^2_{i_2i_1}X^2_{i_1i_4}X^2_{i_4i_5} = \sum_{i_2,i_5} [(X\circ X)^3]_{i_2i_5} = {\bf 1}_n'(X\circ X)^3{\bf 1}_n. 
\end{align*}
We combine the above expressions to get
\begin{align} \label{lem-m3compu-P2}
P_2(X) &= 6\cdot \mathrm{trace}(X^2\circ X^2\circ X^2) + 6\cdot \mathrm{trace}((X\circ X\circ X) X^3)\cr
&\qquad +3\cdot {\bf 1}_n'(X\circ X)^3{\bf 1}_n. 
\end{align}
Consider $P_3(X)$. Note that $S_3$ has ${6 \choose 3}=20$ elements. Case 1: $T_{13}\cap T_{15}\cap T_{35}$. In this case, $i_1=i_3=i_5$. A similar situation happens at $T_{24}\cap T_{26}\cap T_{46}$ (hence, a total of 2 elements in $S_3$ belong to this case). Case 2: $(T_{13}\cap T_{15})\cap T_{24}$. Since $T_{13}\cap T_{15}$ already implies  $i_1=i_3=i_5$,  it does not matter if we change $T_{24}$ to $T_{26}$ or $T_{46}$; similarly, we can change $(T_{13}\cap T_{15})$ to either $(T_{13}\cap T_{35})$ or $(T_{15}\cap T_{35})$; finally, we can always swap the roles of $(i_1, i_3, i_5)$ and $(i_2, i_4, i_6)$. It implies that all the remaining 18 elements in $S$ belong to this case. The analysis yields
\[
P_3(X) = 2\sum_{T_{13}\cap T_{15}\cap T_{35}} h_{i_1i_2i_3i_4i_5i_6}(X) + 18 \sum_{T_{13}\cap T_{15}\cap T_{24}} h_{i_1i_2i_3i_4i_5i_6}(X). 
\]
By direct calculations,
\begin{align*}
\sum_{T_{13}\cap T_{15}\cap T_{35}} h_{i_1\cdots i_6}(X) &= \sum_{T_{13}\cap T_{15}} h_{i_1\cdots i_6}(X)  = \mathrm{trace}(X^2\circ X^2\circ X^2),\cr
\sum_{T_{13}\cap T_{15}\cap T_{24}}  h_{i_1\cdots i_6}(X) &= \sum_{i_1, i_2, i_6}X^4_{i_1i_2}X^2_{i_1i_6}  = \sum_{i_1} [(X\circ X)^2]_{i_1i_1}(X^2)_{i_1i_1} = \mathrm{trace}((X\circ X)^2\circ X^2). 
\end{align*}
As a result, 
\beq \label{lem-m3compu-P3}
P_3(X) = 2\cdot \mathrm{trace}(X^2\circ X^2\circ X^2) + 18\cdot \mathrm{trace}((X\circ X)^2\circ X^2).
\eeq
Consider $P_4(X)$. There are only two cases. In Case 1, we pick two out of $\{(1,3), (1,5), (3,5)\}$ and two from $\{(2,4), (2,6), (4,6)\}$.  It doesn't matter which two are picked from $\{(1,3), (1,5), (3,5)\}$, because all the choices yield $i_1=i_3=i_5$. This covers a total $3\times 3=9$ elements in $S_4$. In Case 2, we either select all three in $\{(1,3), (1,5), (3,5)\}$ and one out of $\{(2,4), (2,6), (4,6)\}$, or we select one in $\{(1,3), (1,5), (3,5)\}$ and all three in $\{(2,4), (2,6), (4,6)\}$. This covers a total of $2\times 3=6$ elements in $S_4$. It follows that 
\begin{align} \label{lem-m3compu-P4}
P_4(X) &= 9\sum_{T_{13}\cap T_{15}\cap T_{24}\cap T_{46}} h_{i_1i_2i_3i_4i_5i_6}(X) + 6 \sum_{T_{13}\cap T_{15}\cap T_{35}\cap T_{24}} h_{i_1i_2i_3i_4i_5i_6}(X) \cr 
&= 9 \sum_{i_1, i_2}X_{i_1i_2}^6 + 6\sum_{T_{13}\cap T_{15}\cap T_{24}}h_{i_1i_2i_3i_4i_5i_6}(X) \cr
&= 9\cdot {\bf 1}_n'[X\circ X\circ X\circ X\circ X\circ X]{\bf 1}_n + 6\cdot \mathrm{trace}((X\circ X)^2\circ X^2),
\end{align}
where in the last line we note that $\sum_{T_{13}\cap T_{15}\cap T_{24}}h_{i_1i_2i_3i_4i_5i_6}(X)$ has already been calculated in the analysis of $P_3(X)$. Finally, $P_5(X)$ and $P_6(X)$ are straightforward to calculate. 
\begin{align} \label{lem-m3compu-P56}
P_5(X) &= 6 \sum_{T_{13}\cap T_{15}\cap T_{35}\cap T_{24}\cap T_{26}} h_{i_1i_2i_3i_4i_5i_6}(X)\cr
&= 6 \sum_{i_1, i_2}X_{i_1i_2}^6 = 6\cdot {\bf 1}_n'[X\circ X\circ X\circ X\circ X\circ X]{\bf 1}_n, \cr
P_6(X) &= \sum_{T_{13}\cap T_{15}\cap T_{35}\cap T_{24}\cap T_{26}\cap T_{46}} h_{i_1i_2i_3i_4i_5i_6}(X)\cr
&= \sum_{i_1, i_2}X_{i_1i_2}^6 = {\bf 1}_n'[X\circ X\circ X\circ X\circ X\circ X]{\bf 1}_n. 
\end{align}
We plug the expressions of $P_1(X), P_2(X), \ldots, P_6(X)$ into \eqref{lem-m3compu-3} to obtain $\zeta(X)$. The claim then follows immediately. \qed

\subsubsection{Proof of Lemma~\ref{lem(m=3):distinct}}
Same as in the proof of Lemma~\ref{lem(m=3):compu}, write $h_{i_1i_2i_3i_4i_5i_6}(X):= X_{i_1i_2}X_{i_2i_3}X_{i_3i_4}X_{i_4i_5}X_{i_5i_6}X_{i_6i_1}$, for any symmetric matrix $X$. 
In Section~\ref{supp(m=3)-var} (the equation above \eqref{supp(m=3)-Var}), we have shown:
\[
\mathrm{Var}(S_n^*) = [1+o(1)]\cdot 768 \sum_{i_1, i_2, i_3, i_4, i_5, i_6 (dist)} h_{i_1i_2i_3i_4i_5i_6}(\Omega). 
\]
Under the null hypothesis, $\Omega=\widetilde{\Omega}=\Theta\Pi P\Pi'\Theta$. In particular, $\Omega_{ij}\leq C\theta_i\theta_j$, and $h_{i_1i_2i_3i_4i_5i_6}(\Omega)\leq C\theta_{i_1}^2\theta_{i_2}^2\theta_{i_3}^2\theta_{i_4}^2\theta_{i_5}^2\theta_{i_6}^2$. 
It follows that
\begin{align*}
\Bigl|\mathrm{trace}(\Omega^6) - \sum_{i_1, i_2, i_3, i_4, i_5, i_6 (dist)} h_{i_1i_2i_3i_4i_5i_6}(\Omega)\Bigr|  \leq C\sum_{i_1=i_2}\theta_{i_1}^2\theta_{i_2}^2\theta_{i_3}^2\theta_{i_4}^2\theta_{i_5}^2\theta_{i_6}^2\leq C\|\theta\|_4^4\|\theta\|^8. 
\end{align*}
At the same time, $\mathrm{trace}(\Omega^6)\geq C^{-1}\|\theta\|^{12}$. Under our assumptions, $\theta_{\max}\to 0$ and $\|\theta\|\to\infty$. It follows that  $\|\theta\|_4^4\|\theta\|^8\leq \theta_{\max}^2\|\theta\|^2\cdot\|\theta\|^8=o(\|\theta\|^{12})$. Combining these results gives
\beq  \label{lem-m3distinct-1}
\mathrm{Var}(S_n^*) \geq [1+o(1)]\cdot \mathrm{trace}(\Omega^6)\geq  C^{-1}\|\theta\|^{12}. 
\eeq
To show the claim, it suffices to show that
\beq  \label{lem-m3distinct-2}
\mathrm{Var}(S_n-S_n^*) =o(\|\theta\|^{12}). 
\eeq

We now show \eqref{lem-m3distinct-2}. Recall that $S_n$ is the sum of $h_{i_1i_2i_3i_4i_5i_6}(W^*)$, over six-tuples such that $(i_1, i_3, i_5)$ are distinct and $(i_2, i_4, i_6)$ are distinct. In $S^*_n$, it requires that all six indices are distinct. The difference between $S_n$ and $S_n^*$ is from cases of $i_1=i_2$, $i_1=i_4$, ..., $i_5=i_6$.  We note that $W^*_{ii}=0$ for $1\leq i\leq n$. Therefore, when $i_1=i_2$ or $i_1=i_6$, $h_{i_1i_2i_3i_4i_5i_6}(W^*)=0$. This implies that $i_1$ can only equal $i_4$. 
Similarly, $i_3$ can only equal $i_6$, and $i_5$ can only equal  $i_2$. We define $T_{jk}=\{
(i_1, i_2, i_3, i_4, i_5, i_6): i_j=i_k \}$ same as in the proof of Lemma~\ref{lem(m=3):compu}. Then, 
\[
S_n-S_n^* = \sum_{(j,k)\in \mathbb{I}^*} h_{i_1i_2i_3i_4i_5i_6}(W^*), \qquad \mbox{where }\mathbb{I}^*=\{(1,4), (2, 5), (3, 6)\}. 
\]
Here, $S_n-S_n^*$ has a similar form as the $\zeta(X)$ in \eqref{lem-m3compu-1}, except that $\mathbb{I}$ is replaced by $\mathbb{I}^*$. We similarly define $P^*_{\ell}$ as in \eqref{lem-m3compu-2}, except that $\mathbb{I}$ is replaced by $\mathbb{I}^*$.
Again, it follows from the inclusion-exclusion principle that 
\[
S_n-S_n^* = P^*_1 - P_2^* + P_3^*. 
\]
Therefore, to show \eqref{lem-m3distinct-2}, it suffices to show that
\beq  \label{lem-m3distinct-3}
\mathrm{Var}(P^*_{\ell}) =o(\|\theta\|^{12}), \qquad\mbox{for }1\leq \ell\leq 3. 
\eeq
We now show \eqref{lem-m3distinct-3}. Using similar analysis as in the proof of Lemma~\ref{lem(m=3):compu}, we see that
\begin{align*}
P_1^*&=3\sum_{T_{14}}h_{i_1i_2i_3i_4i_5i_6}(W^*) = 3\sum_{i_1,i_2,i_3,i_5,i_6 (dist)} W^*_{i_1i_2} W^*_{i_2i_3} W^*_{i_3i_1} W^*_{i_1i_5} W^*_{i_5i_6} W^*_{i_6i_1},\cr
P_2^*&=3\sum_{T_{14}\cap T_{36}}h_{i_1i_2i_3i_4i_5i_6}(W^*) = 3\sum_{i_1, i_2, i_3, i_5 (dist)} W^*_{i_1i_2} W^*_{i_2i_3} (W^*_{i_3i_1})^2 W^*_{i_1i_5} W^*_{i_5i_3},\cr
P_3^*&=\sum_{T_{14}\cap T_{36}\cap T_{25}}h_{i_1i_2i_3i_4i_5i_6}(W^*) = \sum_{i_1, i_2, i_3 (dist)} (W^*_{i_1i_2})^2 (W^*_{i_2i_3})^2 (W^*_{i_3i_1})^2. 
\end{align*}
Here, we put ``(dist)" in the summation because $W^*_{ii}=0$ for all $1\leq i\leq n$. 
To compute the variance of $P_1^*$, write $\eta_{i_1i_2i_3i_5i_6}(W^*):=W^*_{i_1i_2} W^*_{i_2i_3} W^*_{i_3i_1} W^*_{i_1i_5} W^*_{i_5i_6} W^*_{i_6i_1}$. It is associated with a geometric object in a complete graph -  a pentagon with five vertices $(i_1,i_2,i_3,i_5,i_6)$ plus a cross line between vertex $i_1$ and vertex $i_3$. 
We see that $\eta_{i_1i_2i_3i_5i_6}(W^*)$ and $\eta_{j_1j_2j_3j_5j_6}(W^*)$ are correlated if and only if the two corresponding geometric objects are exactly the same (including but not limited to the case of $(i_1, i_2, i_3, i_5, i_6)=(j_1, j_2, j_3, j_5, j_6)$). An important observation is that each  $\eta_{i_1i_2i_3i_5i_6}(W^*)$ can only be correlated with finitely many $\eta_{j_1j_2j_3j_5j_6}(W^*)$. As a result, 
\begin{align*}
\mathrm{Var}(P^*_1) & \leq C \sum_{i_1, i_2, i_3, i_5, i_6 (dist)} \mathrm{Var}(\eta_{i_1i_2i_3i_5i_6}(W^*))  \leq C \sum_{i_1, i_2, i_3, i_5, i_6} \Omega_{i_1i_2}  \Omega_{i_2i_3}  \Omega_{i_3i_1}   \Omega_{i_1i_5}  \Omega_{i_5i_6}  \Omega_{i_6i_1}  \cr
& \leq C \sum_{i_1, i_2, i_3, i_5, i_6} \theta_{i_1}^4 \theta_{i_2}^2 \theta_{i_3}^2 \theta_{i_5}^2 \theta_{i_6}^2\leq C\|\theta\|_4^4\|\theta\|^8 = o(\|\theta\|^{12}). 
\end{align*}
Similarly, for each summand in $P^*_2$, it is correlated with only finitely many other summands. We thus have
\begin{align*}
\mathrm{Var}(P^*_2) & \leq C \sum_{i_1, i_2, i_3, i_5 (dist)} \mathbb{E}(W^*_{i_1i_2} W^*_{i_2i_3} (W^*_{i_3i_1})^2 W^*_{i_1i_5} W^*_{i_5i_3})^2\cr
& \leq C \sum_{i_1, i_2, i_3, i_5} \Omega_{i_1i_2}  \Omega_{i_2i_3} \Omega_{i_1i_3}   \Omega_{i_1i_5}  \Omega_{i_5i_3}   \leq C \sum_{i_1, i_2, i_3, i_5} \theta_{i_1}^3 \theta_{i_2}^2 \theta_{i_3}^3 \theta_{i_5}^2\\
& \leq C\|\theta\|_3^6\|\theta\|^4 = o(\|\theta\|^{12}),
\end{align*}
where in the second line we used $\mathbb{E}[((W^*)_{ij})^4]\leq C\Omega_{ij}$. 
For $P^*_3$, we note that that it has a nonzero mean. Let $B_{ij}=\mathbb{E}[(W^*_{ij})^2]$ and $Z_{ij}=(W^*_{ij})^2-B_{ij}$.  
It is seen that
\[
P^*_3-\mathbb{E}[P^*_3] =  \sum_{i_1, i_2, i_3 (dist)} \bigl[(B_{i_1i_2}+Z_{i_1i_2})(B_{i_2i_3}+Z_{i_2i_3})(B_{i_3i_1}+Z_{i_3i_1})-B_{i_1i_2}B_{i_2i_3}B_{i_3i_1}\bigr].
\]
We note that the summand decomposes into $2^3-1=7$ terms. This yields a decomposition of $P^*_3-\mathbb{E}[P^*_3]$ into 7 terms. It is not hard to verify that the leading term in $\mathrm{Var}(P_3^*)$ comes from the variance of
\[
\sum_{i_1, i_2, i_3 (dist)}Z_{i_1i_2}B_{i_2i_3}B_{i_3i_1} = \sum_{i_1, i_2 (dist)} \Bigl(\sum_{i_3\notin\{i_1,i_2\}}B_{i_2i_3}B_{i_3i_1}\Bigr)Z_{i_1i_3}.
\]
Since $B_{ij}\leq C\Omega_{ij}\leq C\theta_i\theta_j$, we have $\sum_{i_3\notin\{i_1,i_2\}}B_{i_2i_3}B_{i_3i_1}\leq \sum_{i_3}C\theta_{i_2}\theta_{i_3}^2\theta_{i_1}\leq C\|\theta\|^2\theta_{i_1}\theta_{i_2}$. In addition, $\{Z_{i_1i_2}\}_{1\leq i_1<i_2\leq n}$ are independent mean-zero variables, with $\mathrm{Var}(Z_{i_1i_2})\leq \Omega_{i_1i_2}\leq C\theta_{i_1}\theta_{i_2}$. Combining these results, we obtain:
\begin{align*}
\mathrm{Var}(P^*_3) & \leq C \sum_{i_1, i_2 (dist)}(\|\theta\|^2\theta_{i_1}\theta_{i_2})^2\cdot \mathrm{Var}(Z_{i_1i_2})
\leq C\sum_{i_1,i_2} \|\theta\|^4\theta_{i_1}^3\theta_{i_2}^3\cr
&\leq C\|\theta\|^4\|\theta\|_3^6=o(\|\theta\|^{12}). 
\end{align*}
So far, we have studied the variances of $P_1^*$, $P_2^*$ and $P_3^*$, and proved \eqref{lem-m3distinct-3}. It then implies \eqref{lem-m3distinct-2}. This completes the proof. \qed

\section{Additional numerical results} \label{supp:numerical}

In this section, we report some additional numerical results not included in the main paper owing to the space limit.

\subsection{The IBM test statistics for the Enron network} \label{supp:IBM}
In Section~\ref{sec:real} of the main text,  we applied the IBM test to compare the Enron email networks for different time periods. The p-value heat map is shown in Figure~\ref{fig:Enron} there. We now report the values of the IBM test statistics, as a supplement to Figure~\ref{fig:Enron}. These test statistic values are displayed in Table~\ref{tb:Enron-supp}

\begin{table}[htb]
\centering
\spacingset{1}
\caption{The IBM test statistics for the Enron network (a heat map visualization of the corresponding p-values is in Figure~\ref{fig:Enron} of the main paper).} \label{tb:Enron-supp}
\spacingset{1.45}
\vspace{1em}
\addtolength{\tabcolsep}{-4pt}
\scalebox{0.4}{
\begin{tabular}{r|cccccccccccc|cccccccccccc|cccccccccccc|cccccc}
  \hline
 & 99Jan & Feb & Mar & Apr & May & Jun & Jul & Aug & Sep & Oct & Nov & Dec & 00Jan & Feb & Mar & Apr & May & Jun & Jul & Aug & Sep & Oct & Nov & Dec & 01Jan & Feb & Mar & Apr & May & Jun & Jul & Aug & Sep & Oct & Nov & Dec & 02Jan & Feb & Mar & Apr & May & Jun\\ 
  \hline
99Jan & -- & &&&&&&&&&&&&&&&&&&&&&&&&&&&&&&&&&&& \\ 
Feb & 0.00 & -- &  & &&&&&&&&&&&&&&&&&&&&&&&&&&&&&&&&&&& \\ 
Mar & 0.00 & 0.00 & -- &  & &&&&&&&&&&&&&&&&&&&&&&&&&&&&&&&&& \\ 
Apr & 0.00 & 0.00 & 0.00 & -- &  & &&&&&&&&&&&&&&&&&&&&&&&&&&&&&&& \\ 
May & 0.53 & 0.18 & -0.18 & 0.00 &  -- &  & &&&&&&&&&&&&&&&&&&&&&&&&&&&&&&& \\ 
Jun & 0.00 & 0.16 & 0.16 & 0.32 & 0.00 &  -- &  & &&&&&&&&&&&&&&&&&&&&&&&&&&&&&&& \\ 
Jul & 0.25 & 0.37 & 0.00 & 0.37 & 0.20 & 0.00 &  -- &  & &&&&&&&&&&&&&&&&&&&&&&&&&&&&&&& \\ 
Aug & 0.32 & 0.49 & 0.00 & 0.73 & 0.07 & -0.14 & 0.00 & -- &  & &&&&&&&&&&&&&&&&&&&&&&&&&&&&&&&\\ 
Sep & 0.22 & 1.58 & 0.87 & 1.31 & 0.73 & 0.52 & 0.45 & 0.45 & -- &  & &&&&&&&&&&&&&&&&&&&&&&&&&&&&&&&\\ 
Oct & 0.28 & 1.39 & 0.92 & 1.43 & 0.00 & 0.11 & -0.04 & -0.04 & -0.04 &  -- &  & &&&&&&&&&&&&&&&&&&&&&&&&&\\
Nov & 0.58 & 0.51 & 0.44 & 0.66 & -0.07 & -0.13 & -0.06 & 0.00 & 0.09 & 0.00 & -- &  & &&&&&&&&&&&&&&&&&&&&&&&&&&\\  
Dec & 1.12 & 2.14 & 1.66 & 2.36 & 0.27 & 0.50 & 0.12 & 0.18 & 0.32 & 0.39 & 0.15 &  -- &  & &&&&&&&&&&&&&&&&&&&&&&&&&&\\ 
\hline
00Jan & 2.54 & 2.90 & 2.30 & 2.79 & 0.79 & 1.27 & 0.42 & 0.71 & 0.56 & 0.73 & 0.68 & 0.09  &-- &  & &&&&&&&&&&&&&&&&&&&&&&&\\
Feb & 2.99 & 3.34 & 3.14 & 3.20 & 2.32 & 2.14 & 1.68 & 2.26 & 1.98 & 2.14 & 2.05 & 0.98 & 0.60 &-- &  & &&&&&&&&&&&&&&&&&&&&&&&\\
Mar & 1.73 & 1.78 & 1.61 & 1.68 & 1.02 & 0.97 & 0.76 & 1.33 & 0.68 & 0.46 & 0.70 & -0.04 & 1.53 & -0.06 &-- &  & &&&&&&&&&&&&&&&&&&&&&&&\\
Apr & 1.30 & 1.51 & 1.35 & 1.68 & 0.76 & 0.61 & 0.54 & 0.63 & 1.02 & 0.33 & 0.48 & -0.03 & 1.01 & -0.29 & 0.50 &-- &  & &&&&&&&&&&&&&&&&&&&&&&&\\
May & 2.85 & 1.98 & 1.82 & 1.92 & 1.85 & 1.44 & 1.73 & 1.67 & 1.87 & 1.43 & 1.22 & 0.91 & 1.56 & 1.17 & 0.78 & 0.04 &-- &  & &&&&&&&&&&&&&&&&&&&\\
Jun & 2.46 & 2.81 & 2.42 & 2.81 & 2.48 & 2.00 & 1.73 & 1.88 & 1.72 & 1.59 & 1.52 & 0.47 & 0.95 & 1.35 & 0.04 & 0.44 & 0.18 &-- &  & &&&&&&&&&&&&&&&&&&&\\
Jul & 4.48 & 4.65 & 4.40 & 4.63 & 3.73 & 3.97 & 3.62 & 3.61 & 3.74 & 3.10 & 3.17 & 1.86 & 1.76 & 1.27 & 0.84 & 1.12 & 1.04 & 1.22 &-- &  & &&&&&&&&&&&&&&&&&&&\\
Aug & 8.38 & 8.60 & 8.30 & 8.69 & 7.85 & 7.76 & 7.60 & 7.65 & 7.62 & 7.24 & 7.43 & 5.26 & 4.76 & 5.64 & 2.95 & 3.51 & 1.81 & 1.63 & 0.40 &-- &  & &&&&&&&&&&&&&&&&&&&\\ 
Sep & 5.94 & 6.33 & 6.08 & 6.45 & 5.04 & 5.34 & 4.69 & 4.49 & 4.27 & 3.81 & 3.90 & 2.06 & 1.99 & 2.38 & 1.38 & 1.59 & 1.01 & 1.02 & 0.25 & 0.77 &-- &  & &&&&&&&&&&&&&&&&&&& \\ 
Oct & 8.81 & 9.24 & 9.01 & 9.25 & 7.83 & 8.03 & 7.52 & 7.71 & 7.39 & 6.80 & 6.63 & 4.77 & 5.76 & 4.41 & 3.15 & 2.35 & 1.12 & 0.93 & 0.36 & 0.15 & 0.60 &-- &  & &&&&&&&&&&&&\\ 
Nov & 7.44 & 8.45 & 8.17 & 8.37 & 7.18 & 7.28 & 6.74 & 6.94 & 6.45 & 5.70 & 6.17 & 3.89 & 4.37 & 1.97 & 2.06 & 1.58 & 1.38 & 0.48 & 0.23 & 0.34 & 1.06 & 0.15 &-- &  & &&&&&&&&&&&&&&&&\\
Dec & 9.45 & 10.02 & 9.76 & 10.00 & 9.15 & 9.18 & 8.48 & 8.98 & 8.54 & 7.47 & 7.71 & 5.16 & 5.78 & 4.88 & 3.14 & 3.92 & 2.00 & 0.81 & 0.67 & 0.83 & 0.48 & 0.96 & 0.40 &-- &  & &&&&&&&&&&&\\ 
\hline
01Jan & 10.2 & 10.4 & 10.3 & 10.4 & 9.69 & 9.77 & 9.35 & 9.50 & 9.37 & 8.60 & 8.70 & 7.03 & 6.89 & 5.75 & 4.53 & 4.35 & 2.35 & 1.07 & 1.32 & 0.81 & 1.47 & 0.45 & 0.54 & 0.76 &-- &  & &&&&&&&&&&&&&&\\
Feb & 5.60 & 6.89 & 6.44 & 7.00 & 5.34 & 5.41 & 4.62 & 4.70 & 4.45 & 3.96 & 4.24 & 2.87 & 3.63 & 2.60 & 1.41 & 1.72 & 1.49 & 1.03 & 1.01 & 0.92 & 0.60 & 0.53 & 0.45 & 1.23 & 1.31 &-- &  & &&&&&&&&&&&\\ 
 Mar & 9.75 & 10.4 & 10.01 & 10.37 & 8.21 & 8.64 & 7.12 & 7.50 & 7.26 & 6.95 & 6.43 & 5.13 & 5.53 & 4.12 & 3.57 & 3.16 & 2.66 & 1.87 & 1.30 & 1.27 & 0.57 & 0.82 & 0.66 & 0.66 & 0.85 & 0.70 &-- &  & &&&&&&&&&&&\\ 
Apr & 15.0 & 15.3 & 15.2 & 15.2 & 14.2 & 13.5 & 12.2 & 13.9 & 13.7 & 12.6 & 12.2 & 8.94 & 11.31 & 6.01 & 4.94 & 6.43 & 5.32 & 4.50 & 2.31 & 2.22 & 1.98 & 1.62 & 1.37 & 1.36 & 2.33 & 2.56 & 0.78 &-- &  & &&&&&&&&&&&\\  
May & 13.9 & 14.2 & 14.0 & 14.2 & 13.3 & 13.0 & 12.4 & 13.0 & 13.1 & 12.6 & 12.4 & 10.5 & 11.0 & 6.91 & 7.04 & 7.06 & 5.59 & 5.30 & 3.62 & 4.04 & 2.82 & 2.72 & 1.29 & 1.94 & 2.27 & 2.74 & 2.01 & 1.72 &-- &  & &&&&&&&&&&&\\ 
Jun & 6.74 & 7.05 & 6.77 & 7.08 & 4.38 & 5.00 & 3.76 & 3.99 & 4.53 & 4.05 & 3.63 & 3.19 & 4.29 & 3.06 & 2.83 & 2.26 & 2.59 & 2.64 & 2.61 & 3.55 & 2.16 & 2.08 & 1.82 & 1.96 & 3.17 & 2.10 & 2.45 & 3.87 & 3.02 &-- &  & &&&&&&&&  \\ 
July & 10.2 & 10.0 & 10.1 & 10.0 & 9.98 & 9.84 & 8.64 & 10.1 & 10.1 & 8.14 & 8.87 & 7.21 & 7.78 & 5.83 & 5.30 & 6.24 & 6.12 & 5.39 & 5.19 & 6.22 & 4.94 & 5.45 & 4.30 & 4.56 & 5.12 & 5.05 & 5.17 & 4.48 & 5.27 & 2.23 &-- &  & &&&&&&&\\ 
Aug & 18.0 & 17.8 & 17.8 & 17.8 & 17.6 & 17.5 & 16.5 & 17.3 & 17.5 & 17.6 & 17.5 & 16.0 & 16.3 & 15.0 & 13.6 & 16.0 & 13.2 & 13.8 & 14.2 & 13.7 & 13.4 & 14.5 & 13.9 & 12.6 & 13.1 & 14.2 & 12.4 & 12.7 & 10.3 & 10.1 & 4.9 &-- &  & &&&&&&\\  
Sep & 14.0 & 13.9 & 13.9 & 13.8 & 13.8 & 13.7 & 12.7 & 13.6 & 13.7 & 13.0 & 13.8 & 12.5 & 13.0 & 10.3 & 10.9 & 12.3 & 10.1 & 9.73 & 9.45 & 10.3 & 8.86 & 10.8 & 9.47 & 8.12 & 9.01 & 9.73 & 9.30 & 10.1 & 5.27 & 6.87 & 1.74 & 1.78 &-- &  & &&&&&&\\ 
Oct & 25.7 & 25.4 & 25.4 & 25.4 & 25.2 & 25.1 & 24.1 & 25.0 & 25.1 & 24.9 & 25.1 & 20.8 & 23.2 & 19.0 & 19.5 & 20.8 & 18.6 & 17.6 & 16.8 & 15.2 & 15.5 & 15.2 & 12.9 & 12.2 & 12.9 & 14.2 & 13.5 & 11.9 & 9.7 & 11.7 & 8.7 & 6.06 & 2.40 &-- &  & &&&&&&\\ 
Nov & 20.8 & 20.3 & 20.4 & 20.3 & 19.9 & 19.9 & 18.5 & 19.4 & 19.7 & 19.5 & 19.6 & 16.8 & 17.7 & 13.2 & 15.0 & 16.0 & 15.1 & 14.9 & 14.1 & 13.0 & 12.9 & 12.8 & 10.0 & 12.4 & 13.2 & 13.4 & 12.0 & 11.7 & 7.12 & 9.45 & 5.59 & 4.48 & 1.99 & 1.34 &-- &  & &&&&\\  
Dec & 10.3 & 10.1 & 10.1 & 10.1 & 10.1 & 10.1 & 9.48 & 10.2 & 10.1 & 10.1 & 10.0 & 8.79 & 8.73 & 7.10 & 7.59 & 7.73 & 7.64 & 7.21 & 6.94 & 8.26 & 8.01 & 8.74 & 7.61 & 7.94 & 9.12 & 6.69 & 7.60 & 11.26 & 6.63 & 4.35 & 3.32 & 3.58 & 1.48 & 4.79 & 1.24 &-- &  & &&\\ 
\hline 
02Jan & 11.4 & 11.3 & 11.3 & 11.3 & 11.3 & 11.3 & 10.2 & 11.4 & 11.4 & 11.4 & 11.2 & 9.76 & 10.4 & 7.68 & 9.18 & 9.81 & 9.42 & 9.47 & 8.36 & 8.92 & 8.14 & 9.96 & 6.82 & 9.90 & 10.65 & 8.33 & 9.08 & 10.51 & 7.10 & 4.45 & 2.87 & 7.11 & 2.88 & 7.29 & 4.11 & 1.31 &-- &  & &&\\ 
Feb & 11.7 & 11.5 & 11.6 & 11.5 & 11.5 & 11.4 & 10.5 & 11.1 & 11.4 & 11.2 & 11.4 & 10.5 & 10.9 & 8.29 & 9.93 & 10.26 & 10.34 & 10.33 & 10.08 & 11.43 & 10.68 & 11.97 & 8.79 & 12.72 & 13.36 & 9.83 & 11.07 & 13.37 & 13.98 & 6.65 & 7.01 & 13.48 & 8.23 & 16.89 & 7.37 & 3.00 & 2.64 & -- \\ 
Mar & 6.51 & 6.51 & 6.51 & 6.51 & 6.53 & 6.52 & 6.55 & 6.65 & 6.85 & 6.95 & 6.63 & 7.28 & 7.17 & 6.07 & 5.77 & 5.21 & 5.31 & 6.31 & 5.61 & 7.44 & 8.29 & 8.24 & 7.93 & 10.08 & 11.49 & 7.40 & 8.81 & 15.41 & 13.08 & 5.35 & 8.50 & 15.61 & 11.03 & 22.17 & 16.39 & 4.91 & 7.51 & 8.81 & -- \\ 
Apr & 0.00 & 0.00 & 0.00 & 0.00 & 0.71 & 0.79 & 1.00 & 1.54 & 2.29 & 2.70 & 1.71 & 3.42 & 3.56 & 3.77 & 2.48 & 2.45 & 2.44 & 3.51 & 4.96 & 9.03 & 7.02 & 9.72 & 8.89 & 10.44 & 10.66 & 7.81 & 11.04 & 15.6 & 14.3 & 7.88 & 9.89 & 17.71 & 13.8 & 25.3 & 20.2 & 10.00 & 11.3 & 11.5 & 6.49 & --\\ 
May & 1.20 & 1.20 & 1.20 & 1.20 & 1.30 & 1.26 & 1.32 & 1.76 & 2.39 & 2.66 & 1.67 & 3.42 & 3.51 & 3.75 & 2.62 & 2.15 & 2.58 & 3.54 & 5.00 & 8.86 & 6.50 & 9.39 & 8.77 & 10.1 & 10.6 & 7.27 & 10.5 & 15.2 & 13.8 & 6.81 & 10.0 & 17.1 & 12.76 & 25.3 & 19.6 & 10.1 & 11.1 & 11.5 & 6.42 & 0.42 & -- \\ 
Jun & 0.87 & 0.87 & 0.87 & 0.87 & 1.12 & 1.17 & 1.04 & 1.56 & 2.30 & 2.51 & 1.43 & 3.38 & 3.36 & 3.64 & 2.48 & 2.07 & 2.49 & 3.34 & 4.88 & 8.76 & 6.56 & 9.29 & 8.72 & 10.2 & 10.5 & 7.14 & 10.4 & 15.3 & 14.2 & 7.04 & 10.0 & 17.5 & 13.8 & 25.2 & 19.8 & 9.92 & 11.1 & 11.4 & 6.41 & 0.14 & 0.00 & -- \\ 
   \hline
\end{tabular}}
\end{table}

\subsection{Simulations of the IBM test with $m=3$} \label{supp:IBM-higher-order}

We applied the IBM test with $m=3$ to some cases in Experiment 1 of Section~\ref{sec:numer}. The notations are the same as Experiment 1.

\begin{itemize}
\item \textit{Case 1: Different degree parameters}. We let $\widetilde{\Omega}=\widetilde{\Theta}\Pi P\Pi'\widetilde{\Theta}$, where $(\Pi, P)$ are the same as those in $\Omega$ and $\tilde{\theta}$'s are generated as follows: $\tilde{\theta}_i=\beta_n\times \tilde{\theta}_i^u/\|\tilde{\theta}^u\|$, for $1\leq i\leq n$, where $\tilde{\theta}_i^{u}\stackrel{iid}{\sim} 0.95\delta_1+0.05\delta_3$ with $\delta_a$ representing a point mass at $a$. We fix $(n, K)=(1000, 5)$ and let $\beta_n$ range from 6 to 10.5 with a step size 1.5. As $\beta_n$ increases, the network becomes less sparse. For each value of $\beta_n$, we select $b_n$ (the off-diagonal elements of $P$) such that the SNR defined in \eqref{DefineSNR} is fixed at $3.75$. 

\item \textit{Case 4: Different mixed membership vectors}. We fix $(n, K)=(1000, 2)$ and generate $(\Theta, P)$ in the same way as before (see the paragraph above Case 1 in Experiment 1 of Section~\ref{sec:numer}). We then generate ${\pi}_i\stackrel{iid}{\sim} \text{dir}(1.6,0.4)$ and $\tilde{\pi}_i\stackrel{iid}{\sim} \text{dir}(1,1)$, $1\leq i\leq n$. Let $\Omega=\Theta\Pi P\Pi'\Theta$ and $\widetilde{\Omega}=\Theta\widetilde{\Pi}P\widetilde{\Pi}'\Theta$. Let $\beta_n$ range from 6 to 15 with a step size 3, where for each value of $\beta_n$ we select $b_n$ such that the SNR is equal to 1. 
\end{itemize}

\spacingset{1.2}
\begin{figure}[tb!]
    \centering
    \includegraphics[height=7.2cm]{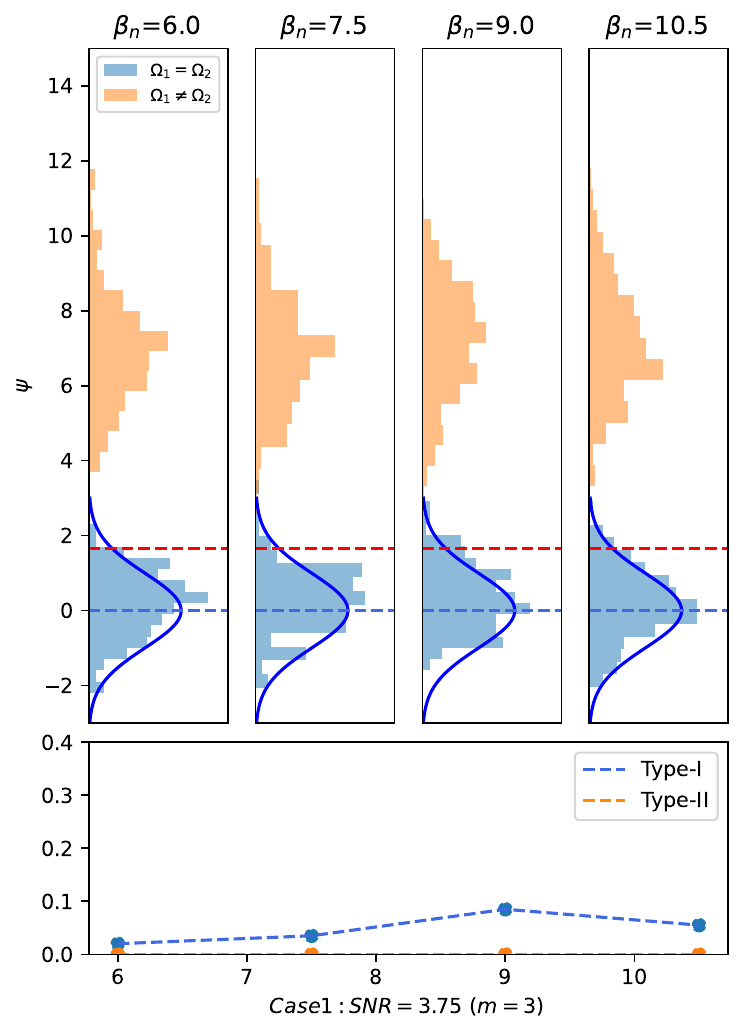}
    \includegraphics[height=7.2cm]{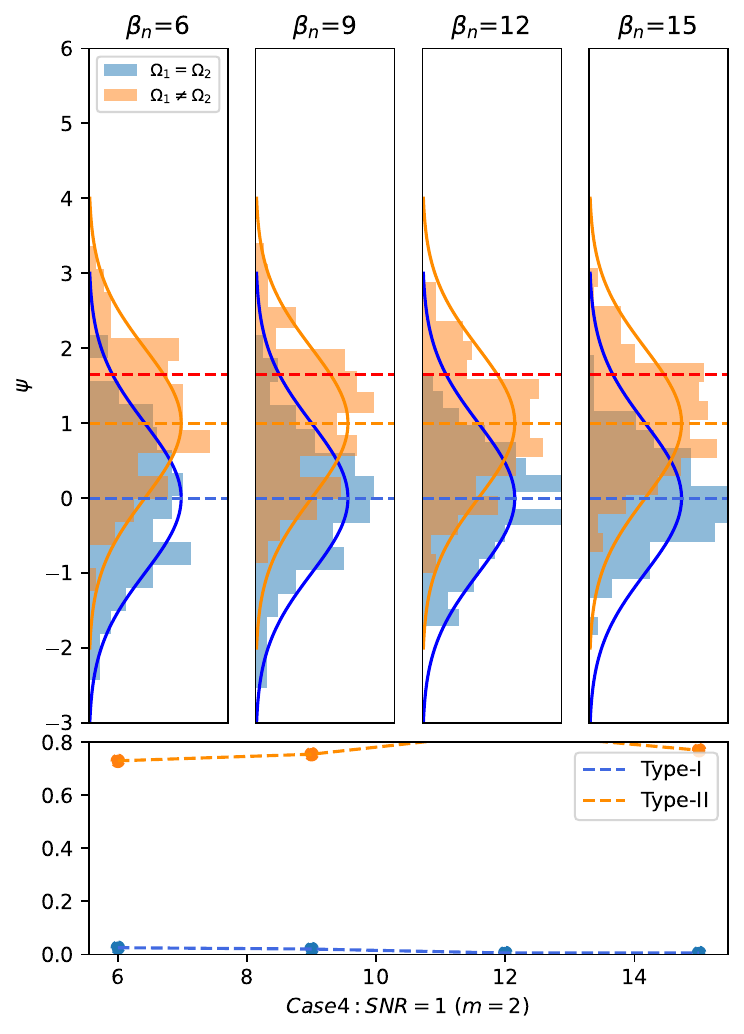}
    \includegraphics[height=7.2cm]{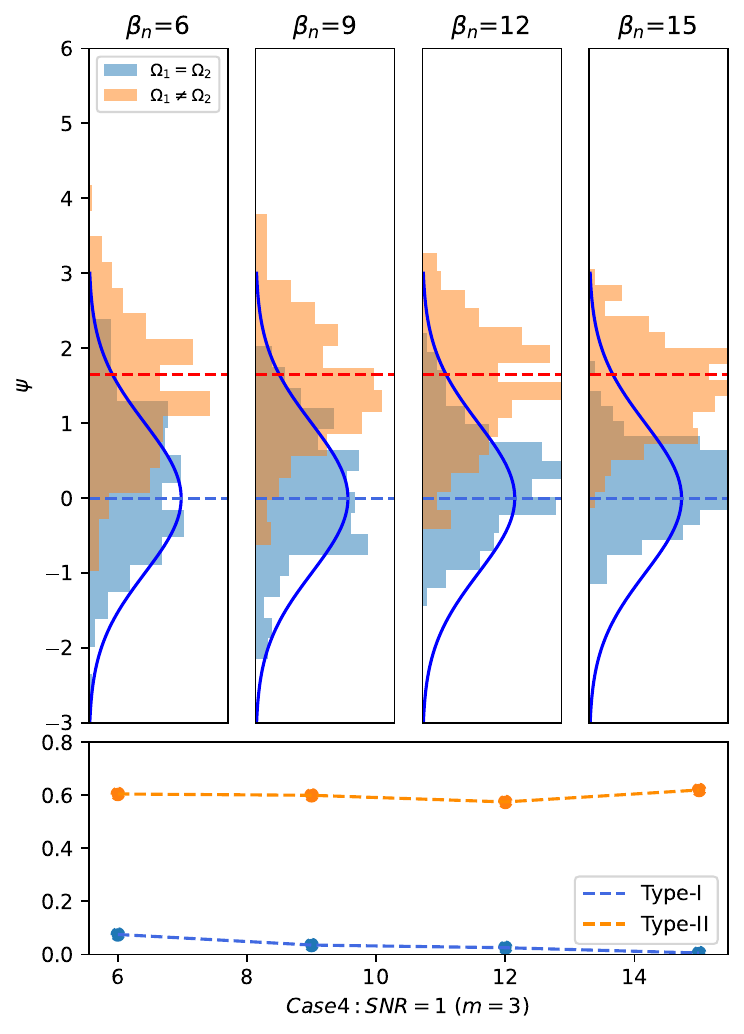}
    \caption{\small The order-3 IBM test for undirected networks, where top panels show the histograms of $\phi_n$ (see \eqref{TestStat(m=3)}) and bottom panels show the testing errors.  For each case, $\beta_n$ controls network sparsity. As $\beta_n$ varies, we keep the SNR in \eqref{DefineSNR} unchanged. The red dashed line is the cut-off of level-95\% IBM test.}  \label{fig(m=3):Exp-Undirected}
\end{figure}
\spacingset{1.45}

For Case 1, we compare the order-3 IBM test (left panel of Figure~\ref{fig(m=3):Exp-Undirected}) with the order-2 IBM test (left panel of Figure~\ref{fig:Exp-Undirected}). Both tests work reasonably well. We observe that the separation between the null and alternative histograms is a little better when $m=3$. This suggests that using a higher-order IBM test may improve power (in finite-sample).  However, $m=3$ also has disadvantages: We frequently observe that the alternative variance of the order-3 test statistic is different from its null variance (this doesn't happen for $m=2$),  and we also observe that the convergence to the limiting distribution is much slower than the order-2 test statistic. As a result, the order-2 test has a better type-I error control. 

For Case 4, the results from the order-3 and order-2 IBM tests are presented in middle and right panels of Figure~\ref{fig(m=3):Exp-Undirected}. 
We observe that the order-3 test does reduce the Type-II error rate from 0.8 to 0.6 when controlling Type-I error rate at 0.05.
However, the order-3 test has a slower convergence of the null distribution, especially when $\beta_n$ is large. 

Finally, we make a remark about the computational costs for order-2 and order-3 m=2 tests. According to Lemma~\ref{lemma:comp} and Lemma~\ref{lem(m=3):compu}, the term-by-term differences in the formula are something like computing $A^4$ versus $A^6$. And the number of terms to be evaluated also increases as $m$ increases from $2$ to $3$. In our implementation, when $n=1000$, the computing time of  $q(A)$ is 27.8ms for $m=2$ and $98.5ms$ for $m=3$. When $n=5000$, the computing time is 1.95s for $m=2$ and $ 7.95s$ for $m=3$. 

\end{document}